\documentclass[cernpreprint,atlasdraft=true,texlive=2020,UKenglish,texmf]{atlasdoc}
 
\usepackage{atlaspackage}
\usepackage{atlasbiblatex}
 
\usepackage{atlasphysics}
 
\graphicspath{{logos/}{figures/}}
 
\usepackage{ANA-TRIG-2018-08-PAPER-defs}
 
\usepackage{multirow}
\usepackage{booktabs}
\usepackage{xtab}
\usepackage{enumitem}
\usepackage{pdfpages}
\usepackage{rotating}
\usepackage{pdflscape}
\usepackage{tabulary}
\usepackage{lscape}

\AtlasTitle{Configuration and performance of the \\ATLAS $b$-jet triggers in Run 2}
 
\AtlasAbstract{
Several improvements to the ATLAS triggers used to identify jets
containing $b$-hadrons ($b$-jets) were implemented for data-taking
during Run~2 of the Large Hadron Collider from 2016 to 2018. These
changes include reconfiguring the $b$-jet trigger software to
improve primary-vertex finding and allow more stable running in
conditions with high pile-up, and the implementation of the
functionality needed to run sophisticated taggers used by the
offline reconstruction in an online environment.  These improvements
yielded an order of magnitude better light-flavour jet rejection for
the same $b$-jet identification efficiency compared to the
performance in Run~1 (2011--2012).  The efficiency to identify
$b$-jets in the trigger, and the conditional efficiency for $b$-jets
that satisfy offline $b$-tagging requirements to pass the trigger
are also measured.  Correction factors are derived to calibrate the
$b$-tagging efficiency in simulation to match that observed in data.
The associated systematic uncertainties are substantially smaller
than in previous measurements.  In addition, $b$-jet triggers were
operated for the first time during heavy-ion data-taking, using
dedicated triggers that were developed to identify semileptonic
$b$-hadron decays by selecting events with geometrically overlapping
muons and jets.}
 
\author{The ATLAS Collaboration}

\PreprintIdNumber{CERN-EP-2021-032}
 
\AtlasJournalRef{Eur. Phys. J. C 81 (2021) 1087}
\AtlasDOI{10.1140/epjc/s10052-021-09775-5}

\hypersetup{pdftitle={ATLAS document},pdfauthor={The ATLAS Collaboration}}
 
\begin{document}
 
\maketitle
 
\tableofcontents
 
\section{Introduction}
 
Techniques to identify jets containing $b$-hadrons (\bjets) are widely
used in ATLAS \cite{PERF-2007-01}, both in searches for new physics
and in measurements of Standard Model processes, including properties
of the Higgs boson.  The ability to select events containing \bjets at
the trigger level is crucial when studying or searching for processes
containing \bjets, especially those that do not provide any other
distinguishing characteristics that are easier to identify, such as
high transverse momentum (\pT) light leptons (electrons or muons) or
missing transverse momentum.  In particular, for measurements of
processes such as $HH\rightarrow b\bbar b\bbar$
\cite{HDBS-2018-18,EXOT-2016-31}, $H\rightarrow b\bbar$\ produced via
vector-boson fusion (VBF) \cite{HIGG-2016-30,Aad:2020vbr}, or
all-hadronic $\ttbar H(H\rightarrow b\bbar$) \cite{HIGG-2015-05}, or
for searches for bottom squarks \cite{ATLAS-CONF-2020-031} or
$b\phi(\phi\rightarrow b\bbar)$ \cite{HIGG-2016-32}, efficient \bjet
triggers are crucial for the success of the analyses.  In heavy-ion
collisions, heavy-flavour jets are considered to be an important
signature for understanding the flavour-dependence of radiative quark
energy loss in the quark--gluon plasma \cite{Dokshitzer:2001zm}.
 
\begin{figure}[htbp]
\centering
\includegraphics[width=0.8\textwidth]{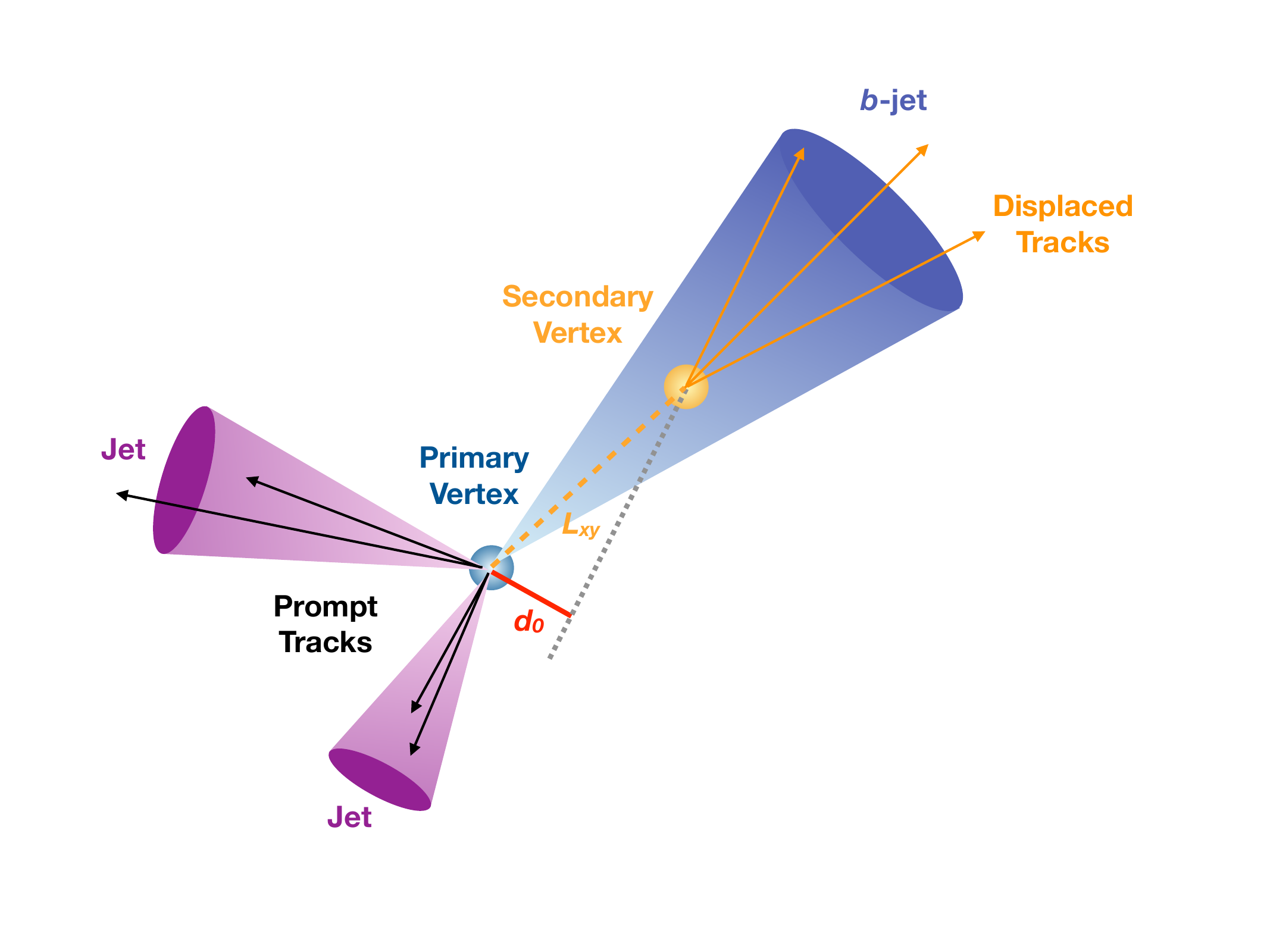}
\caption{A schematic diagram of an interaction producing two
light-flavour jets and one \bjet, shown in the transverse plane.
The lifetime of $b$-hadrons corresponds to a transverse decay
length, $L_{xy}$\ (typically a few mm), and produces displaced
tracks originating from a secondary vertex.  The distance of closest
approach of a displaced track to the primary vertex is defined as
the transverse impact parameter, $d_{0}$, and typically is large for
tracks originating from the decay of $b$-hadrons.  Conversely, jets
initiated by light-flavour quarks or gluons will not exhibit these
features and typically contain mostly prompt tracks originating from
the primary vertex.}
\label{fig:bjet}
\end{figure}

Discriminating a \bjet\ from charm ($c$) and light-flavour ($(u, d,
s)$-quark- or gluon-initiated) jets relies on exploiting the
properties of $b$-hadrons, which have a relatively long lifetime, of
the order of 1.5~ps.  This leads to a displaced (secondary) vertex,
typically a few millimetres from the hard-scatter interaction
(primary) vertex.  Tracks from the $b$-hadron decay typically have a
large transverse impact parameter, $d_{0}$, defined as the distance of
closest approach to the primary vertex in the
$r$--$\phi$\ projection.\footnote{ATLAS uses a right-handed coordinate
system with its origin at the nominal interaction point (IP) in the
centre of the detector and the $z$-axis along the beam pipe. The
$x$-axis points from the IP to the centre of the LHC ring, and the
$y$-axis points upward. Cylindrical coordinates $(r,\phi)$ are used
in the transverse plane, $\phi$ being the azimuthal angle around the
$z$-axis. The pseudorapidity is defined in terms of the polar angle
$\theta$ as $\eta=-\ln\tan(\theta/2)$.  Transverse momenta and
energies are defined as $\pt=p\sin\theta$ and $\ET=E\sin\theta$,
respectively.  Angular distance is measured in units of $\Delta R =
\sqrt{(\Delta \eta)^{2} +(\Delta \phi)^{2}}$.}  A large
longitudinal impact parameter, $z_{0}$, defined as the distance of
closest approach along the $z$-axis, is also a characteristic property
of \bjets.  Both $d_{0}$\ and $z_{0}$\ are defined to have a positive
sign if the track crosses the jet axis in front of the primary vertex
with respect to the jet direction of flight, and negative
otherwise. Additionally, $b$-hadrons can decay semileptonically
(either promptly, or via the decay of a subsequent c-hadron decay),
to electrons or muons, with a branching ratio of $\sim$20\% each, in
which case they can be characterised by the presence of a relatively
low \pt\ lepton that is geometrically matched to a jet.  A schematic
diagram of an interaction producing a \bjet plus two light-flavour
jets is shown in Figure~\ref{fig:bjet} and illustrates some of the
features that can be used to identify \bjets.
 
The identification of \bjets requires precise tracking information in
order to accurately reconstruct secondary vertices and measure the
impact parameters of tracks relative to the primary vertex.  When
\btagging is performed offline, precision tracking information is
available for the entire detector, but the CPU requirements of this
approach are prohibitively large for the trigger where the average
time per event for data retrieval and processing and data retrieval
per event must not exceed 500~ms.  Identifying \bjets in the trigger
therefore poses particular challenges, so the software is designed to
use the available resources in an optimal way in order to provide the
best possible performance.
 
The \bjet\ trigger software can be broadly considered to consist of
two steps:
\begin{enumerate}
\item{Identifying the coordinates of the hard-scatter interaction
point (primary-vertex finding).}
\item{Reconstructing secondary vertices and assessing the
probability that a given jet originated from a $b$-hadron decay
(\btagging).}
\end{enumerate}
 
Jets passing the specified transverse energy (\ET) requirements are
used as seeds to identify which regions of the detector should be
further processed in the trigger.  One \bjet trigger can make use of
several different jet-\ET\ thresholds, by using all jets with
\ET~$>$~30~\GeV\ for primary-vertex finding and variable
\ET\ thresholds for jets to be evaluated for \btagging.  Jet
reconstruction and identification in the trigger is described in
Section~\ref{sec:jets}.
 
Two different tracking configurations are used in \bjet triggers and
are presented in Section~\ref{sec:tracking}: a `Fast Tracking'
algorithm for primary-vertex finding, and `Precision Tracking' for
\btagging.  Different track-\pt thresholds (e.g.\ hard tracks for
vertexing, softer tracks for \btagging) are also required.
 
Offline algorithms are used for primary-vertex
finding~\cite{PERF-2015-01} and \btagging~\cite{FTAG-2018-01} in order
to maximise the correlation between the trigger and the offline
reconstruction, since this provides the best overall performance for
physics analyses where both components are required. In particular,
the use of the same \btagging algorithms in both the offline and
online environments significantly increases the overall efficiency for
physics analyses that depend on \bjet triggers because the same events
are more likely to be accepted both by the trigger and offline than if
different taggers are used.  The offline taggers are also the most
sophisticated taggers developed by the ATLAS Collaboration and
therefore provide the best available signal selection and background
rejection.  The \btagging of jets is described in
Section~\ref{sec:bjets}, where the performance of the \bjet triggers
is also shown.
 
ATLAS successfully used \bjet\ triggers throughout the Run~1
data-taking campaign, and several improvements to the \bjet\ triggers
were implemented during the long shutdown period (2013--2014) to
further improve performance for Run~2 (2015--2018) data-taking.  The
new \bjet triggers were commissioned during 2015, while the Run-1-style
\bjet triggers (i.e.\ those that used the same software and \btagging
algorithms as were used in Run~1 but benefited from other upgrades to
the ATLAS detector and trigger system) were the primary triggers for physics
analyses using the data taken that year.  The new triggers were
deployed online as the primary triggers from 2016 onward and these
form the focus of this paper.  The evolution of the \bjet trigger menu
(i.e.\ triggers that were run online) from 2016 to 2018 is described in
Section~\ref{sec:menu}.
 
The efficiency of the \bjet triggers is evaluated in simulation and
measured in data using the same likelihood-based
method~\cite{FTAG-2018-01} that is used to evaluate the performance
of the offline flavour-tagging.  This calibration of the \bjet
triggers and their performance relative to offline flavour-tagging is
described in Section~\ref{sec:calibrations}.
 
Specially designed \bjet\ triggers were implemented for running during
lead ion (\PbPb) collisions provided by the Large Hadron Collider
(LHC)~\cite{LHC} in 2018, to preferentially select semileptonic decays
of the $b$-hadrons, characterised by the presence of a low-\pT\ muon
matched to a jet.  This approach provided a mechanism to study
\bjets\ in \PbPb\ collisions, where the high rates and high CPU cost
of running tracking algorithms on all jets meant that it was
unfeasible to run the standard \bjet triggers.  The muon-jet triggers
used during \PbPb\ data-taking are presented in
Section~\ref{sec:muon_jets}.

\section{ATLAS detector and trigger system}

The ATLAS detector at the LHC covers nearly the
entire solid angle around the collision point.  It consists of an
inner tracking detector surrounded by a thin superconducting solenoid,
electromagnetic and hadronic calorimeters, and a muon spectrometer
incorporating three large superconducting toroidal magnets.
 
The inner-detector system is immersed in a \SI{2}{\tesla} axial
magnetic field and provides charged-particle tracking in the range
\(|\eta| < 2.5\). The high-granularity silicon pixel detector covers the vertex region
and typically provides four measurements per track, the first hit
normally being in the insertable B-layer installed before
Run~2~\cite{ATLAS-TDR-2010-19,PIX-2018-001}. It is followed by the
silicon microstrip tracker which usually provides eight
measurements per track.
These silicon detectors are complemented by
the transition radiation tracker (TRT), which enables radially
extended track reconstruction up to \(|\eta| = 2.0\).  The TRT also
provides electron identification information based on the fraction of
hits (typically 30 in total) above a higher energy-deposit threshold
corresponding to transition radiation.
 
The calorimeter system covers the pseudorapidity range \(|\eta| <
4.9\).  Within the region \(|\eta|< 3.2\), electromagnetic calorimetry
is provided by barrel and endcap high-granularity lead/liquid-argon
(LAr) calorimeters, with an additional thin LAr presampler covering
\(|\eta| < 1.8\) to correct for energy loss in material upstream of
the calorimeters.  Hadronic calorimetry is provided by the
steel/scintillator-tile calorimeter, segmented into three barrel
structures within \(|\eta| < 1.7\), and two copper/LAr hadronic endcap
calorimeters.  The solid angle coverage is completed with forward
copper/LAr and tungsten/LAr calorimeter modules optimised for
electromagnetic and hadronic measurements respectively.
 
The muon spectrometer comprises separate trigger and high-precision
tracking chambers measuring the deflection of muons in a magnetic
field generated by the superconducting air-core toroids. The field
integral of the toroids ranges between \num{2.0} and
\SI{6.0}{\tesla\metre} across most of the detector. A set of precision
chambers covers the region \(|\eta| < 2.7\) with three layers of
monitored drift tubes, complemented by cathode-strip chambers in the
forward region, where the background is highest.  The muon trigger
system covers the range \(|\eta| < 2.4\) with resistive-plate chambers
in the barrel, and thin-gap chambers in the endcap regions.
 
The trigger and data aquisition system is responsible for selecting,
processing, and storing interesting events for offline data analysis.
Events are selected using a two-stage trigger system which is
described in detail in Refs.~\cite{TRIG-2016-01} and
\cite{TRIG-2019-04}.  The first-level (L1) trigger system uses
coarse-granularity signals from the calorimeters and the muon system
with a 2.5~$\mu$s fixed latency and accepts events from the
\SI{40}{\MHz} bunch crossings at a rate below \SI{100}{\kHz}.
Regions-of-interest (RoIs) from the L1 trigger are used to define 3D
spatial regions of the detector.  The L1 trigger decision is formed by
the Central Trigger Processor (CTP), which is also responsible for
applying preventative deadtime, limiting the time between accepted
events to be within the detector read-out
latency~\cite{Bertelsen_2016}.  The peak inefficiency due to this
deadtime was approximately 1\% in Run 2.  When an event is selected by
the L1 trigger, data from the front-end electronics of all detector
subsystems is read out.  After some initial processing and formatting
of the data, events are buffered in the ReadOut System (ROS) before
being sent to the second stange of the trigger, the high-level trigger
(HLT).
 
The HLT is a software-based trigger, making use of dedicated
reconstruction algorithms to further refine the event selection
decision process.  Only the RoIs selected by the L1 trigger are
processed in the HLT, in order to minimise algorithm execution times
and computing costs.  Events accepted by the HLT are transferred to
local storage and exported to the Tier-0 facility at CERN to be fully
reconstructed offline.  An extensive software
suite~\cite{ATL-SOFT-PUB-2021-001} is used in the reconstruction and
analysis of real and simulated data, in detector operations, and in
the trigger and data acquisition systems of the experiment.

\section{Datasets and simulated events}
\label{sec:datasets}
 
The results presented here use data from proton--proton (\pp)
collisions with a centre-of-mass energy $\sqrts = 13$\,\TeV, collected
during Run~2 of the LHC, between 2016 and 2018.
 
The \bjet triggers were monitored during the on-going runs as an
early-warning alert mechanism to spot problems and improve data
quality.  Monitored variables include the jet and track
multiplicities, the primary- and secondary-vertex positions, the
variables used as inputs to the \btagging algorithms, and the output
discriminants of the taggers.  Histograms of these variables were
compared with reference histograms, using an automated evaluation
system and checked by shift-personnel in the ATLAS Control Room.  A
more in-depth evaluation of the data quality was performed offline,
soon after the data were recorded, and used as the input to a
per-luminosity-block\footnote{A luminosity block is defined as a
period of time where the instantaneous luminosity and detector
conditions can be assumed to be constant.} evaluation of the
suitability of the data for use in physics analysis.  Data quality
monitoring in the ATLAS trigger system is described in
Ref.~\cite{Aad:2020wji}.
 
Large discrepancies between data and simulation were observed in \bjet
trigger efficiencies (compared with the offline \btagging) at the
start of the 2016 data-taking campaign.  The cause of this was found
to be that the performance of the algorithm used to determine the
hard-scatter primary-vertex position depended on the nominal online
beamspot position (the centre of the region where the two proton
bunches cross in the detector).  The nominal beamspot position is
estimated online by averaging the primary-vertex position over many
events~\cite{Aad:2020wji}.  The track reconstruction in the trigger
uses the nominal online beamspot position while the online
primary-vertex position is defined relative to the detector origin,
($x=0$, $y=0$, $z=0$).  A mismatch in the handling these two
coordinate systems resulted in the online primary-vertex-finding
algorithm failing to efficiently reconstruct the vertex position in
cases where the mean $z$-position of the interaction region
($z_{\mathrm{beamspot}}^{\mathrm{online}}$) was far from the nominal
$z=0$\ origin used elsewhere in ATLAS software.  The problem was
resolved during 2016 data-taking and a \bjet-trigger-aware Good Run
List (GRL)\footnote{GRLs define the portion of the collected dataset
which is suitable for use in analyses, based upon a set of
data-quality requirements determined by each detector subsystem and
the trigger.}  is provided to reject events with
$|z_{\mathrm{beamspot}}^{\mathrm{online}}| > 2$~mm in the affected
data.  Further information is available in Ref.~\cite{LaurieThesis}.
The application of the \bjet-trigger-aware GRL reduces the integrated
luminosity of the 2016 dataset from 32.9\,\ifb\ to 24.6\,\ifb.  A more
stringent GRL is provided for use in precision measurements and
tightens this requirement so as to reject events with
$|z_{\mathrm{beamspot}}^{\mathrm{online}}| > 1$~mm, which reduces the
integrated luminosity further to 20.6\,\ifb.  In all years, luminosity
blocks at the start of each run associated with an out-of-date or
invalid beamspot position are discarded.  This additional requirement
reduces the integrated luminosity of the 2017 and 2018 datasets by
approximately 1.5\% compared with the baseline ATLAS GRL.
 
The maximum instantaneous luminosity, and therefore the average number
of \pp interactions per bunch crossing under constant beam
conditions, \pileupmu, commonly referred to as `pile-up', increased by
a factor of four during Run~2.  This information, together with the
integrated luminosity of the datasets after requiring stable beam
conditions and the \bjet-trigger-aware GRL described above, is
summarised for each year of Run~2 data-taking in Table~\ref{tab:dataset}.
Uncertainties in the integrated luminosity are
obtained using the methods discussed in Ref.~\cite{ATLAS-CONF-2019-021}
and the LUCID-2 detector \cite{LUCID2} for the primary
luminosity measurements.
 
\begin{table}[htbp]
\begin{center}
\renewcommand{\arraystretch}{1.2}
\caption{The maximum instantaneous luminosity (\lumi), the peak
pile-up (\pileupmu), the average \pileupmu, and integrated
luminosity ($\int$\lumi) per year, after applying the
\bjet-trigger-aware GRL for each year of \pp collision
data-taking.\vspace{2mm} }
\begin{tabular}{ c c c c c } \toprule
Year    & Peak \lumi\ [\lumiunit]    & Peak \pileupmu & Average \pileupmu & $\int$\lumi\ [\ifb]  \\ \midrule
2016    & 1.4 $\times 10^{34}$       & 45             & 25.1              & 24.6 $\pm$ 0.5 \\ 
2017    & 2.1 $\times 10^{34}$       & 80             & 37.8              & 43.7 $\pm$ 1.0 \\ 
2018    & 2.1 $\times 10^{34}$       & 60             & 36.1              & 57.7 $\pm$ 1.2 \\ \bottomrule 
\end{tabular}
\label{tab:dataset}
\end{center}
\end{table}
 
Monte Carlo (MC) simulations of top-quark pairs (\ttbar) produced in
\pp collisions are used throughout this paper to provide a sample of
simulated $b$-, $c$-, and light-flavour jets.  The production of
\ttbar\ events was modelled using the
\POWHEGBOX~v2~\cite{Frixione:2007nw,Nason:2004rx,Frixione:2007vw,Alioli:2010xd}
generator at next-to-leading order with the
\nnpdfnlo~\cite{Ball:2014uwa} parton distribution function (PDF) set
and the \hdamp\ parameter\footnote{The \hdamp\ parameter is a
resummation damping factor and one of the parameters that controls
the matching of \POWHEG matrix elements to the parton shower and
thus effectively regulates the high-\pt\ radiation against which the
\ttbar\ system recoils.} set to
1.5~$m_{\mathrm{top}}$~\cite{ATL-PHYS-PUB-2016-020}.  The events were
interfaced to \PYTHIA~8.230~\cite{Sjostrand:2014zea} to model the
parton shower, hadronisation, and underlying event, with parameter
values set according to the A14 tune~\cite{ATL-PHYS-PUB-2014-021} and
using the \nnpdftwo set of PDFs~\cite{Ball:2012cx}. The decays of
bottom and charm hadrons were performed by
\EVTGEN~1.6.0~\cite{Lange:2001uf}.  The \ttbar\ sample was normalised
to a cross-section of $\mathrm{832\pm51~pb}$, corresponding to the
prediction at next-to-next-to-leading order in QCD including the
resummation of next-to-next-to-leading logarithmic soft-gluon terms
calculated using
\textsc{Top++2.0}~\cite{Beneke:2011mq,Cacciari:2011hy,Baernreuther:2012ws,Czakon:2012zr,Czakon:2012pz,Czakon:2013goa,Czakon:2011xx}.
At least one top quark was required to decay into a final state with a
lepton.  Other MC processes used in the \bjet trigger efficiency
measurement and calibration described in
Section~\ref{sec:calibrations}\ are the same as those used in
Ref.~\cite{FTAG-2018-01}.
 
For certain studies (for example, the hybrid tuning described in
Section~\ref{subsec:btagging_algorithms}), a sample of high-\ET
simulated \bjets was required.  In these cases, simulated $Z^{\prime}
\to q\bar{q}$\ events are used, where the $Z'$\ boson has a mass of
1\,\TeV\ and has equal branching fractions to light-, $c$-, and
$b$-flavour quark--antiquark pairs.  The samples were generated using
\PYTHIA~8.165 with the \nnpdftwo PDF set and the A14 set of tuned
parameters.
 
The effect of multiple $pp$ interactions per bunch crossing, as well
as the effect on the detector response due to interactions from bunch
crossings before or after the one containing the hard interaction, was
modelled by overlaying the hard-scatter interactions with events from
the \PYTHIA~8.160 generator, using the \nnpdftwo PDF set
and the A3 parameter tune~\cite{ATL-PHYS-PUB-2016-017}.
Simulated events were then processed through the ATLAS detector
simulation~\cite{SOFT-2010-01} based on $\GEANT$~\cite{Agostinelli:2002hh}.
 
Jets in simulations are assigned labels based on geometric matching to
particle-level information in the MC event record.  Jets with radius
$R$ = 0.4 that are matched to a weakly decaying $b$-hadron with $\pt
\ge 5\,\GeV$ within $\Delta R = 0.3$ of the jet axis are labelled as
\bjets.  If the \bjet labelling requirements are not satisfied then
the procedure is repeated for charm hadrons and then $\tau$-leptons.
Any remaining jets are labelled as light-flavour.
 
The LHC also operates a heavy-ion physics programme, providing
lead--lead (\PbPb), and proton--lead (\pPb) collisions.  Specially
modified \bjet triggers, designed to select semileptonic $b$-hadron
decays characterised by a muon geometrically matched to a jet, were
operated during the 2018 \PbPb~run where 1.7~nb$^{-1}$ of data with a
nucleon--nucleon centre-of-mass energy $\sqrt{s_{_\text{NN}}} =
5.02\,\TeV$ and a peak luminosity of $6.2\times
10^{27}$~\lumiunit\ were collected.
\section{Trigger jets}
\label{sec:jets}
 
The \btagging of jets online (i.e.\ at the trigger level) requires
that jets must first have been reconstructed by the trigger and
required to pass a given transverse energy threshold, initially at L1,
and subsequently in the HLT~\cite{TRIG-2012-01}.  In general, only
calorimeter information is used to identify and measure the properties
of jets at the trigger level and they are characterised by their \ET.
This is in contrast to the offline environment~\cite{PERF-2015-09},
where information from the tracking detectors is available for all
jets and they are described in terms of their transverse momentum.

\subsection{L1 jet reconstruction}
 
Jets are identified by the L1 calorimeter trigger~\cite{ATLAS_L1Calo,
TDAQ-2019-01} in an $8 \times 8$\ trigger-tower cluster that
includes a $2 \times 2$\ local maximum that defines the RoI's
coordinates. Trigger towers are formed independently for the
electromagnetic and hadronic calorimeter layers with a finer
granularity of approximately $\Delta\eta\times\Delta\phi = 0.1\times0.1$
in the central $|\eta|<2.5$ part of the detector and a coarser
granularity for $|\eta|>2.5$.  The summed energy of deposits in both
the electromagnetic and hadronic calorimeters is required to pass the
minimum \ET\ requirements of a given trigger item.  Jets can be
identified at L1 out to $|\eta| = 4.9$, although usually only jets
out to $|\eta| = 3.2$\ are considered for \bjet trigger chains (and
\btagging is only run on jets out to $|\eta| = 2.5$).  For the
multi-\bjet triggers that have low \ET thresholds, jets are required
to be within the acceptance of the tracking detectors (i.e.\ $|\eta| <
2.5$) in order to lower the rates at L1.  Requirements are placed on the
L1 jets to select events for further processing in the HLT, and also
to seed HLT jet reconstruction.  A new topological trigger
(L1Topo)~\cite{TRIG-2016-01} that uses field-programmable gate arrays
(FPGAs) was installed and commissioned in 2016.  L1Topo provides the
functionality to make selections based on geometric or kinematic
matching between different L1 objects and refine the selection
criteria used at L1.
 
\subsection{HLT jet reconstruction}
 
Jets are reconstructed in the HLT using the \antikt jet clustering
algorithm \cite{anti_kT,Fastjet}.  Only jets with radius parameter $R =
0.4$\ were considered for \btagging during \pp\ data-taking, although
jets with radii of 0.2 or 0.3 were also used during the
\PbPb\ data-taking in 2018.  The calorimeter topoclusters~\cite{PERF-2014-07} that are
used as inputs to the HLT jet algorithm are reconstructed from the
full set of calorimeter cell information and calibrated at the
electromagnetic scale.  The jets then are calibrated using a procedure
similar to that used for offline jets~\cite{JETM-2018-05}, by
subtracting contributions to the jet energy from pile-up and applying
\ET- and $\eta$-dependent calibration factors derived from
simulations.

Two sets of jets are used in the \bjet trigger.  As a first step, all
jets with \ET$~>~$30\,\GeV\ are used to find the primary vertex of the
event, as described in Section~\ref{subsubsec:super_roi}.  In the second
step, RoIs are constructed for jets passing the specific
\ET\ threshold(s) of that trigger, as described in
Section~\ref{subsubsec:bjet_rois}.

\subsubsection{Super-RoI approach for primary-vertex finding}
\label{subsubsec:super_roi}
 
While the usual approach of sequentially processing individual RoIs is
acceptable in `quiet' events where only a few RoIs are selected, in
events with significant activity, e.g.\ those with large jet
multiplicities and/or higher pile-up, this approach can lead to the
same regions of the detector being processed multiple times, as
illustrated in Figure~\ref{subfig:standardROI}. In addition to the
clear downside of wasting CPU resources, this approach has the added
disadvantage of potentially biasing the primary-vertex finding
(described in Section~\ref{subsec:pv_finding}) by double-counting
tracks in overlapping regions.  An alternative approach is to consider
an amalgamation of the individual RoI constituents, with each
corresponding to a single jet, and removing any overlapping regions so
that these are only processed once (as illustrated in
Figure~\ref{subfig:superROI}).  This `super-RoI' functionality
provides a means to perform primary-vertex finding (along the
beamline) in a uniform way, regardless of the jet thresholds
fulfilled.
 
This approach was used for primary-vertex finding in the \bjet
triggers from 2016 onward, by consolidating all HLT jets with $\ET >
30\,\GeV$ and $|\eta| < 2.5$ into a super-RoI.  The individual jet RoI
constituents which constitute the super-RoI were defined with spatial
dimensions of 0.2 for the $\eta$\ and $\phi$ half-width (half of the
full width) during 2016.  In 2017 and 2018 these were reduced to 0.1
in both directions with negligible loss of \bjet trigger performance.
No constraint in the $z$-direction is applied and the RoI covers the
full range in $z$ of the detector ($\pm 225$~mm around $z = 0$).
 
\begin{figure}[htbp]
\centering \subfloat[]{
\includegraphics[width=0.48\textwidth]{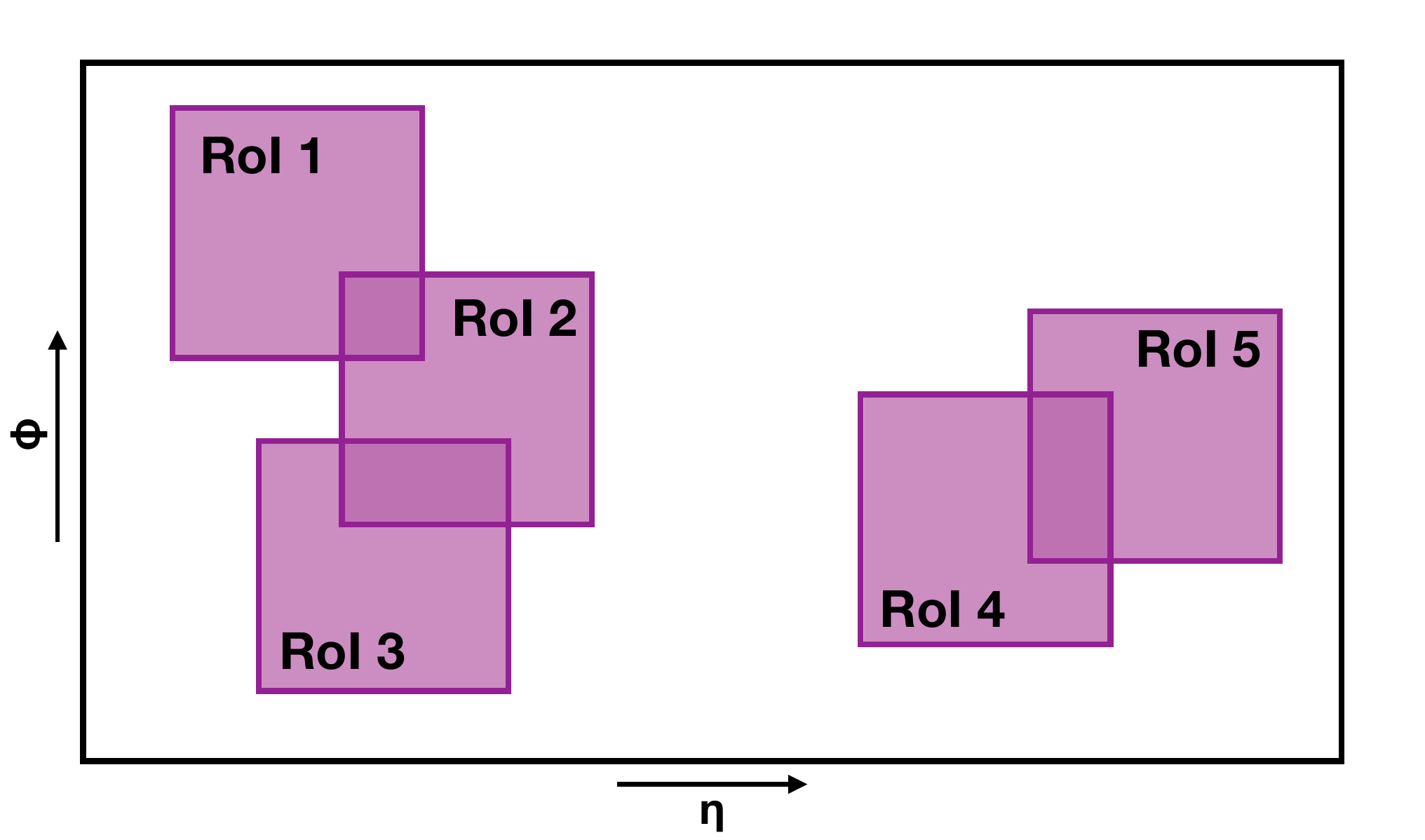}
\label{subfig:standardROI}
} \subfloat[]{
\includegraphics[width=0.48\textwidth]{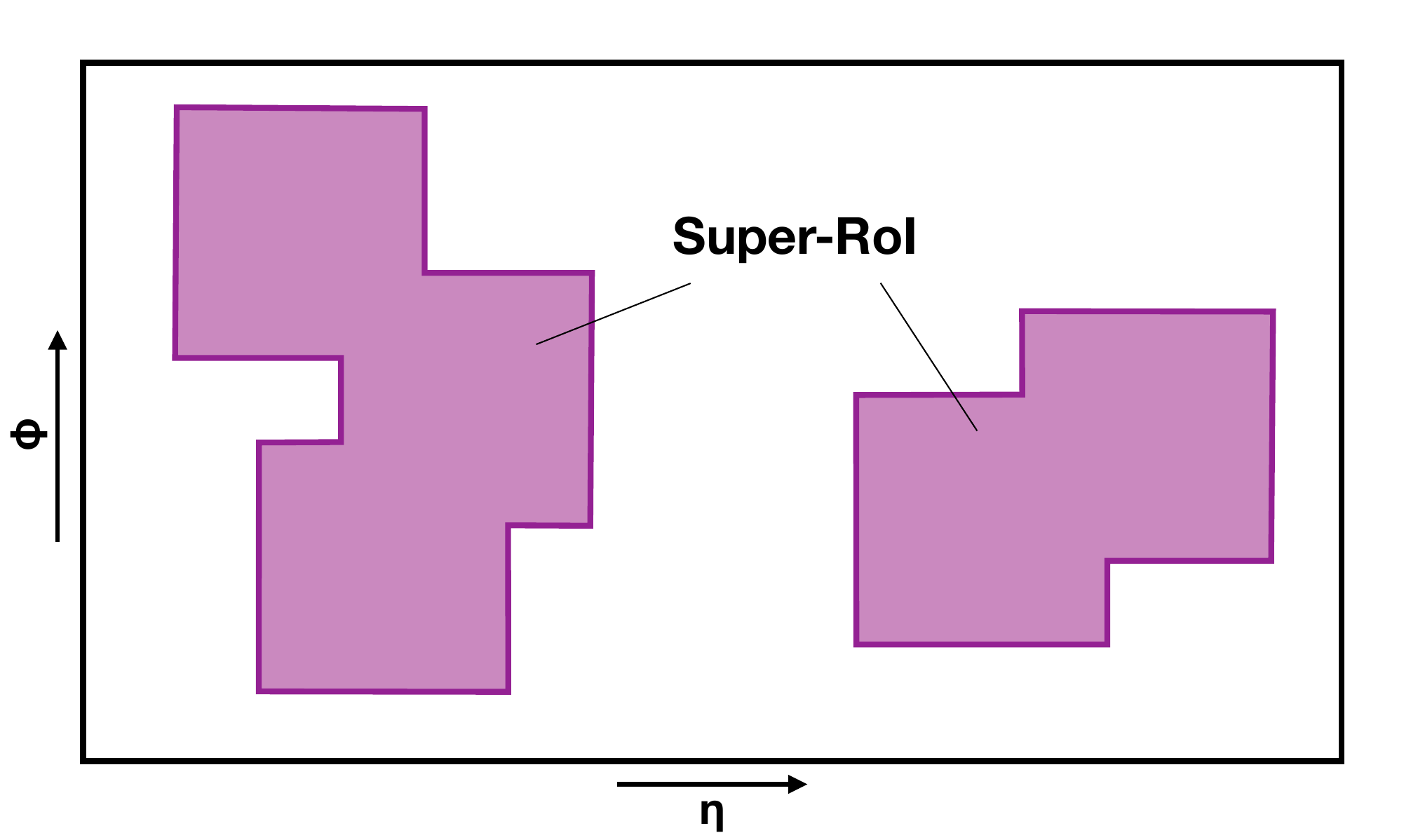}
\label{subfig:superROI}
}
\caption{A representation of the two different approaches to
processing RoIs in the detector.  In the standard approach (a),
each RoI is treated separately, resulting in overlapping regions
of the detector being processed multiple times.  In the super-RoI
approach (b) the different RoIs are amalgamated into a single complex
region of detector space, thus avoiding the problems associated
with processing the same detector region multiple times.}
\end{figure}
 
\subsubsection{RoIs for \btagging jets}
\label{subsubsec:bjet_rois}
 
The jets that will be considered for \btagging are formed from RoIs
with $|\eta| < 2.5$ and a half-width in the $\eta$\ and
$\phi$\ directions of 0.4 around the jet axis, with the apex centred
on the primary-vertex position.  A schematic diagram illustrating the
RoI defined for a single jet (passing the relevant \ET\ requirements
for each step) and used in the trigger is shown in
Figure~\ref{fig:ROIVolumes}.  The width along the $z$-direction was
conservatively constrained to be $\pm$20~mm either side of the primary
vertex during 2016, and optimised to $\pm$10~mm in 2017 and 2018 with
negligible loss of performance.  This requirement dramatically reduces
the volume that the tracking must be run on and makes the choice of an
RoI $\eta$--$\phi$ half-width of 0.4 affordable in terms of the CPU
processing time of the trigger software.  This RoI $\eta$--$\phi$
half-width of 0.4 is comparable to the radius parameter of 0.4
used for \antikt\ jets in the offline reconstruction and ensures
that the jet is fully contained within the RoI volume.  This provides
better tagging performance, particularly for softer jets, than
the $\eta$--$\phi$ half-width of 0.2 that was used for \bjet
triggers in Run~1.  Jets selected for \btagging are also required to
pass the specific \ET\ thresholds of that particular trigger.  If
these \ET\ requirements are not satisfied then the \bjet trigger
algorithms are terminated and no further processing is performed.
 
\begin{figure}[htbp]
\centering
\includegraphics[width=0.9\textwidth]{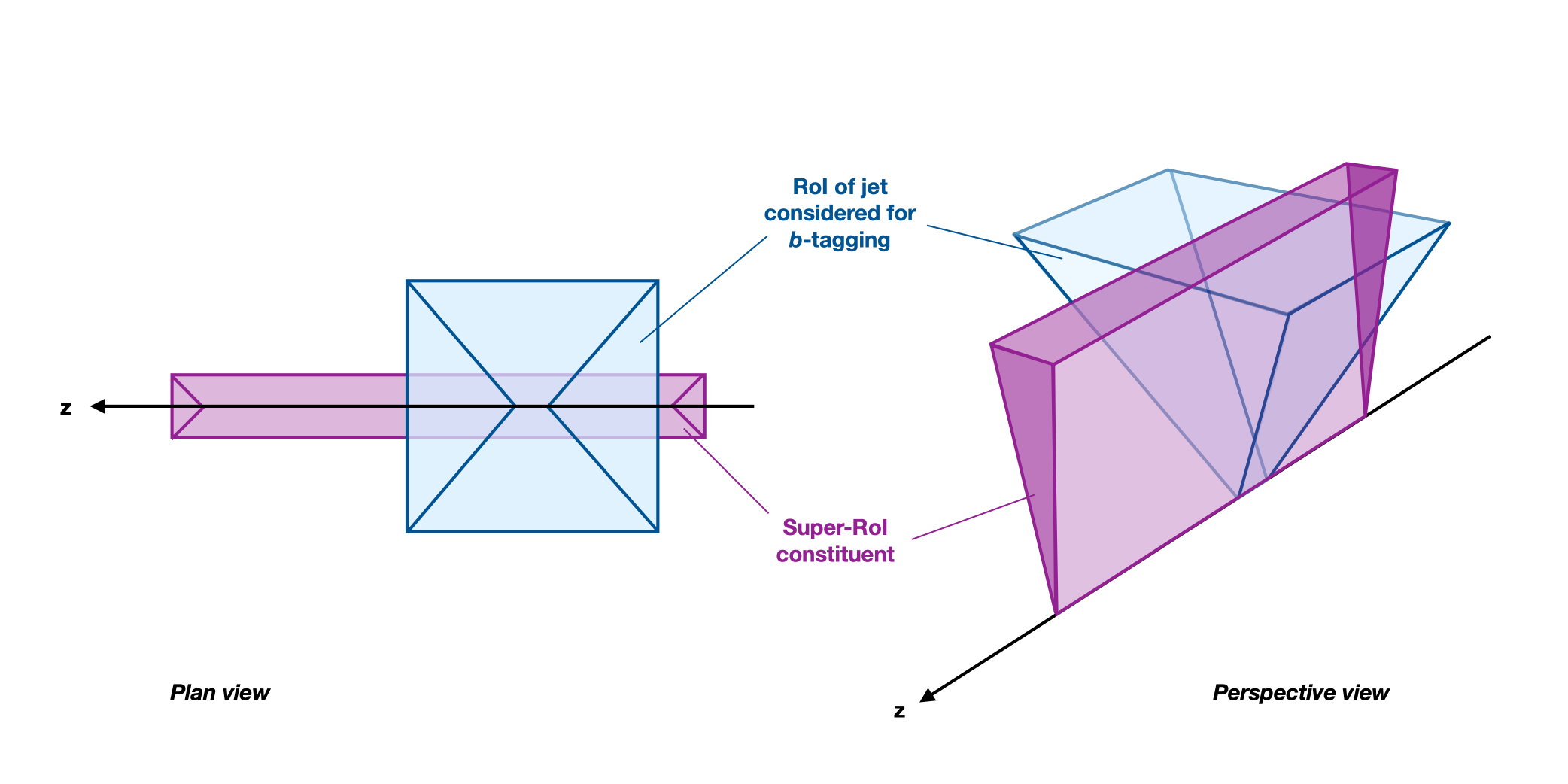}
\caption{A schematic illustration of one jet RoI that is a
constituent of the super-RoI (used for the primary-vertex-finding
stage of the \bjet trigger) and the RoI constructed around a given
jet for which the \btagging algorithms are evaulated, shown in plan view ($x,z$\ plane,
left), and in perspective view (right).  The beamline is along the
$z$-axis.  The first stage of the \bjet trigger constructs a
super-RoI from an amalgamation of RoIs that are narrow in
$\eta$\ and $\phi$ ($\eta$--$\phi$ half-width of 0.2 in 2016, and
reduced to 0.1 in 2017--2018) but cover the full range in $z$ of
the detector ($\pm 225$~mm around $z = 0$).  The second stage of
the \bjet trigger constructs RoIs that are wider in $\eta$ and
$\phi$ (0.4 half-width) but are constrained to have a half-width
in $z$ of 20~mm (10~mm) in 2016 (2017--2018).
}
\label{fig:ROIVolumes}
\end{figure}

\subsection{Global sequential jet calibration}
\label{subsec:gsc}
 
An improved jet energy calibration scheme, the global sequential jet
calibration (GSC)~\cite{PERF-2014-02,JETM-2018-05} was introduced for
2017 data-taking in order to improve the jet energy resolution in the
HLT.  The GSC uses information about the longitudinal shower shapes of
jets, and characteristics of associated tracks, to correct the energy
scale of jets.  The GSC profits from the availability of the primary
vertex and precision tracking information already provided by the
\bjet trigger (described in Section \ref{sec:tracking}).  Using the
calibrated jet \ET\ measurement from the GSC, a tighter jet selection
can subsequently be applied to the jets evaluated for \btagging in the
\bjet trigger, resulting in better efficiency turn-on curves.  The GSC
is also used to improve the trigger efficiency turn-on curves for
inclusive jet triggers.
\section{Tracking and vertex finding}
\label{sec:tracking}
 
Tracking must be run inside the RoI of HLT jets in order to find the
primary and secondary vertices, and extract information
about the jet properties, including the likelihood that they
originate from a heavy-flavour hadron decay.
 
The HLT tracking was redesigned for Run~2 in order to fully benefit
from the merging of the two stages of the high-level trigger that had been
used in Run~1~\cite{TRIG-2011-02,TRIG-2016-01,TrigID2020}.
Information about hits in the silicon detectors is extracted for each
RoI and a custom fast-tracking stage is used which generates triplets
of hits that are then used to seed track candidates. The track
candidates are then extended into the rest of the silicon detector
using the offline combinatorial track-finding tool \cite{newt}.  A
fast Kalman filter~\cite{l2kalman} is subsequently used to define
track candidates.  These steps comprise the `Fast Tracking' algorithm
that is used by the \bjet trigger for primary-vertex finding
(described in Section~\ref{subsec:pv_finding}).  These tracks
typically have a resolution of better than $\sim$100$~\mu$m for
their $z$-position along the beamline.
 
Precision Tracking is also available in the HLT.  The Fast Tracking
algorithm is run as a first step, and tracks are subsequently passed
to the offline ambiguity-solving algorithm \cite{newt}\ that (among
other functions) removes duplicate tracks, and are extended into the
TRT.  This second stage greatly improves the resolution of the track
parameters and removes many fake track candidates produced by the Fast
Tracking, which is optimised for efficiency rather than purity.  In
the \bjet trigger, the Precision Tracking is run on all jets that pass
the minimum \ET thresholds to be further considered for \btagging
(discussed in Section \ref{subsec:bTagging}).
 
\subsection{Primary-vertex finding}
\label{subsec:pv_finding}
 
Precisely determining the position of the primary vertex of the event
is the crucial first step when evaluating the probability that a jet
is a \bjet (the `\btagging weight').  Only by knowing the
primary-vertex position, can secondary vertices then be reconstructed
and evaluated to determine the final \btagging weight.
 
The Fast Tracking algorithm is run for all regions of the detector
encompassed by the super-RoI, described in
Section~\ref{subsubsec:super_roi}, and the found tracks are used as
inputs to the primary-vertex-finding algorithm.  The same iterative
primary-vertex-finding algorithm that is used
offline~\cite{PERF-2015-01} was used in the \bjet\ trigger from 2016
onward.  The algorithm looks for combinations of tracks that have
compatible $z$-positions and the primary vertex is chosen to be the
one with the highest $\Sigma p_{\mathrm{T}}^{2}$ of associated tracks.
This improves the precision with which the primary vertex is
reconstructed by approximately 10\% (in each direction) compared with
an alternative histogram-based approach used during Run~1 and in
2015~\cite{TrigID2020}.  For the histogramming approach, the
$z$-coordinate positions of all tracks in an event, relative to the
centre of the beamspot, were weighted by their \pt\ and used to
populate a histogram with a 1~mm bin width.  The centre of the most
populated bin was taken to be the primary-vertex $z$ coordinate with
the online beamspot position then used to define the $x$\ and
$y$\ coordinates.  A comparison of the performance of the
histogram-based and iterative primary-vertex-finding algorithms used
in the trigger is shown in Figure~\ref{fig:PVResolution}, which
displays the differences between primary-vertex coordinates found
online and offline in simulated \ttbar events.  The performance of
primary-vertex-finding algorithms in the trigger is presented in
detail in Ref.~\cite{TrigID2020}.
 
\begin{figure}[htbp]
\centering
\subfloat[]{\includegraphics[width=0.5\textwidth]{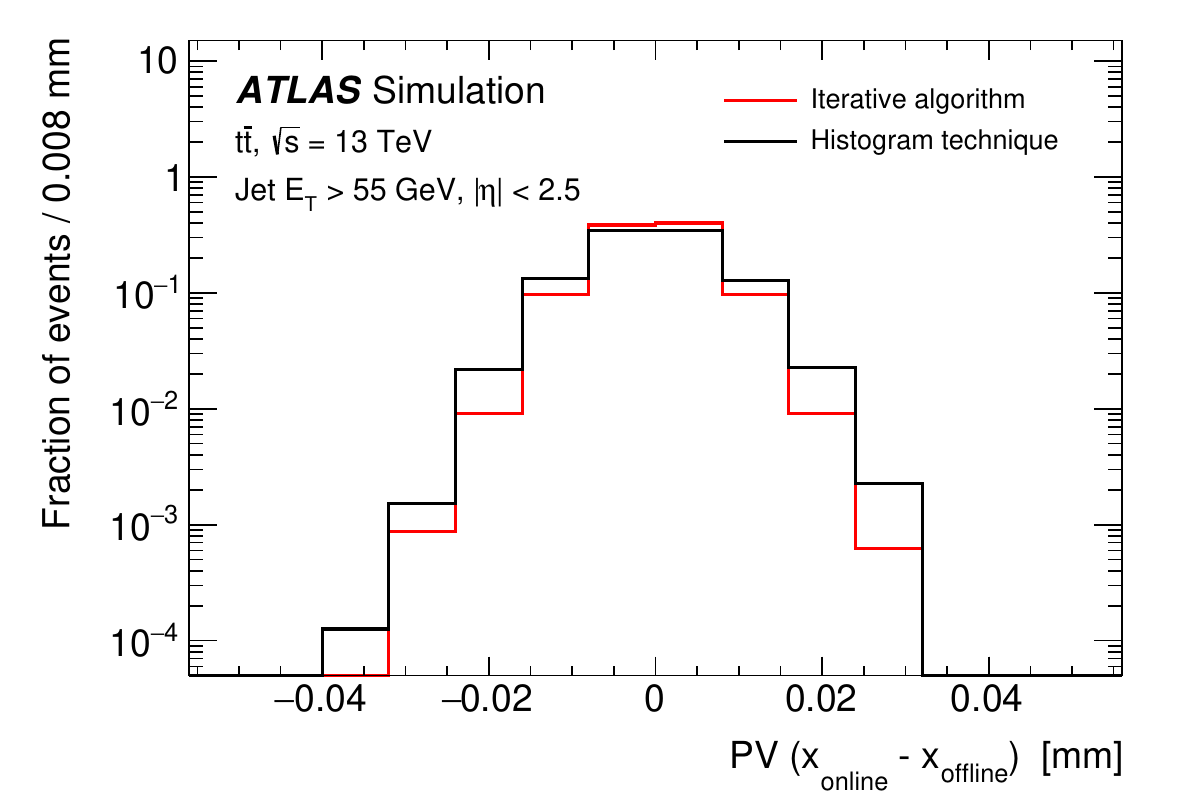}}
\subfloat[]{\includegraphics[width=0.5\textwidth]{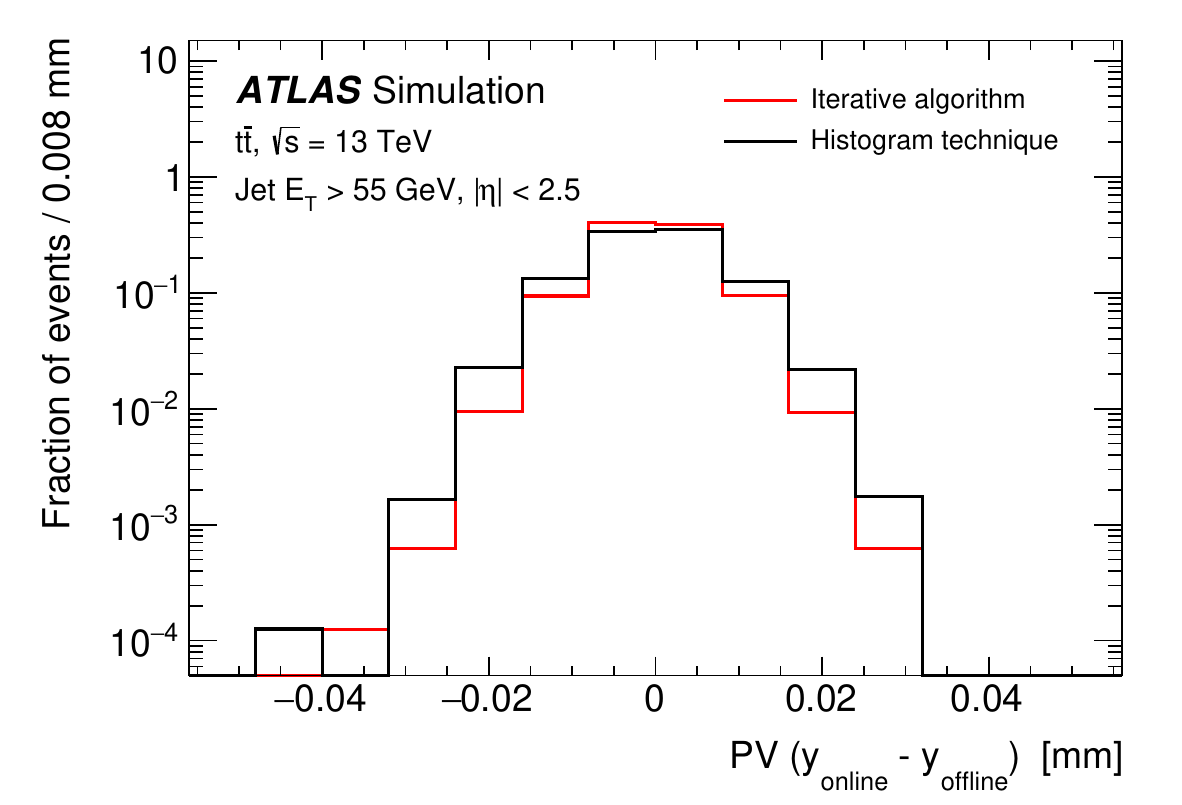}}\\
\subfloat[]{\includegraphics[width=0.5\textwidth]{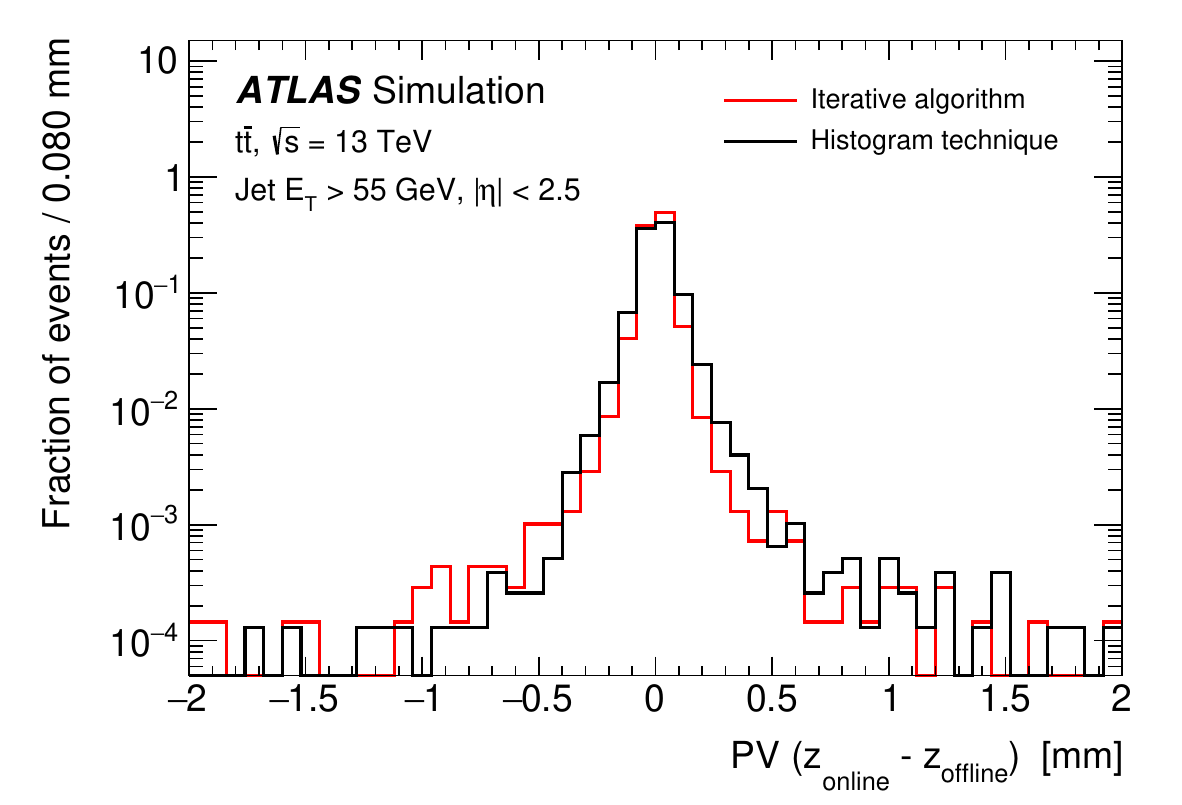}}
\caption{Comparison of the difference between the online and offline
primary-vertex positions in the (a) $x$, (b) $y$, and (c) $z$
directions when using the histogramming approach and the iterative
primary-vertex-finding algorithm in the \bjet trigger.  Selected
events must pass a trigger requiring a single jet with $\ET >
55\,\GeV$.  Tracks from all jets in the event that satisfy the Super-RoI
requirements described in Section~\ref{subsubsec:super_roi}\ are
considered as inputs to the primary-vertex-finding algorithms.}
\label{fig:PVResolution}
\end{figure}
 
In Run~1 and 2015--2016, tracks with $\pt > 1\,\GeV$\ were considered
for primary-vertex finding.  In 2017 and 2018 this threshold was
raised to 5\,\GeV, to reduce the CPU cost of primary-vertex
finding (and its associated tracking) by a factor of five, with a
negligible effect on the primary-vertex-finding efficiency or
\bjet\ trigger efficiencies.
 
\subsection{Tracking for secondary-vertex finding and \btagging}
\label{subsec:bTagging}
 
For each trigger, jets are selected for further processing if they
pass the lowest $\et$ threshold.  Precision Tracking, consisting of
the Fast Tracking plus ambiguity-solving steps, is run in the RoIs
corresponding to these jets and all tracks with $\pt > 1\,\GeV$ are
kept.  The tracks found at the primary-vertex-finding stage cannot be
reused as the Fast Tracking inputs to the ambiguity-solving step of
the Precision Tracking for \btagging, since different regions of the
detector are considered for the two stages.
 
The tracks in the RoI are used together with information about the jet
direction and the primary vertex as inputs to the \btagging algorithms
(described in Section~\ref{sec:bjets}).
 
\subsection{Tracking performance in \bjet triggers}
 
To evaluate the performance of the tracking used in \bjet triggers,
offline tracks are selected and matched to online tracks using the
procedure described in Ref.~\cite{TrigID2020}.  The efficiencies of
the Fast and Precision Tracking algorithms used in the \bjet triggers
relative to the offline tracking are shown as a function of both the
offline track transverse momentum and pseudorapidity in
Figure~\ref{fig:tracking_eff}.  The $d_{0}$\ and $z_{0}$\ resolutions
are shown in Figure~\ref{fig:tracking_res}.  Both figures show results
for the Fast Tracking within the super-RoI discussed in
Section~\ref{subsubsec:super_roi}\ that is used to find the primary
vertex, and also results for the Fast and Precision Tracking that is
used for secondary-vertex finding and \btagging within the individual
jets.  Results are produced by using dedicated `\bjet performance
triggers' that require jet \ET thresholds of 55\,\GeV\ or
150\,\GeV\ and run the full tracking and \btagging software, but do
not place any requirements on the \btagging weight of the jet.  These
provide an unbiased estimate of the tracking efficiency. Both triggers
were prescaled during the data-taking period (meaning that not every
event that satisfied the trigger requirements was recorded for further
processing).  The 150\,\GeV\ threshold trigger was run with a lower
prescale factor, and correspondingly improved statistical precision,
compared with the 55\,\GeV\ trigger, particularly at high transverse
momenta.  The data used were collected during a single run in 2018.
The average \pT\ of tracks in the RoI is correlated with the jet
\ET\ threshold of the trigger.  The 150\,\GeV\ jet trigger therefore
has a higher proportion of high-\pT\ tracks compared with the trigger
that requires a 55\,\GeV\ jet.  These differences in the track
\pt\ spectra mean that the track reconstruction efficiency at low
track \pt\ appears slightly worse in the 55\,\GeV\ trigger than in the
150\,\GeV\ trigger, as within a single bin, the former contains
relatively more tracks at low \pt\ and the efficiency of some bins is
therefore skewed by the steeply falling \pT\ distribution.  Tracks
selected by the lower \ET\ chain are therefore more sensitive to
threshold effects when performing the matching to offline tracks,
which also causes the integrated efficiency to be slightly lower.  The
$d_{0}$\ and $z_{0}$\ resolution distributions are largely insensitive
to the jet \ET\ threshold of the trigger and so are only shown here
for the data collected using the trigger with a 55\,\GeV\ threshold.
 
\begin{figure}[htbp]
\centering
\subfloat[]{\includegraphics[width=0.5\textwidth]{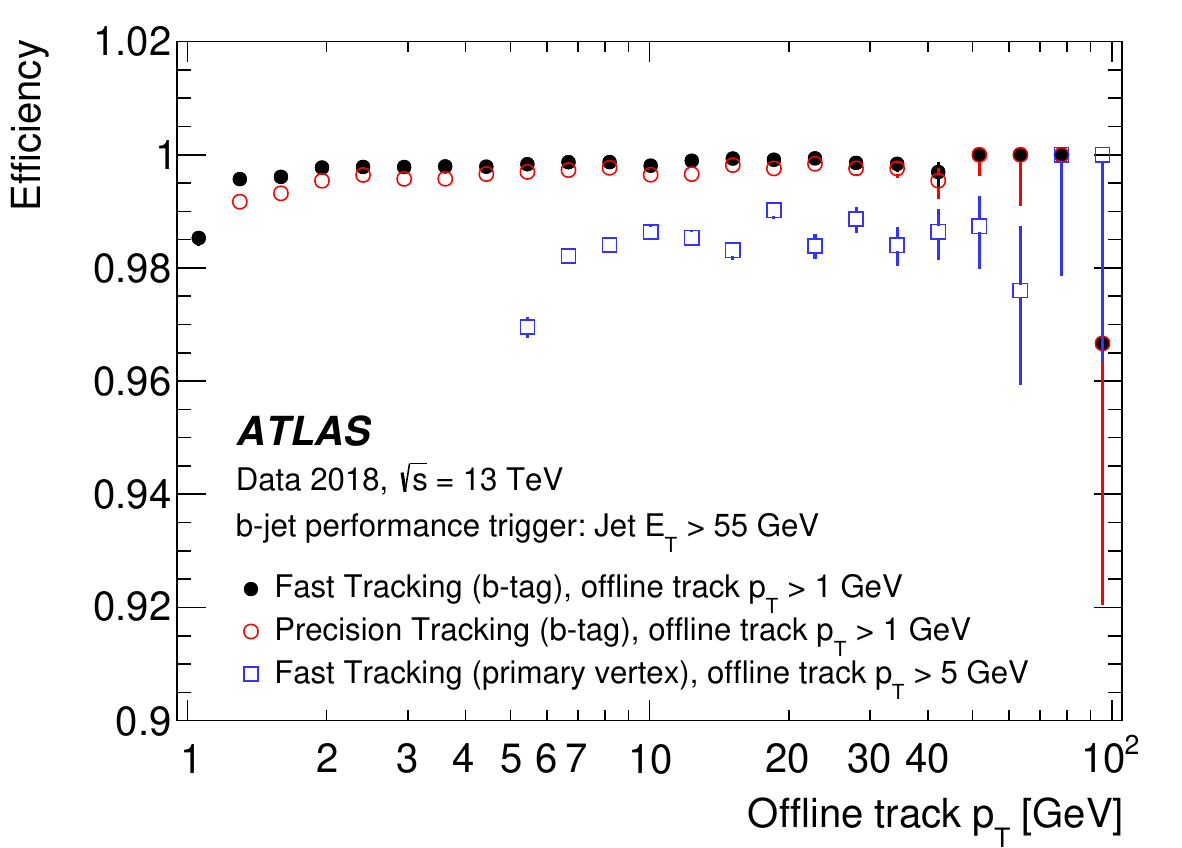}}
\subfloat[]{\includegraphics[width=0.5\textwidth]{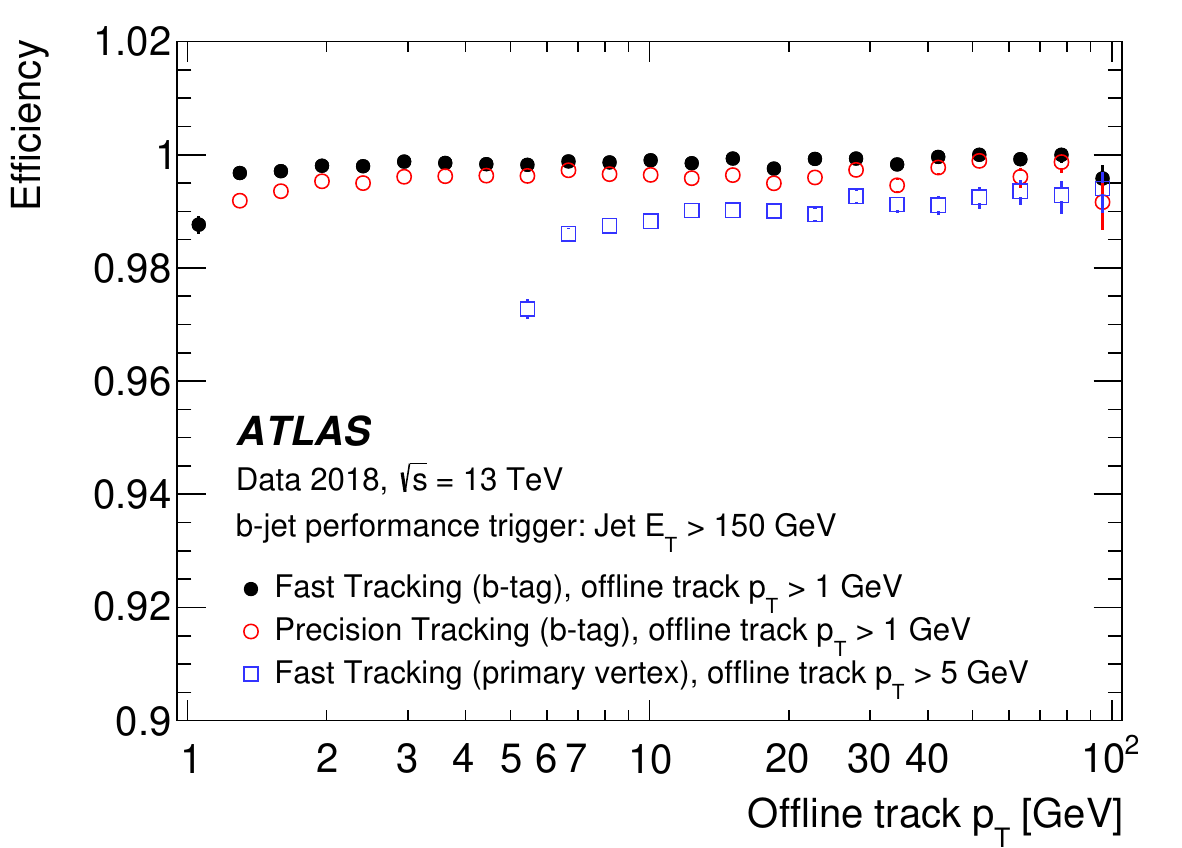}} \\
\subfloat[]{\includegraphics[width=0.5\textwidth]{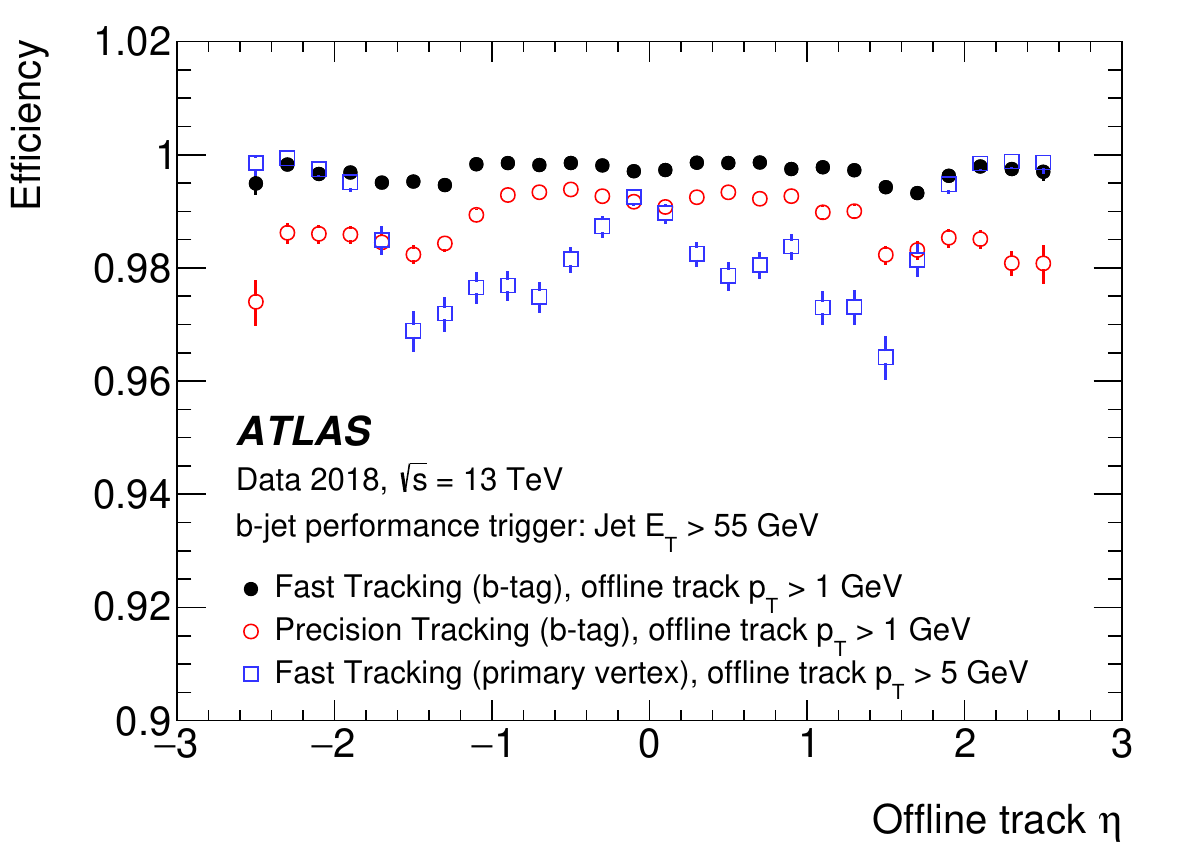}}
\subfloat[]{\includegraphics[width=0.5\textwidth]{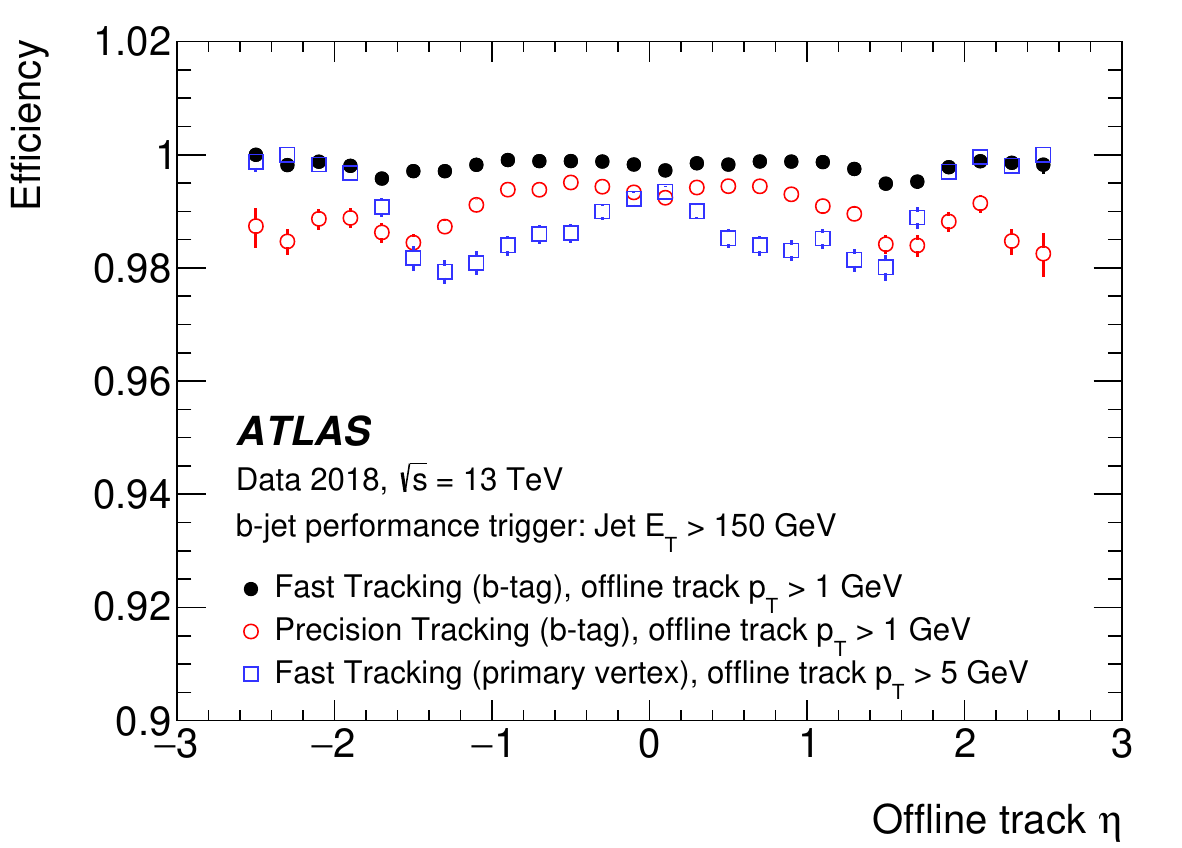}}
\caption {Efficiency of the tracking algorithms used by the \bjet
trigger, as a function of offline track \pt (top) and
$\eta$\ (bottom), for data taken during 2018 using \bjet
performance triggers with a jet \ET\ threshold of 55~\GeV\ (left)
and 150\,\GeV\ (right).  The Precision Tracking efficiency between
1\,\GeV\ and 1.2\,\GeV\ drops to 84\% due to a tight selection in
the transverse momentum of the candidates used by the ambiguity
solver, which is needed to reduce the execution time.  For that
reason, this efficiency point is not seen in the first bin in (a)
and (b).}
\label{fig:tracking_eff}
\end{figure}
 
\begin{figure}[htbp]
\centering
\subfloat[]{\includegraphics[width=0.5\textwidth]{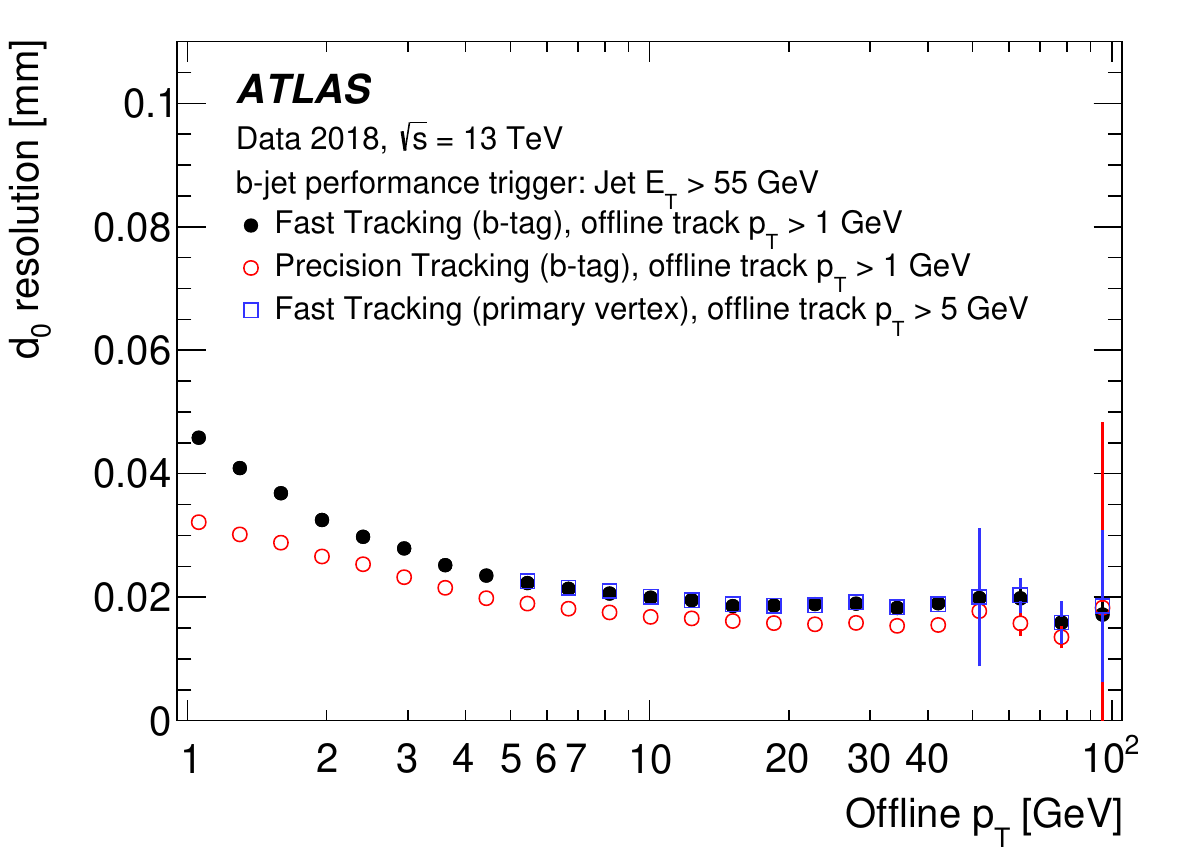}}
\subfloat[]{\includegraphics[width=0.5\textwidth]{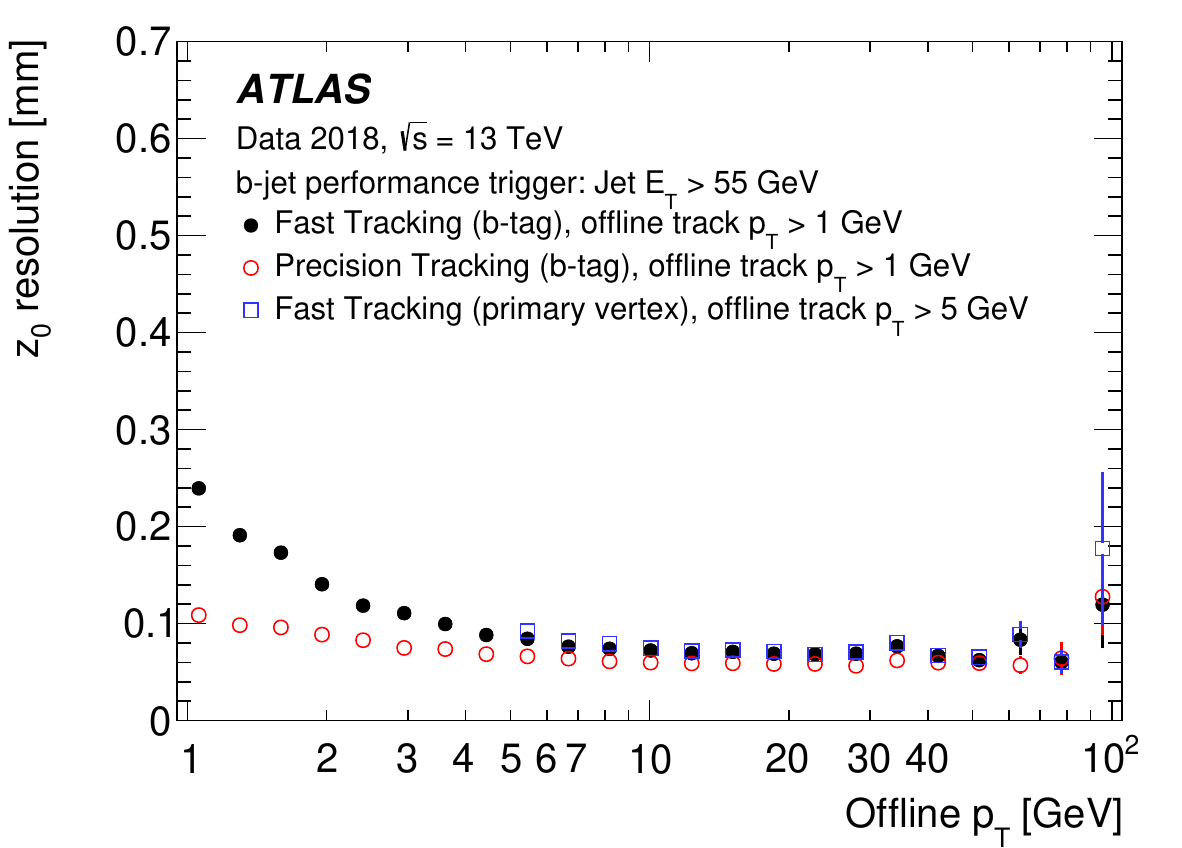}}\\
\subfloat[]{\includegraphics[width=0.5\textwidth]{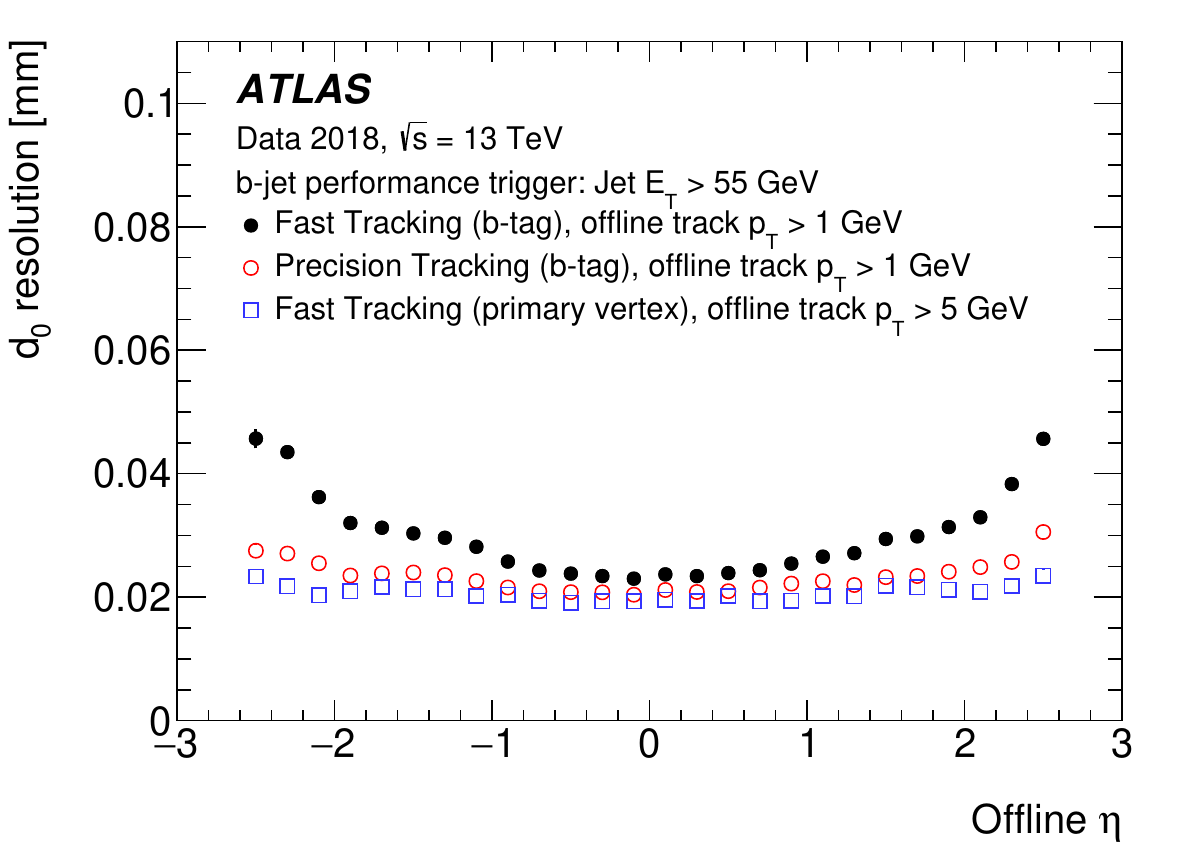}}
\subfloat[]{\includegraphics[width=0.5\textwidth]{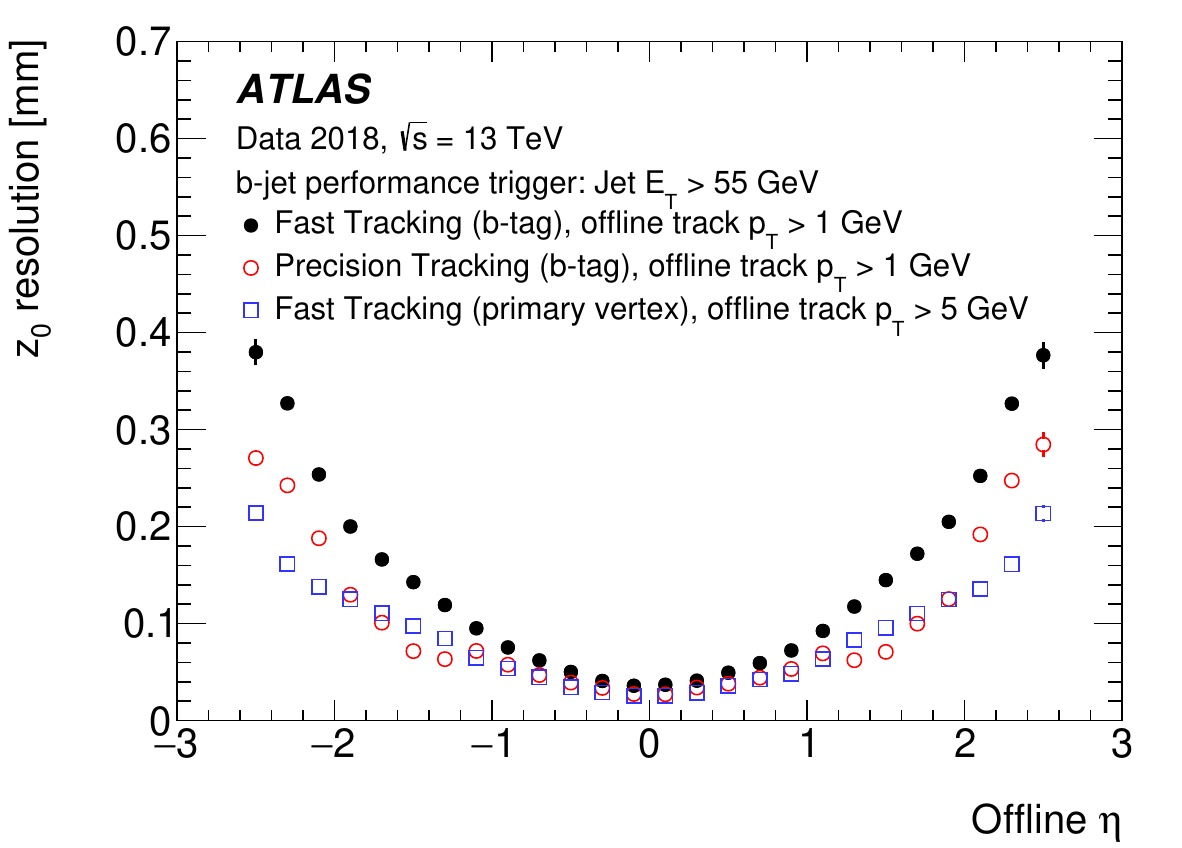}}
\caption {The $d_{0}$ resolution (left) and $z_{0}$\ resolution
(right) of the tracking algorithms used by the \bjet trigger with
respect to the offline measurement, as a function of offline track
\pt (top) and $\eta$\ (bottom), for data taken during 2018 using a
\bjet performance trigger with a jet \ET\ threshold of 55~\GeV.
Similar performance is observed when the data are selected using
triggers with a higher jet \ET\ threshold.}
\label{fig:tracking_res}
\end{figure}
 
The Fast Tracking for the primary vertex is configured only to
reconstruct tracks with \pt\ above 5\,\GeV, and so the efficiencies
and resolutions are only evaluated for offline tracks that fulfil the
same requirement. For the Fast and Precision Tracking used for the
\btagging, the efficiencies and resolutions are calculated relative to
offline tracks with transverse momentum above 1\,\GeV.  The
requirement of \pT~$>$~5\,\GeV\ applied during pattern recognition in
the Fast Tracking used for primary-vertex reconstruction means that
the track-finding efficiency is very sensitive to the track momentum
resolution around the offline track \pT\ threshold of 5\,\GeV, and
also slightly reduces the track reconstruction efficiency at higher
\pT.  Partly as a consequence of this track \pt\ threshold, the
presence of inactive pixel modules has the potential to affect the
reconstruction of a large fraction of tracks in the super-RoI
constituent; the narrowness of the individual RoIs means that the
width of the individual constituent RoIs in both $\eta$\ and
$\phi$\ may often span no more than a single module for the innermost
pixel layers.  The primary-vertex tracking at all transverse momenta
is therefore very sensitive to inactive modules in these inner layers,
and a reduction in the efficiency of up to a few percent is observed
in some regions of $\phi$.  This results in a lower overall tracking
efficiency when compared with either the Fast or Precision Tracking
when executed in a wider region of interest.  Since the purpose of the
vertex tracking is only to identify the $z$-position of the primary
vertex for the second-stage Precision Tracking, the reduced track
reconstruction efficiency does not lead to any significant performance
loss in the trigger.
 
The efficiency is generally better than 99\% at higher \pt\ but is
somewhat lower for Precision Tracking near the 1\,\GeV\ track
\pt\ threshold.  The Precision Tracking efficiency in this first bin
between 1\,\GeV\ and 1.2\,\GeV\ drops to 84\% due to a tight selection
in the transverse momentum of the candidates used by the ambiguity
solver, which is needed to reduce the execution time.  For that
reason, this efficiency point is not seen in
Figure~\ref{fig:tracking_eff}.  This reduced efficiency near the
threshold is the primary reason for the slightly lower efficiency seen
in the Precision Tracking as a function of track pseudorapidity.
 
The $z_0$\ and $d_0$\ resolutions improve at higher transverse
momenta to approximately 70~$\mu$m and 20~$\mu$m respectively, taking
the mean across the full pseudorapidity range, and with a
$z_0$\ resolution as low as 40~$\mu$m for tracks
perpendicular to the beamline. The deterioration of the tracking
resolution at large $|\eta|$ as the tracks traverse more material at
large angles can be seen clearly.  An improvement of the $z_0$
resolution by a factor of two at low \pt\ and by nearly 100~$\mu$m in the
endcap is observed for the Precision Tracking compared with the Fast
Tracking. For $d_0$\ the improvement is 10~$\mu$m at low \pt\
compared with the Fast Tracking, and is approximately 5~$\mu$m at large
\pt\ and central pseudorapidities.

\section{HLT \bjet identification}
\label{sec:bjets}
 
A schematic overview of the complete sequence of algorithms that form the \bjet
trigger is shown in Figure~\ref{fig:overview}.  The final stage of the
\bjet trigger is to assess the probability that jets that passed the
required \ET\ thresholds originated from a $b$-hadron decay.  The
output of the \btagging algorithm is evaluated for each individual jet,
and the requirements of the trigger are assessed.  If these are
satisfied, the event is kept, otherwise it is discarded.
 
\begin{figure}[htbp]
\centering
\includegraphics[width=1.0\textwidth]{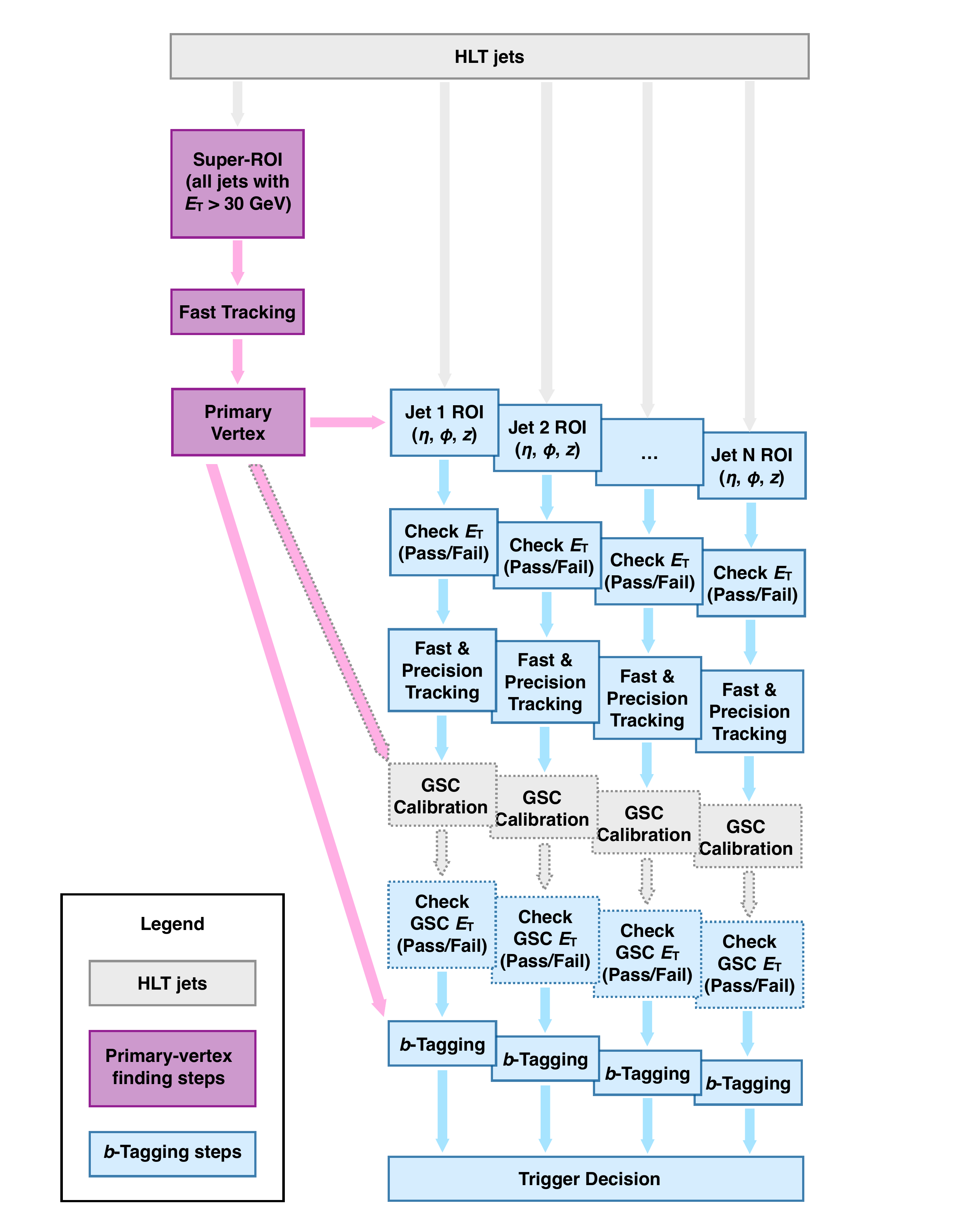}
\caption{A schematic overview of the different components of the
\bjet trigger sequence.  HLT jets (grey boxes) are used as inputs
to the primary-vertex finding (pink boxes) and \btagging of jets
that point towards the primary vertex (blue boxes).  The GSC
(dashed outline), as provided by the HLT jets (described in
Section~\ref{subsec:gsc}) can be applied as an optional step and
in this case a second requirement is placed on the jet \ET, using
the calibrated value.}
\label{fig:overview}
\end{figure}
 
\subsection{\btagging algorithms}
\label{subsec:btagging_algorithms}
 
The probability that a given jet originated from a $b$-hadron decay is
assessed by using low-level algorithms to match tracks to jets,
reconstruct secondary vertices, and identify tracks with large impact
parameters relative to the primary vertex.  The same `shrinking cone'
algorithm that is used offline~\cite{FTAG-2018-01} is employed for
matching tracks to jets.  The outputs of these low-level \btagging
algorithms are then used as inputs to multivariate algorithms that
provide excellent discrimination between \bjets and light-flavour jets
or $c$-jets.
 
Four low-level algorithms that exploit different
features of $b$-hadron decays are used in ATLAS:
\begin{itemize}
\item{IP2D: Uses the signed transverse impact parameter significance
(defined as $d_{0}/\sigma_{d_{0}}$, where $\sigma_{d_{0}}$\ is the
uncertainty on the reconstructed $d_{0}$) of tracks associated
with a jet~\cite{ATL-PHYS-PUB-2017-013}.  Reference histograms
derived using MC simulations provide probability density functions
that are used to calculate the probabilities that a given track
originated from a $b$-jet, $c$-jet, or light-flavour jet.  The
ratios of the per-track probabilities for each jet-flavour
hypothesis are calculated, and their logarithms summed for all
tracks to provide a per-jet probability of the jet's flavour
origin.  Three separate discriminants are defined, separating
$b$-jets from light-flavour jets, $c$-jets from light-flavour
jets, and $b$-jets from $c$-jets.}
\item{IP3D: Uses a log-likelihood-ratio discriminant similar to
those in IP2D, but uses both the transverse and longitudinal
signed impact parameter significances to construct the track flavour
origin probability density functions~\cite{ATL-PHYS-PUB-2017-013}.
The longitudinal impact parameter significance is defined as
$z_{0}/\sigma_{z_{0}}$, where $\sigma_{z_{0}}$\ is the uncertainty
on the reconstructed $z_{0}$.}
\item{SV1: Creates two-track secondary vertices for all combinations
of tracks associated with the jet~\cite{ATL-PHYS-PUB-2017-011}.  The
secondary vertices are identified using a Kalman filter
\cite{Kalman} that uses the Billoir method \cite{Billoir}.  Tracks
compatible with decays of long-lived particles (\KzeroS\ or
$\Lambda$), photon conversions, or hadronic interactions with the
detector are rejected.  The algorithm iterates over all of the
two-track vertices, trying to fit a single secondary vertex.  At
each iteration the fit is evaluated using a $\chi^2$\ test, and
the track with the largest $\chi^2$\ is removed.  The fit
continues until the secondary vertex has an acceptable $\chi^2$,
and the invariant mass of the track system associated with the
vertex is less than 6\,\GeV.  Discriminating variables are
used as inputs to the higher-level taggers.  When used
as a stand-alone \btagging\ algorithm, the secondary-vertex mass,
the ratio of the sum of the transverse momenta (\pt) of tracks
associated with the secondary vertex to the sum of the \pt\ of all
tracks in the jet
($\Sigma(\pt^{\mathrm{SV\,tracks}})/\Sigma(\pt^{\mathrm{All\,tracks}})$),
and the number of two-track vertices are used to determine
probability density functions for each jet flavour hypothesis.
The probabilities are used as inputs to log-likelihood-ratio
discriminants that separate $b$-jets from light-flavour jets,
$c$-jets from light-flavour jets, and $b$-jets from $c$-jets.  }
\item{JetFitter: Exploits the topology of the $b$/$c$-hadron decay
chain ($b \rightarrow c \rightarrow X$) inside jets and uses a
Kalman filter to find a common line consistent with the primary,
$b$-hadron decay, and $c$-hadron decay vertices~\cite{ATL-PHYS-PUB-2018-025}.
The $b$-hadron flight path and vertex positions are approximated,
and with this approach it is possible to resolve the $b$- and
$c$-hadron decay vertices, even in cases where there is only a
single track associated with them. }
\end{itemize}
 
The final \btagging discriminant used during Run~1 and 2015 was based
on the output of the IP3D and SV1 taggers, which were combined into a
final weight and referred to as `IP3D+SV1'.  From 2016 onward it was
possible to deploy the MV2 \btagging algorithm (described in detail in
Ref.~\cite{FTAG-2018-01}) that was developed for offline
flavour-tagging in ATLAS, in the online environment.  MV2 combines the
outputs of the low-level IP2D, IP3D, SV1 and JetFitter algorithms into
a boosted decision tree (BDT).
 
The transverse and longitudinal track impact parameters and their
corresponding significances are key inputs to all of the \btagging
algorithms described above and are shown in
Figure~\ref{fig:tagger_d0_z0}\ for light-flavour jets and \bjets, when
computed online and offline. Distributions of selected jet-level
variables related to the IP3D, SV1 and JetFitter \btagging algorithms
are shown in Figure~\ref{fig:tagger_inputs}.  The distributions are
shown for jets with $\ET > 55\,\GeV$ and $|\eta| < 2.5$ in simulated
\ttbar events.  Good separation between light-flavour jets and \bjets
is observed.  The differences in the distributions between HLT and
offline quantities clearly motivate the necessity of reoptimising and
retraining the multivariate algorithms for the online environment, and
substantially improved performance is observed with dedicated
reoptimisations.
 
\begin{figure}[htbp]
\centering
\subfloat[]{\includegraphics[width=0.35\textwidth]{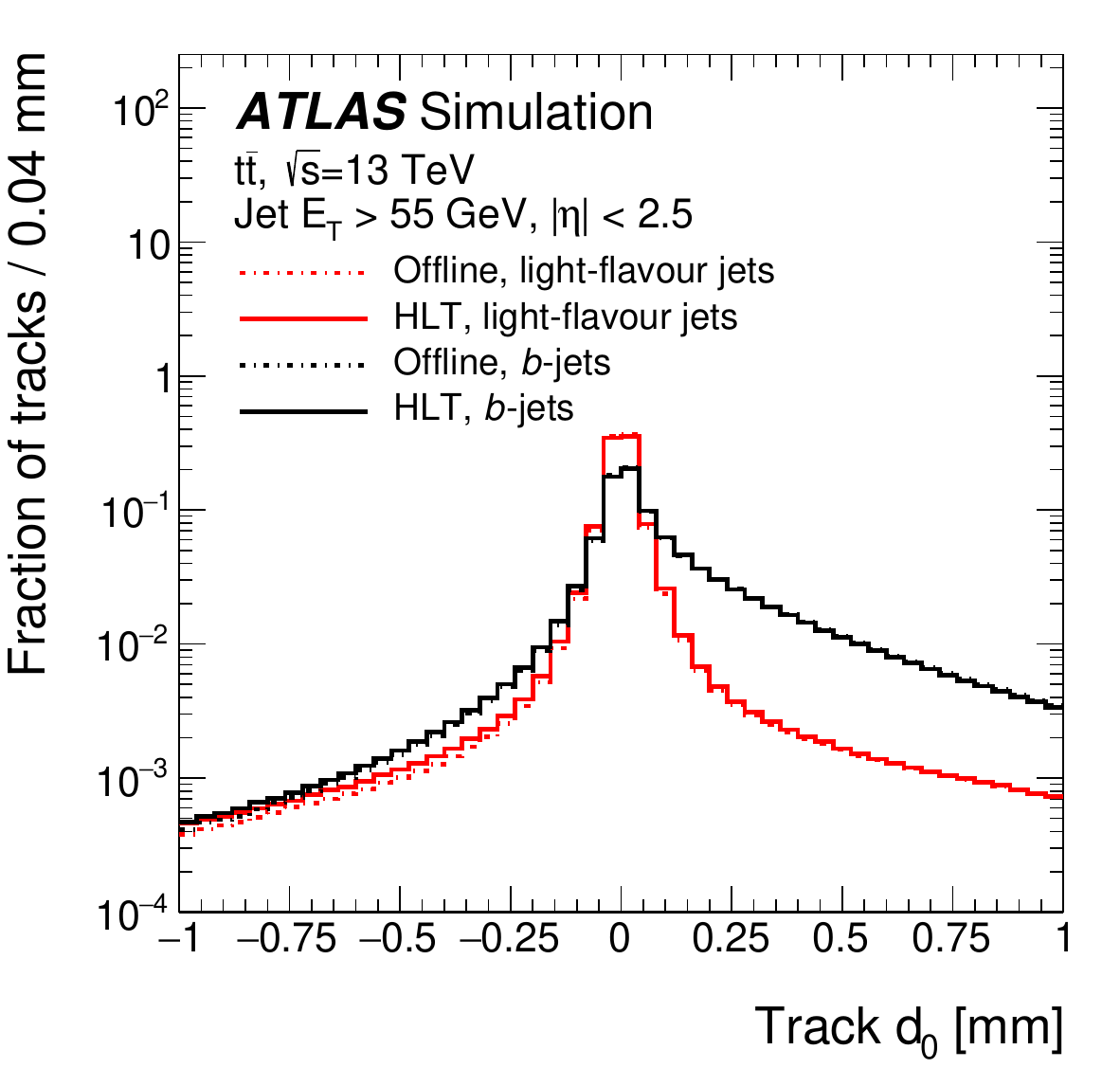}}
\subfloat[]{\includegraphics[width=0.35\textwidth]{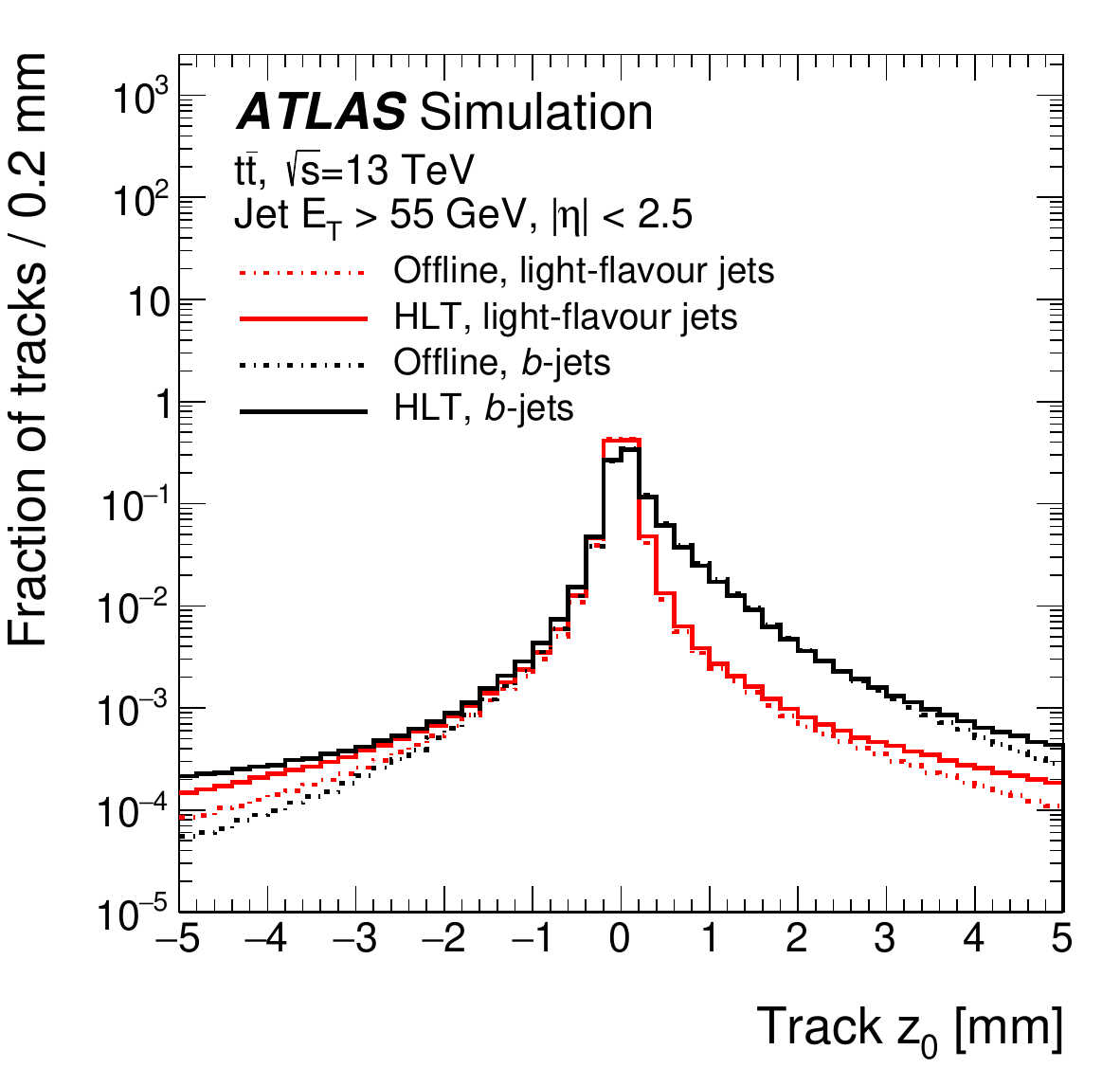}}\\
\subfloat[]{\includegraphics[width=0.35\textwidth]{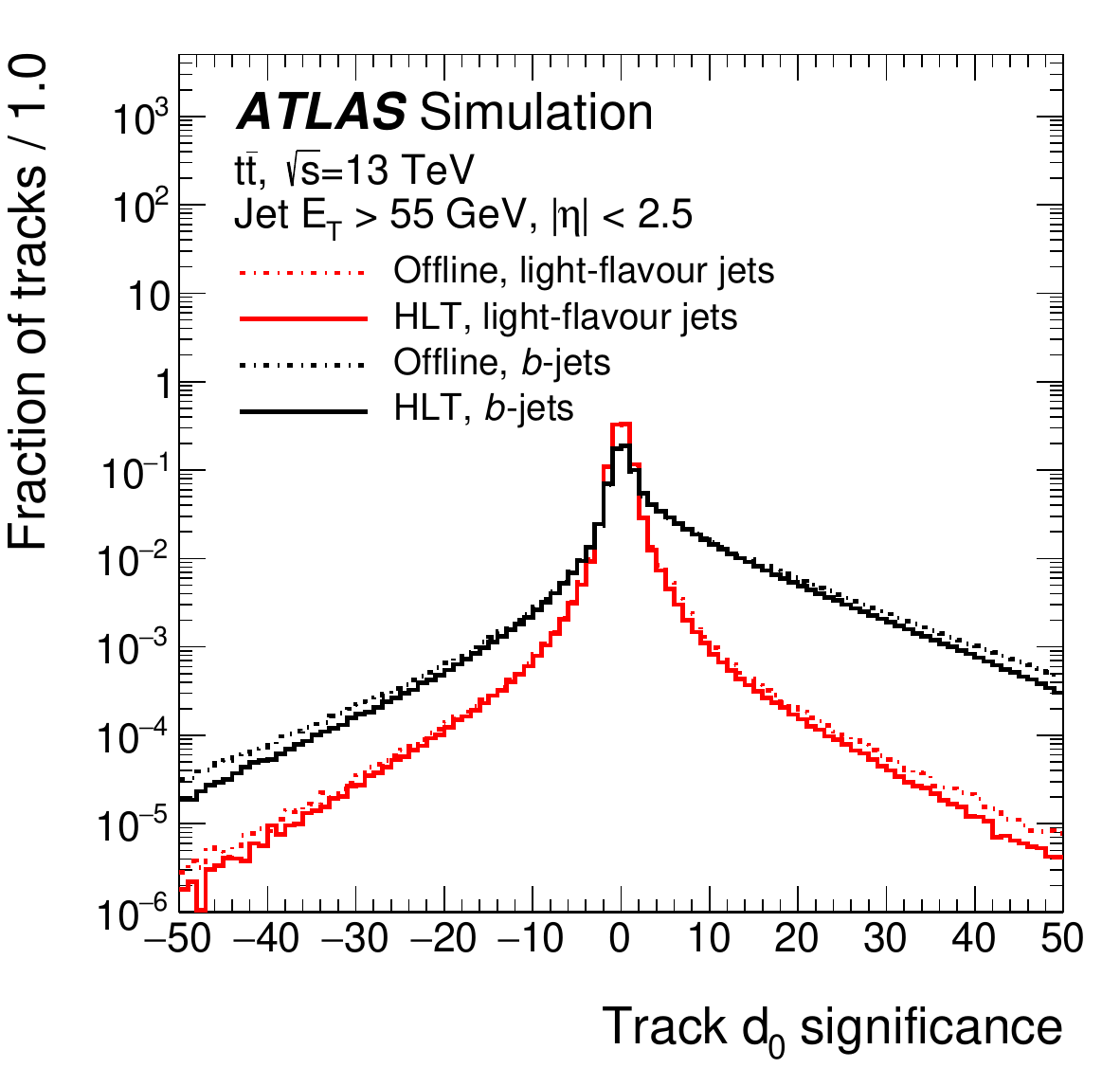}}
\subfloat[]{\includegraphics[width=0.35\textwidth]{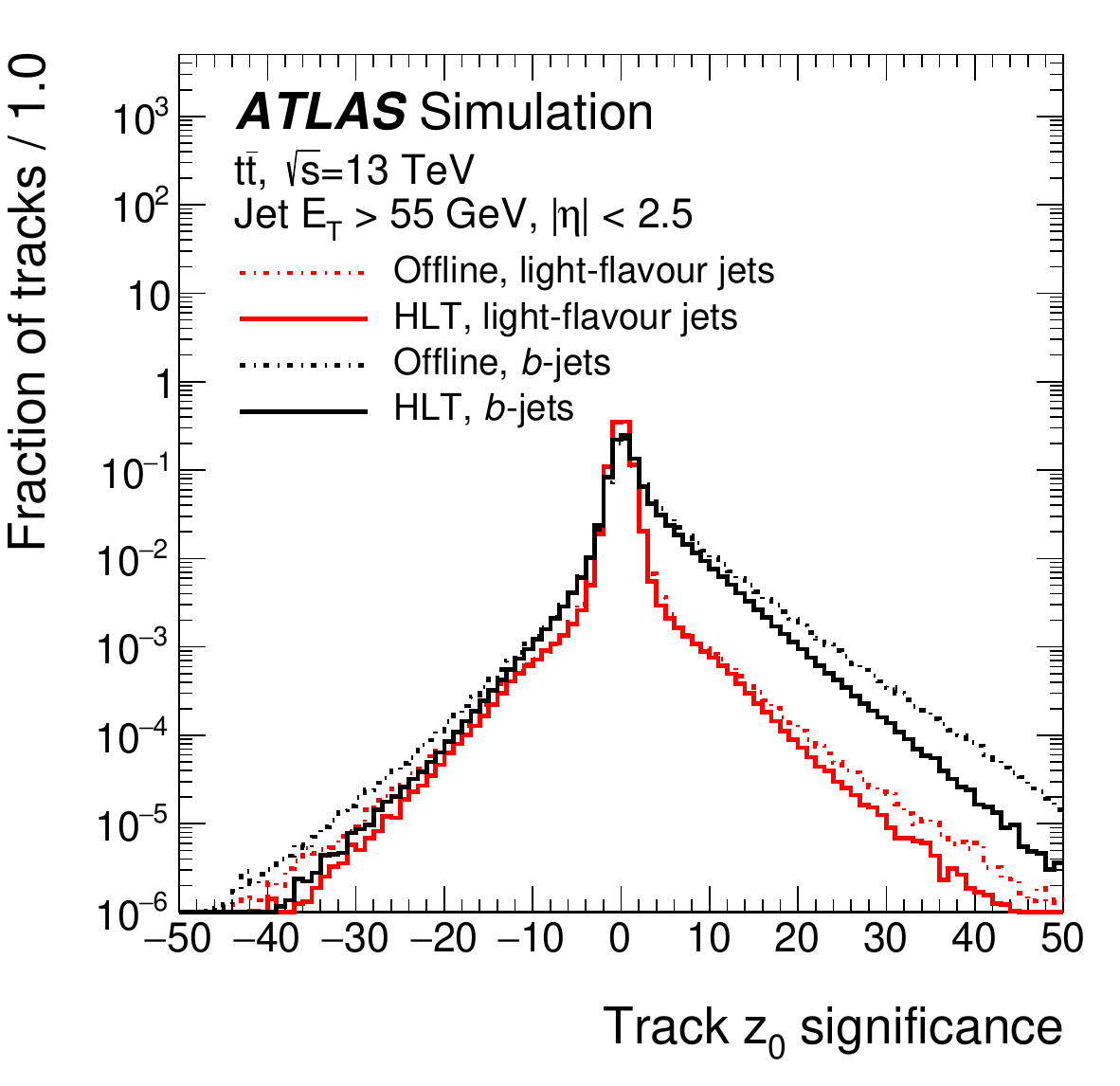}}
\caption {The (a) transverse, and (b) longitudinal track impact
parameters, and (c,d) their corresponding significances for $b$-jets
and light-flavour jets.  The distributions show both the online and
offline versions of these quantities.}
\label{fig:tagger_d0_z0}
\end{figure}
 
\begin{figure}[htbp]
\centering
\subfloat[]{\includegraphics[width=0.35\textwidth]{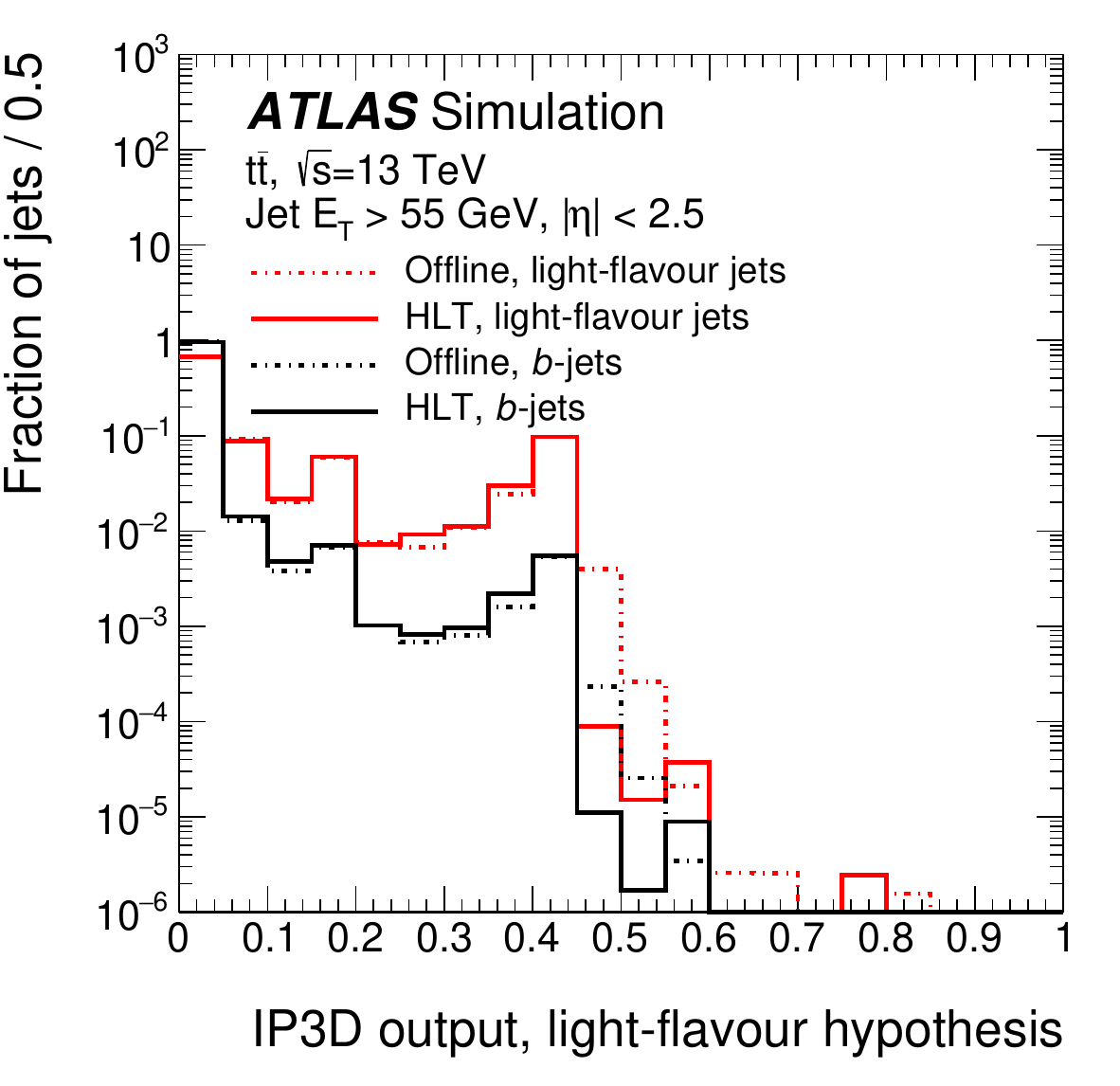}}
\subfloat[]{\includegraphics[width=0.35\textwidth]{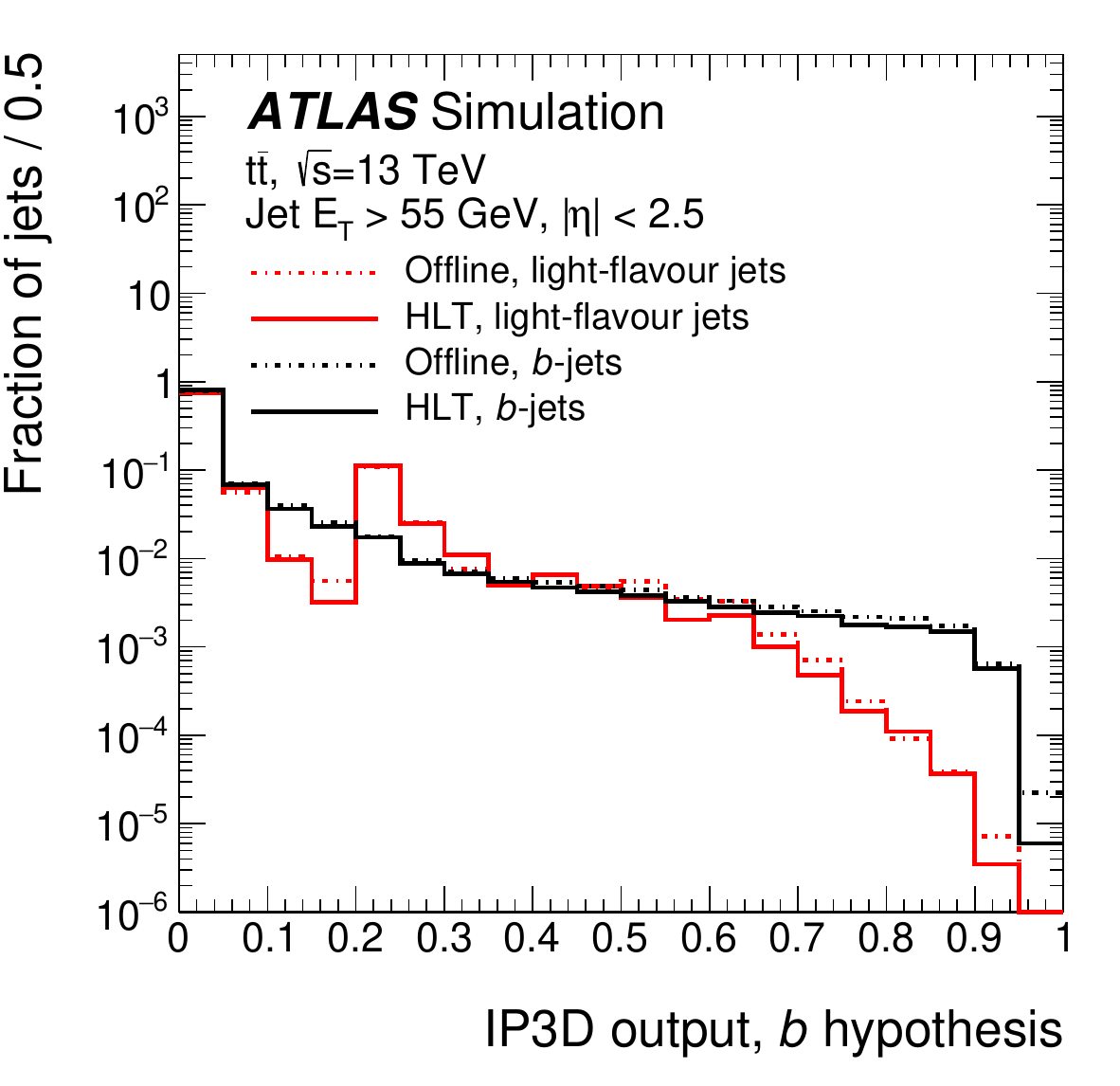}}\\
\subfloat[]{\includegraphics[width=0.35\textwidth]{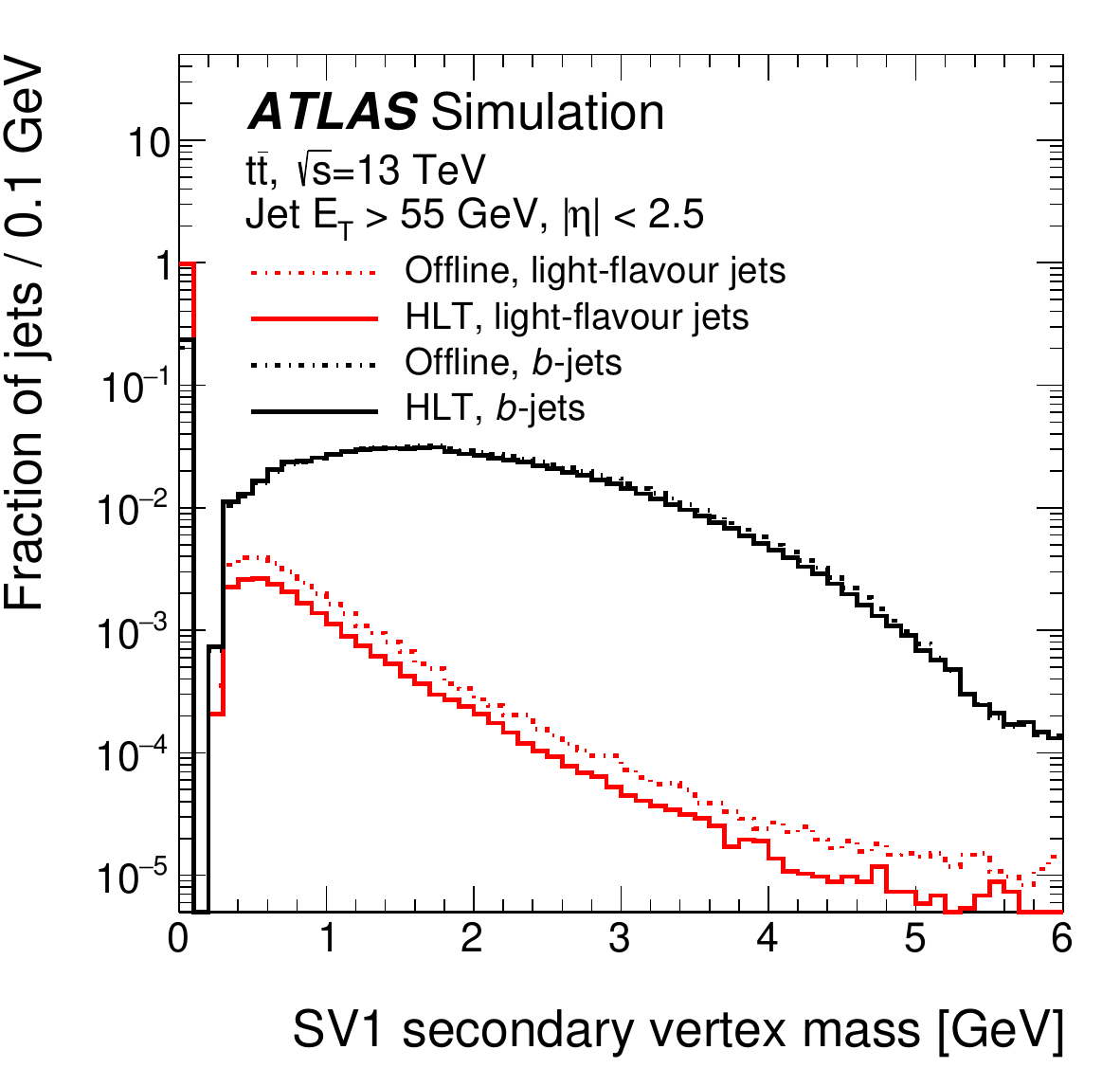}}
\subfloat[]{\includegraphics[width=0.35\textwidth]{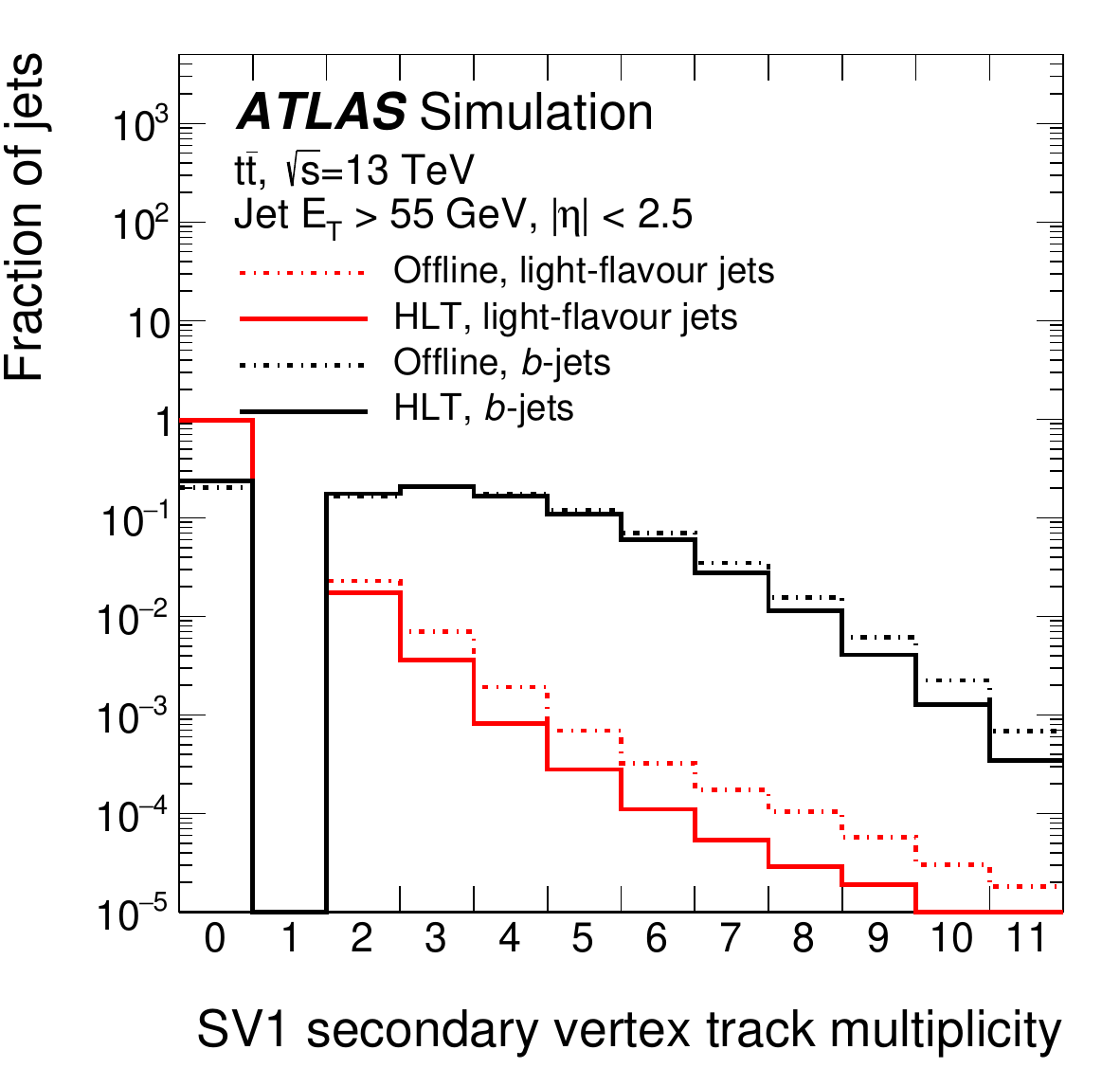}} \\
\subfloat[]{\includegraphics[width=0.35\textwidth]{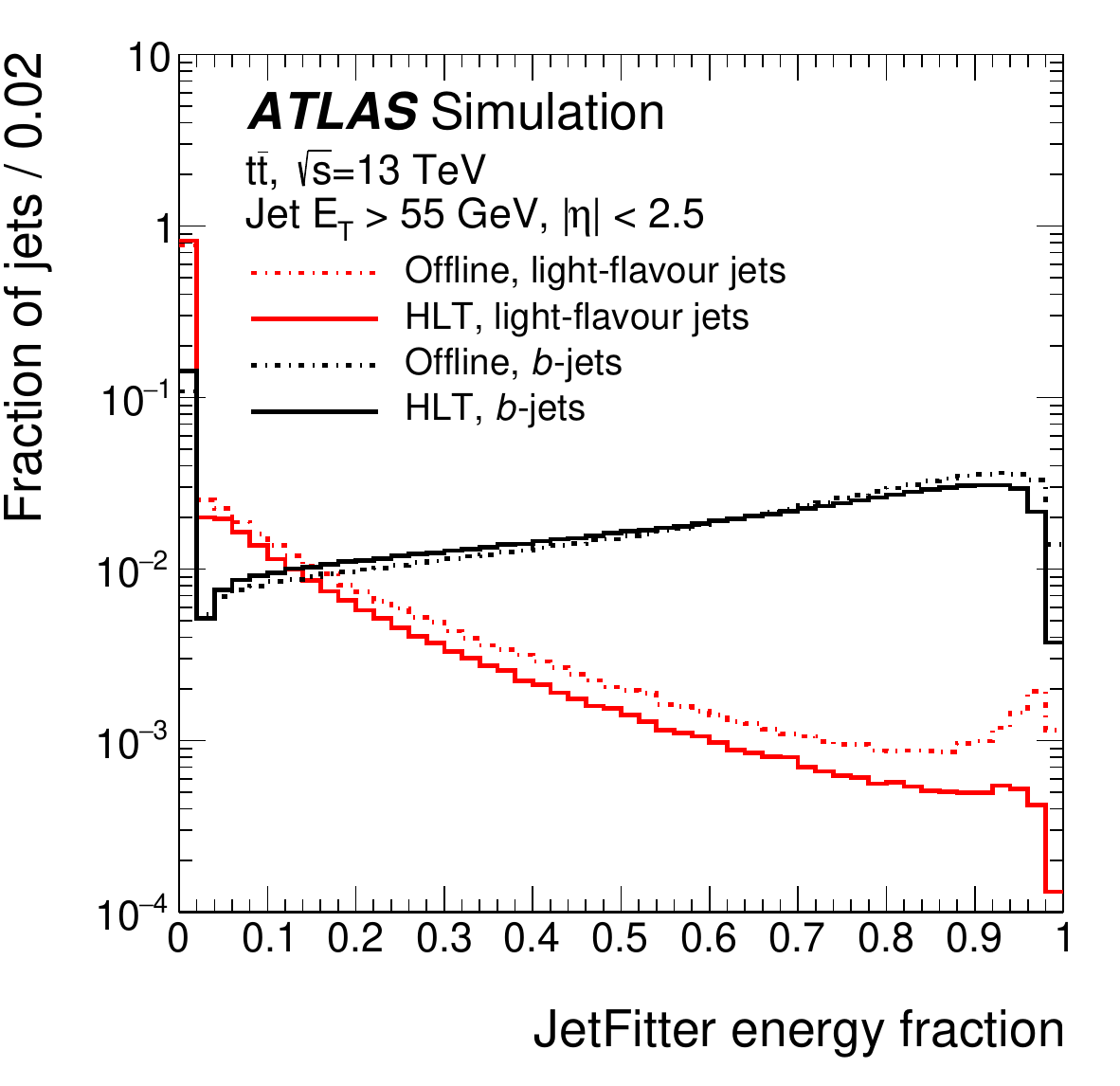}}
\subfloat[]{\includegraphics[width=0.35\textwidth]{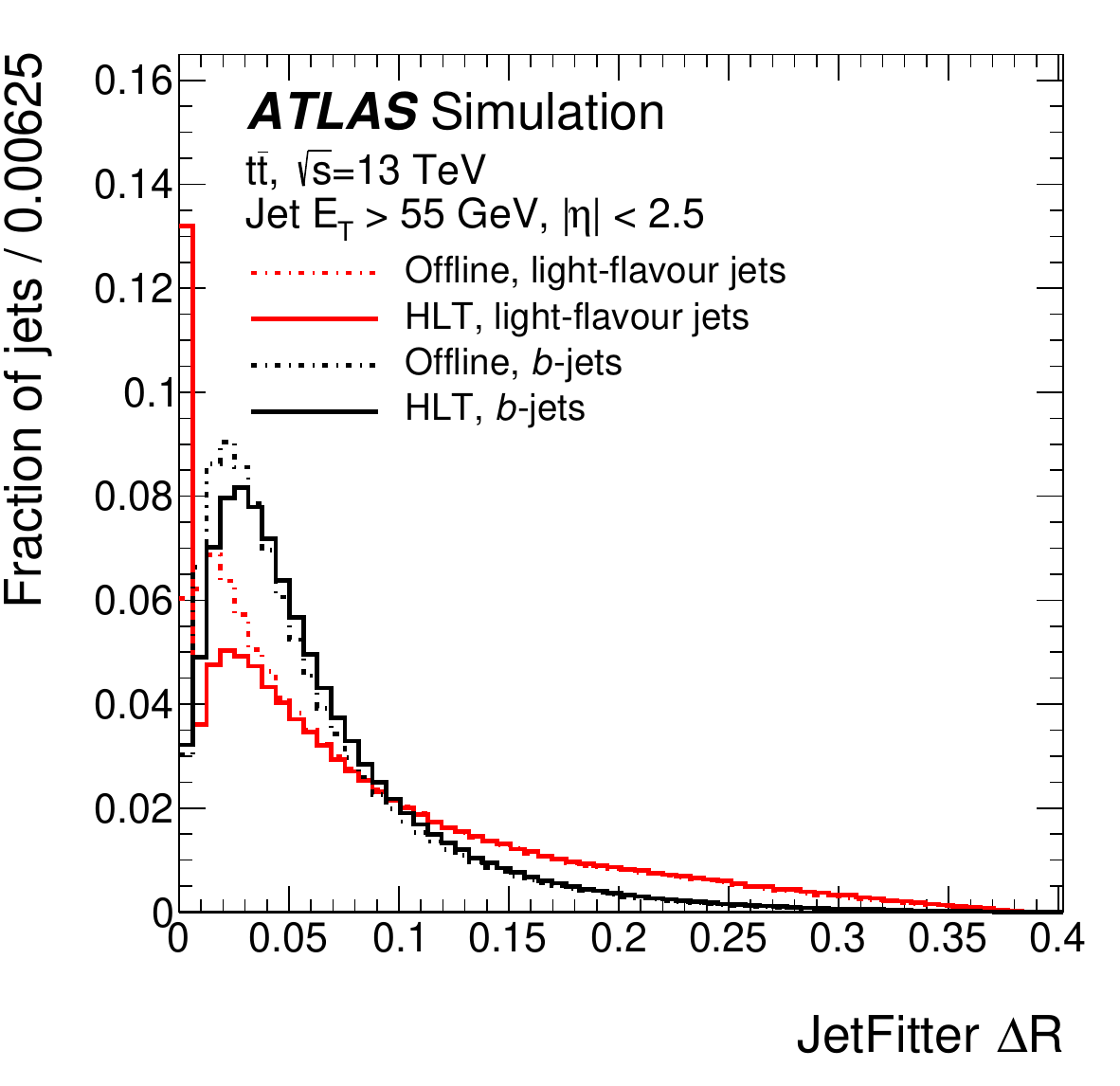}} \\
\caption {Selected variables related to the IP3D (top row), SV1
(middle row), and JetFitter (bottom row) algorithms.  The
distributions are shown for both $b$-jets and light-flavour jets,
and when computed using HLT and offline quantities.  The
distributions show the probability for a jet to originate from (a) a
light-flavour quark and (b) a $b$-quark, as computed by the IP3D
algorithm, (c) the mass of the jet associated with the secondary
vertex, (d) the number of tracks associated with the secondary
vertex (e) the fraction of the total jet energy carried by tracks
associated with displaced vertices, and (f) the $\Delta R$\ between
the jet axis and the vector sum of the momenta of all tracks
attached to displaced vertices.  The SV1 (JetFitter) variables have
a default value of zero in cases where no secondary (displaced)
vertex was found.}
\label{fig:tagger_inputs}
\end{figure}
 
The MV2 algorithms (and the low-level algorithms that form the inputs
to MV2) were retrained for the online environment on simulated
\ttbar\ events and using HLT tracks and \btagging information to
provide a discriminant to assess whether an individual jet arises from
the hadronisation of a bottom or charm quark, or light-flavour quark
or gluon.  Tunings were performed using the same procedures adopted
for offline flavour-tagging in ATLAS~\cite{FTAG-2018-01}, further
harmonising the procedures used in the trigger with those used
offline.  In 2016 a version of this tagger was used that was trained
to identify \bjets\ using a background sample composed of 80\%
light-flavour jets and 20\% $c$-jets and is denoted `MV2c20'.  In 2017
and 2018 the fraction of $c$-jets in the background sample was reduced
to 10\% to mirror the evolution of the offline \btagging
\cite{PERF-2016-05}\ and the algorithm is therefore denoted `MV2c10'.
 
Working points for the MV2 algorithms were designed that mirror the
offline working points providing 60\%, 70\%, 77\%, and 85\%
\bjet\ tagging efficiencies for \bjets in the simulated
\ttbar\ sample.  In addition, working points providing selection
efficiencies of 40\% and 50\% for \bjets were included in order to
provide triggers with lower jet \ET\ thresholds.  Requiring that jets
are \btagged at the trigger level means that the jet \ET\ thresholds
can be lowered significantly.  For example, including the requirement
that jets pass the MV2c10 tagger at a 40\% (70\%) working point allows
the \ET\ threshold of single-\bjet triggers to be reduced to 225
(300)\,\GeV, from a threshold of 420\,\GeV\ when no \btagging
requirements are applied.  Requiring more than one \btagged jet in a
trigger allows jet \ET\ thresholds to be lowered even further.
Four-jet triggers required \ET thresholds of 115\,\GeV\ when no
\btagging requirements were applied, but these thresholds could be
reduced to as low as 35\,\GeV\ when two of the jets are required to be
\btagged (details of these triggers are provided in
Section~\ref{sec:menu}).
 
The total processing time for the \bjet triggers is dominated by the
jet-finding and tracking that are used as inputs to the \btagging
algorithms.  The mean time to evaluate the \btagging weight of a
single jet is 16.2 ms (for $<\mu> = 52$), once the jet and tracks have
been found.  Optimising the software throughout Run~2 in order to
reduce the CPU cost of the \bjet triggers meant that the rates rather
than the CPU processing time were always the determining factor for
the \ET\ threshold of triggers used for physics analysis.
 
MV2 was superseded in 2019 by the DL1r algorithm (described in
Ref.~\cite{ATL-PHYS-PUB-2017-013}), which uses a deep feed-forward neural
network to provide a multidimensional output corresponding to the
probabilities for a jet to be a $b$-jet, $c$-jet, or light-flavour
jet, and is now the default for offline physics analyses in ATLAS.
This algorithm was not available in time to be used in the online
environment, but provides the baseline against which the
\bjet\ trigger performance is measured (as described in
Section~\ref{sec:calibrations}).
 
\subsection{\bjet trigger performance}
 
The performance of the \bjet triggers is quantified by the probability
of tagging a \bjet (\bjet efficiency, $\varepsilon_{b}$) and by the
rejection power against $c$-jets and light-flavour jets, where the
rejection is defined as the inverse of their efficiency to pass the
\btagging requirements.  Jets are categorised as $b$-jets, $c$-jets or
light-flavour jets following the particle-level definitions described
in Section~\ref{sec:datasets}.  Figure~\ref{fig:bJetPerf2016} shows
the expected performance of the \bjet trigger in terms of
light-flavour jet and $c$-jet rejection of the MV2c20 tagger together
with the performance of the IP3D+SV1 tagger that was used during
Run~1. The tuning is performed on simulated \ttbar events with
$\sqrt{s} = 13$\,\TeV. Jets used are required to have $\ET > 55\,\GeV$
and $|\eta| < 2.5$.  An order of magnitude improvement in
light-flavour jet rejection for the same \bjet selection efficiency
was achieved in 2016 compared with 2012 (Run~1).  This performance
increase is attributed to the installation of the insertable B-layer
for Run~2, in conjunction with all of the software and algorithmic
improvements described in this work.  An additional factor $\sim$1.5
improvement in light-flavour jet rejection was attained in 2017 and
2018 by further optimising the use of the MV2 algorithm in the HLT.
These improvements made it feasible to operate triggers with lower
\ET\ thresholds and/or higher-efficiency working points than would
have been affordable otherwise.
 
\begin{figure}[htbp]
\begin{center}
\centering
\subfloat[]{\includegraphics[width=0.5\textwidth]{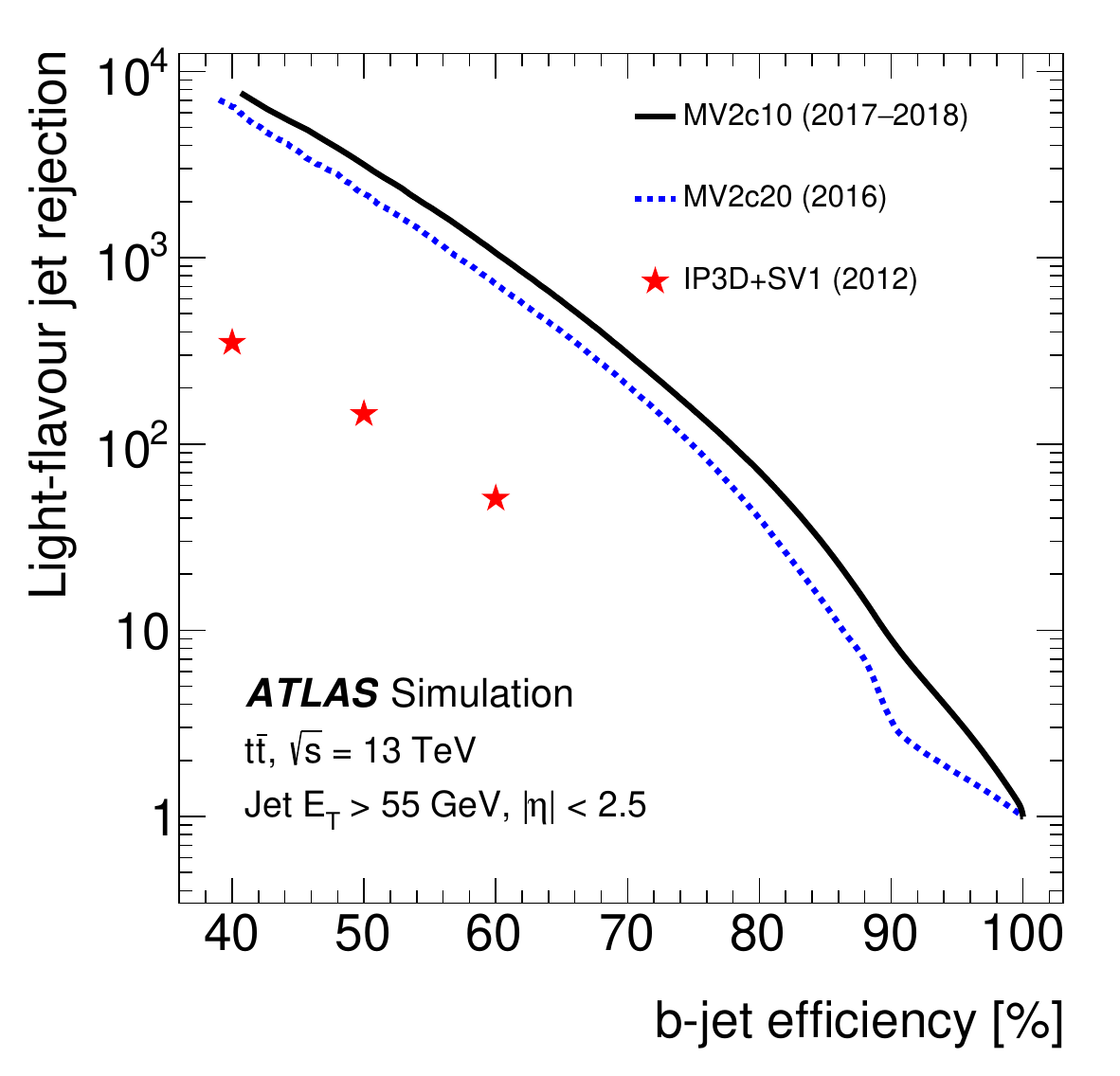}}
\subfloat[]{\includegraphics[width=0.5\textwidth]{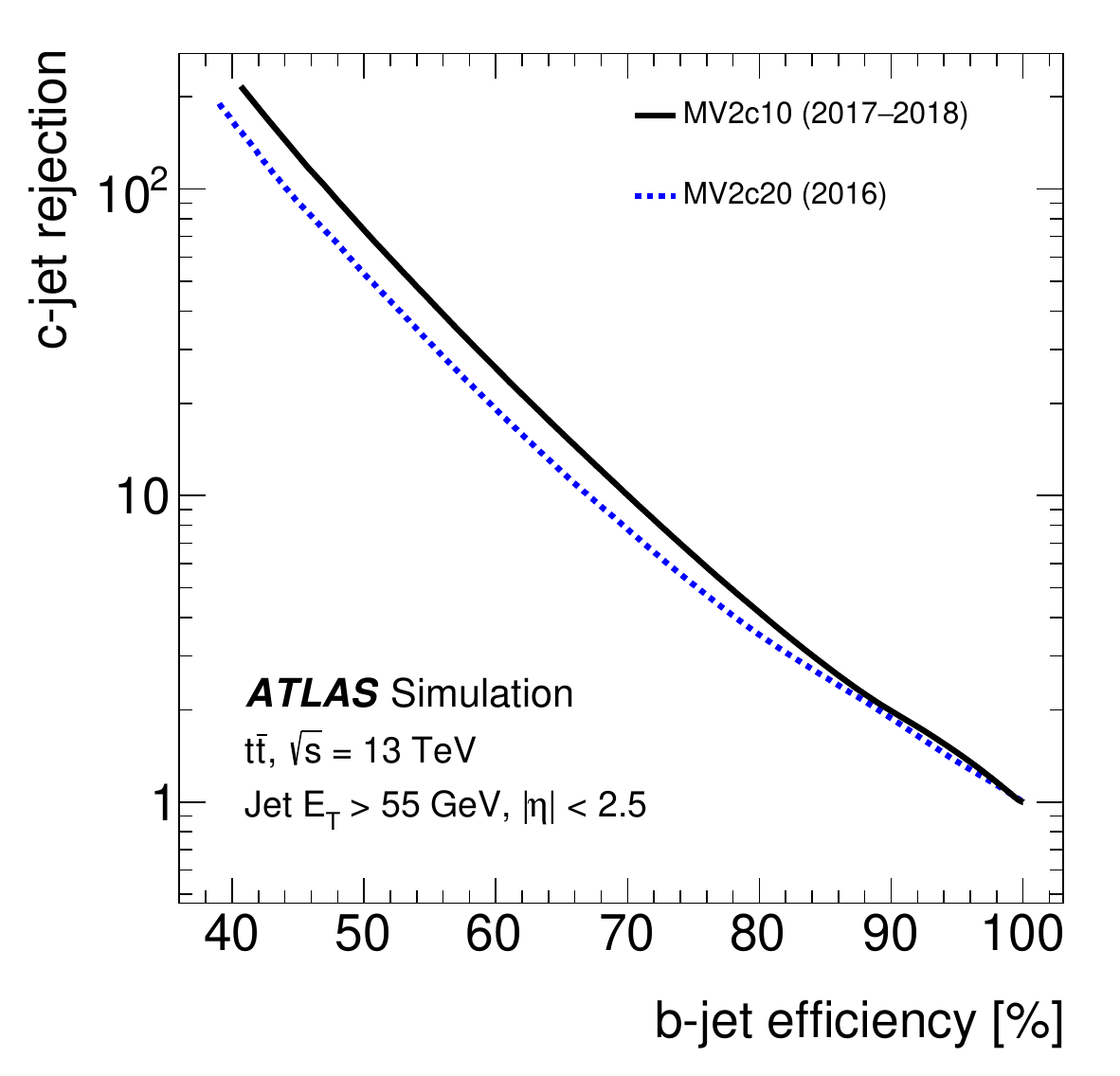}}
\caption{The expected online performance in terms of (a) light-flavour jet
and (b) $c$-jet rejection by the \bjet triggers for the MV2c10
tagger (used during data-taking in 2017 and 2018), and the
MV2c20 tagger (used during 2016).  The performance of the
\bjet trigger during Run~1 (stars) is also shown
in terms of its light-flavour jet rejection.}
\label{fig:bJetPerf2016}
\end{center}
\end{figure}
 
The baseline configuration of \bjet triggers in 2018 used the same
tuning of MV2c10 that was deployed during the 2017 data-taking period.
This was possible due to the general similarity between the running
conditions in these two years.  However, the \bjet trigger menu
included several triggers that used a dedicated tuning of MV2c10
intended to improve the performance of the \btagging algorithms at
high-\ET (e.g.\ $\ET \gtrsim 250$~\GeV) where it becomes harder to
identify \bjets.  Following the same approach as is used for offline
\btagging in ATLAS, the \ttbar sample used for the baseline tuning was
interleaved with a $Z^{\prime} \to q\bar{q}$ sample, which has a much
larger proportion of jets at high \ET\ and therefore increases the
attention of the BDT to these jets during training.  The heavy vector
boson ($Z^{\prime}$) is generated with a mass of 1\,\TeV\ with a flat
\pT spectrum, and decays at equal rates into light-, $c$-, and
$b$-flavour quark--antiquark pairs.  This process, referred to as the
`hybrid tuning', provides the BDT with consistent exposure to both
high- and low-$\ET$ jets.
 
The performance of the baseline 2018 tuning (which uses only \ttbar
simulation in the training) and the hybrid tuning is compared in
Figure~\ref{fig:hybridROCs}.  Little difference is
observed between the online 2018 baseline and hybrid approaches in a
sample dominated by low-\ET jets (\ttbar).  However, for the sample
dominated by high-\ET\ jets ($Z^{\prime}\rightarrow q\bar{q}$) the online
hybrid tuning provides better rejection against light-flavour jets.
 
\begin{figure}[htbp]
\centering
\subfloat[]{\includegraphics[width=0.5\textwidth]{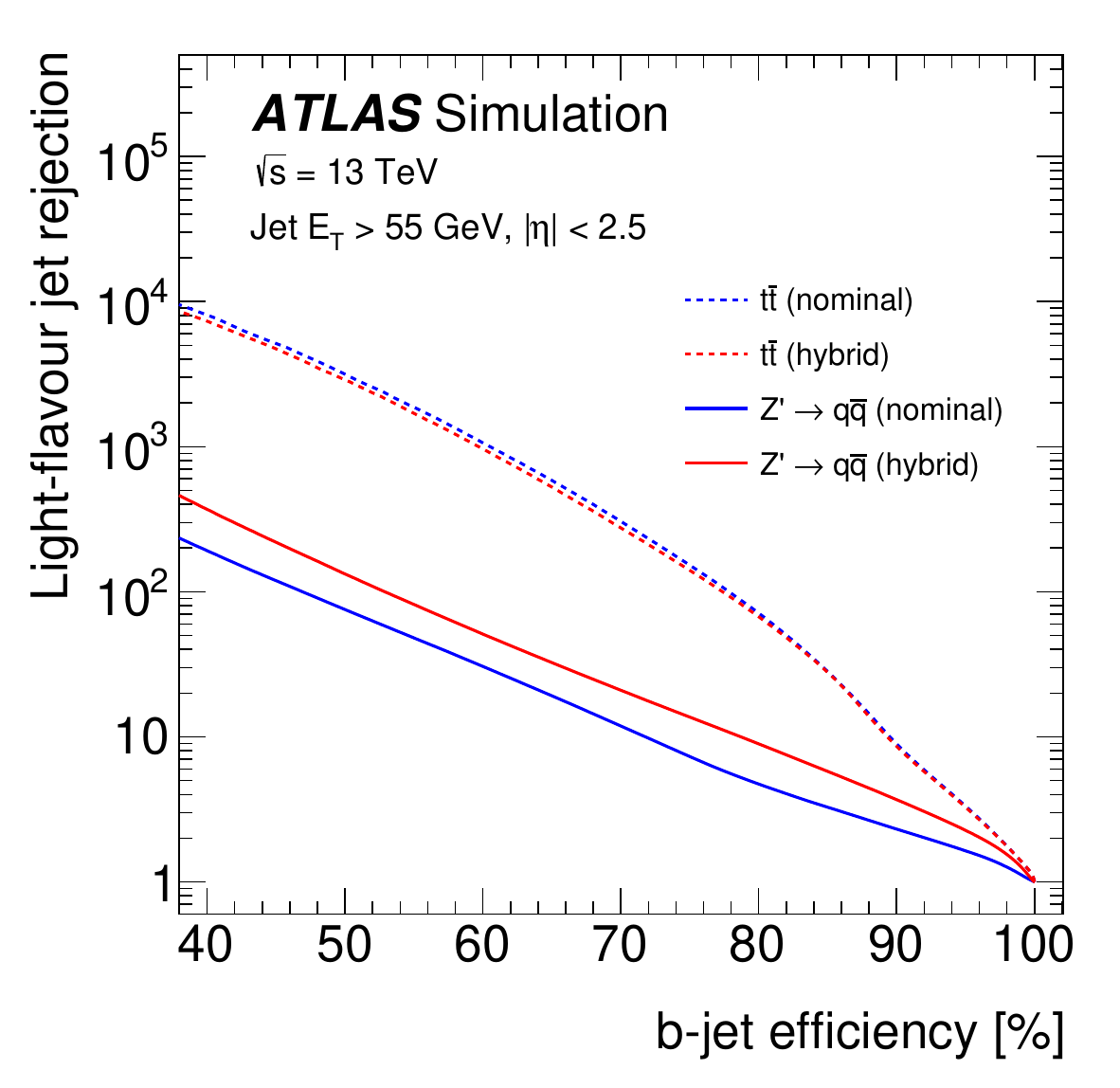}}
\subfloat[]{\includegraphics[width=0.5\textwidth]{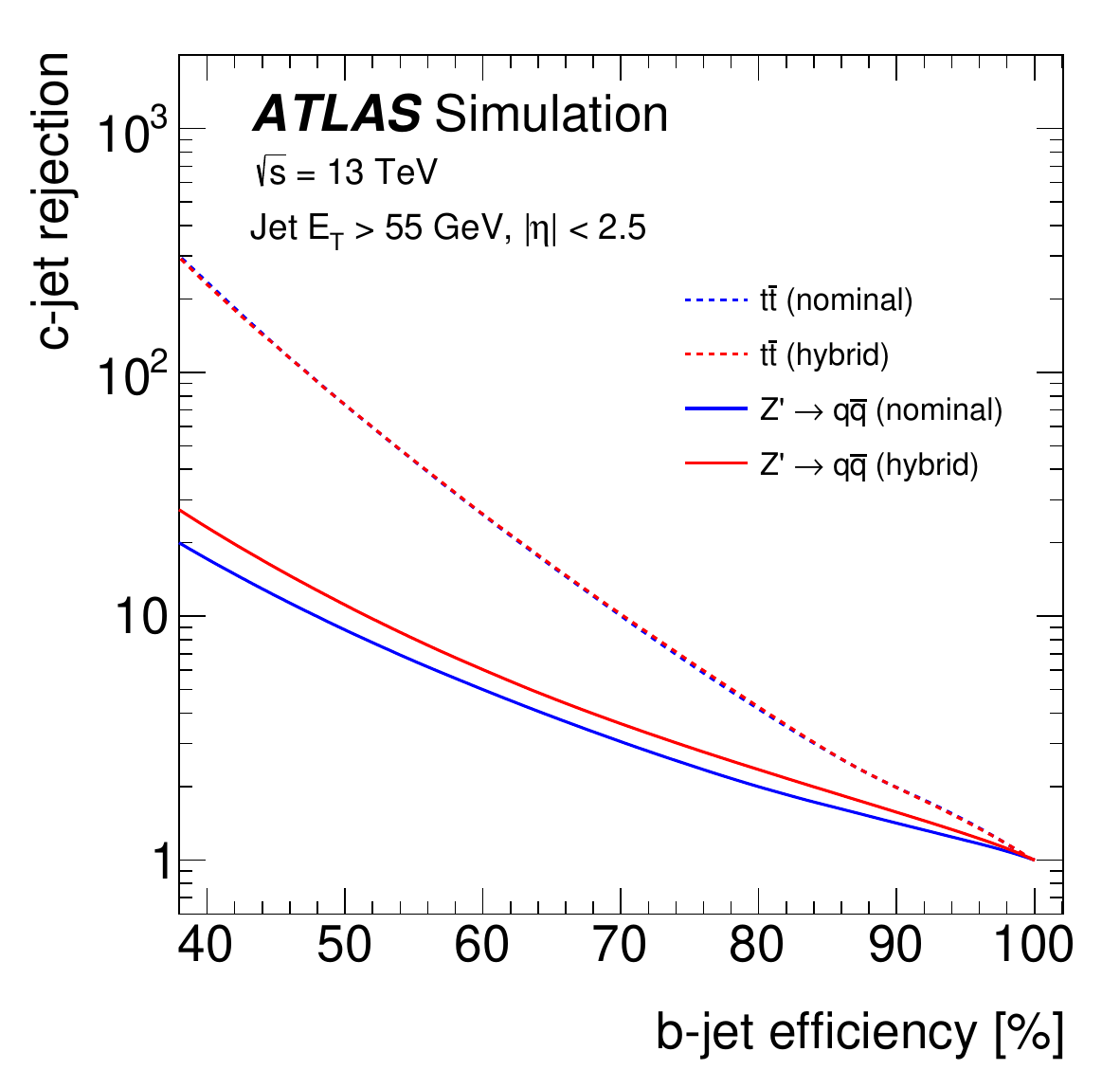}}
\caption{The expected performance of \btagging algorithms using the nominal
and hybrid tunings, using simulated \ttbar and $Z^{\prime}\rightarrow q\bar{q}$
events.  The (a) light-flavour jet and (b) $c$-jet rejection factors are shown as
a function of \bjet selection efficiency.}
\label{fig:hybridROCs}
\end{figure}

\section{\bjet trigger evolution during Run~2}
\label{sec:menu}
 
Several different types of \bjet triggers were operational throughout
Run~2, where the \ET\ thresholds and \btagging\ requirements evolved
in response to the increasing instantaneous luminosity during this
time.  Different combinations of jet and \bjet multiplicities, with
different \ET\ thresholds, with and without GSC calibrations
(described in Section~\ref{subsec:gsc}), and different \btagging
algorithms and working points were used to provide optimal coverage
for the different analyses using \bjet triggers within the allocated
trigger acceptance rate.  The total rate for the full suite of $b$-jet
triggers was up to 180 Hz at peak luminosity. Triggers that place requirements on the
scalar sum of the \ET of hadronic objects in the event (\HT) were also
provided.  This set of \bjet triggers was designed to provide optimal
acceptance for processes targeted in current analyses, as well as to
be general enough to provide good acceptance for yet-to-be-considered
physics analyses.
 
The parameters defining the \bjet triggers -- including the ($b$-)jet
multiplicity, \ET, and $\eta$ requirements, and the \btagging algorithm and
working point(s) -- are summarised for single-\bjet triggers in
Table~\ref{tab:menu_single_chains}, di-\bjet triggers in
Table~\ref{tab:menu_asymmetric_2b_chains}, jet+di-\bjet triggers with
asymmetric \ET thresholds in
Table~\ref{tab:menu_asymmetric_j2b_chains}, di-$b$-jet+di-jet triggers
in Table~\ref{tab:menu_2b2j_chains}, and di-$b$-jet+\HT triggers in
Table~\ref{tab:menu_2b_HT_chains}.

\begin{table}[htb]
\centering
\renewcommand{\arraystretch}{1.2}
\caption{Details, by year, of the lowest-threshold unprescaled
single-\bjet triggers.  The minimum ($b$-)jet \ET, $|\eta|$, and
$b$-tagging requirements are specified for each item.  HLT \bjets
are required to be within $|\eta| < 2.5$.  \ET thresholds denoted
$^{*}$ indicate that the GSC was applied and the value quoted is
the calibrated \ET threshold.  \vspace{2mm}}
\begin{tabular}{ l l l l } \toprule
Year                        & L1 jet                              & HLT \bjet                                                  \\ \toprule
2016                        & \multirow{5}{*}{\shortstack[l]{
1 $\times$\ $\ET > 100$\,\GeV,\\
$|\eta|<3.2$}}                       & 1 $\times$\ $\ET > 175$\,\GeV,  MV2c20, $\eff=40\%$        \\ \cmidrule(lr){1-1} \cmidrule(lr){3-3}
\multirow{4}{*}{2017--2018} &                                     & 1 $\times$\ $\ET > 225$\,\GeV$^{*}$, MV2c10, $\eff=40\%$    \\ \cmidrule(lr){3-3}
&                                     & 1 $\times$\ $\ET > 275$\,\GeV$^{*}$, MV2c10, $\eff=60\%$    \\ \cmidrule(lr){3-3}
&                                     & 1 $\times$\ $\ET > 300$\,\GeV$^{*}$, MV2c10, $\eff=70\%$    \\ \cmidrule(lr){3-3}
&                                     & 1 $\times$\ $\ET > 360$\,\GeV$^{*}$, MV2c10, $\eff=77\%$    \\ \bottomrule
\end{tabular}
\label{tab:menu_single_chains}
\end{table}

\begin{table}[htb]
\centering
\renewcommand{\arraystretch}{1.2}
\caption{Details, by year, of the lowest-threshold unprescaled
di-\bjet triggers.  The minimum ($b$-)jet \ET, $|\eta|$, and
$b$-tagging requirements are specified for each item.  HLT \bjets
are required to be within $|\eta| < 2.5$.  \ET thresholds denoted
$^{*}$ indicate that the GSC was applied and the value quoted is
the calibrated \ET threshold.  \vspace{2mm}}
\begin{tabular}{ l l l }  \toprule
Year                        & L1 jet                              & HLT \bjets                                                    \\ \toprule
\multirow{2}{*}{2016}       & \multirow{4}{*}{\shortstack[l]{
1 $\times$\ $\ET > 100$\,\GeV,\\
$|\eta|<3.2$}}                       & \multirow{2}{*}{\shortstack[l]{
1 $\times$\ $\ET > 150$\,\GeV, MV2c20, $\eff=60\%$ \\
1 $\times$\ $\ET > 50$\,\GeV, MV2c20, $\eff=60\%$}}         \\
 
&                                     &                                                              \\ \cmidrule(lr){1-1}\cmidrule(lr){3-3}
\multirow{2}{*}{2017--2018} &                                     & \multirow{2}{*}{\shortstack[l]{
1 $\times$\ $\ET > 175$\,\GeV, MV2c10, $\eff=60\%$ \\
1 $\times$\ $\ET > 60$\,\GeV, MV2c10, $\eff=60\%$}}          \\
&                                     &                                                              \\ \bottomrule
\end{tabular}
\label{tab:menu_asymmetric_2b_chains}
\end{table}

\begin{table}[htb]
\centering
\renewcommand{\arraystretch}{1.2}
\caption{Details, by year, of the lowest-threshold unprescaled
triggers requiring a high-\pT~jet plus two \bjets signature, such as
might arise from the process where a particle decaying into two
\bjets is accompanied by a jet from initial- or final-state
radiation.  No \btagging requirements are applied to this
'additional jet' in the event. The minimum ($b$-)jet \ET,
$|\eta|$, and $b$-tagging requirements are specified for each
item.  HLT \bjets are required to be within $|\eta| < 2.5$.  \ET
thresholds denoted $^{*}$ indicate that the GSC was applied and
the value quoted is the calibrated \ET threshold.  \vspace{2mm}}
\begin{tabular}{ l l l l } \toprule
\multirow{2}{*}{Year}    & \multirow{2}{*}{L1}                              & \multicolumn{2}{c}{HLT}                                                                                          \\ \cmidrule(lr){3-4}
&                                                  & \bjets                                     & Additional jet                                                      \\ \toprule
\multirow{2}{*}{2016}    & \multirow{2}{*}{\shortstack[l]{
1 $\times$\ $\ET > 70$\,\GeV, $|\eta| < 3.2$ \\
2 $\times$\ $\ET > 20$\,\GeV, $|\eta| < 3.2$}}   & \multirow{2}{*}{\shortstack[l]{
2 $\times$\ $\ET > 55$\,\GeV, \\
MV2c20, $\eff=60\%$}}                     & \multirow{2}{*}{1 $\times$\ $\ET > 100$\,\GeV, $|\eta| < 3.2$}       \\
&                                                  &                                            &                                                                      \\ \cmidrule(lr){1-4}
\multirow{2}{*}{2017}    & \multirow{4}{*}{\shortstack[l]{
1 $\times$\ $\ET > 85$\,\GeV, $|\eta| < 3.2$ \\
2 $\times$\ $\ET > 30$\,\GeV, $|\eta| < 3.2$}}   & \multirow{2}{*}{\shortstack[l]{
2 $\times$\ $\ET > 55$\,\GeV$^{*}$, \\
MV2c10, $\eff=77\%$}}                     & \multirow{2}{*}{1 $\times$\ $\ET > 150$\,\GeV$^{*}$, $|\eta| < 3.2$}  \\
&                                                  &                                            &                                                                       \\ \cmidrule(lr){1-1}\cmidrule(lr){3-4}
\multirow{2}{*}{2018}  &                                                  & \multirow{2}{*}{\shortstack[l]{
2 $\times$\ $\ET > 55$\,\GeV$^{*}$, \\
MV2c10, $\eff=70\%$}}                     & \multirow{2}{*}{1 $\times$\ $\ET > 150$\,\GeV$^{*}$, $|\eta| < 3.2$}   \\
&                                                  &                                            &                                                                        \\ \bottomrule
\end{tabular}
\label{tab:menu_asymmetric_j2b_chains}
\end{table}

\begin{table}[htb]
\centering
\renewcommand{\arraystretch}{1.2}
\caption{Details, by year, of the lowest-threshold unprescaled
triggers requiring two \btagged jets plus an additional two jets
with no \btagging requirements.  The minimum ($b$-)jet \ET,
$|\eta|$, and $b$-tagging requirements are specified for each
item.  HLT \bjets are required to be within $|\eta| < 2.5$.
Additional HLT jets are accepted up to $|\eta| < 3.2$\ but in
practice are mostly limited to be within $|\eta| < 2.5$ because of
the L1 requirements. \ET thresholds denoted $^{*}$\ indicate that
the GSC was applied and the value quoted is the calibrated \ET
threshold. The \btagging requirements were tightened to use a 60\%
efficiency working point for part of the data-taking during
2016. \vspace{2mm} }
\begin{tabular}{ l l l l } \toprule
\multirow{2}{*}{Year}      & \multirow{2}{*}{L1}                           & \multicolumn{2}{c}{HLT}                                                                     \\ \cmidrule(lr){3-4}
&                                               & \bjets                                                   & Additional jets                  \\ \toprule
2016                       & \multirow{7}{*}{\shortstack[l]{
4 $\times$\ $\ET > 15$\,\GeV,\\
$|\eta|<2.5$}}                & 2 $\times$\ $\ET > 35$\,\GeV, MV2c20, $\eff=70\%$         & 2 $\times$ $\ET>35$\,\GeV, $|\eta| < 3.2$        \\ \cmidrule(lr){1-1}\cmidrule(lr){3-4}
\multirow{3}{*}{2017}      &                                               & 2 $\times$\ $\ET > 35$\,\GeV$^{*}$, MV2c10, $\eff=40\%$   & 2 $\times$ $\ET>35$\,\GeV$^{*}$, $|\eta| < 3.2$   \\  \cmidrule(lr){3-4}
&                                               & 2 $\times$\ $\ET > 45$\,\GeV$^{*}$, MV2c10, $\eff=50\%$   & 2 $\times$ $\ET>45$\,\GeV$^{*}$, $|\eta| < 3.2$   \\  \cmidrule(lr){3-4}
&                                               & 2 $\times$\ $\ET > 55$\,\GeV$^{*}$, MV2c10, $\eff=60\%$   & 2 $\times$ $\ET>55$\,\GeV$^{*}$, $|\eta| < 3.2$   \\  \cmidrule(lr){1-1}\cmidrule(lr){3-4}
\multirow{3}{*}{2018}      &                                               & 2 $\times$\ $\ET > 35$\,\GeV, MV2c10, $\eff=60\%$         & 2 $\times$ $\ET>35$\,\GeV, $|\eta| < 3.2$        \\ \cmidrule(lr){3-4}
&                                               & 2 $\times$\ $\ET > 45$\,\GeV$^{*}$, MV2c10, $\eff=50\%$   & 2 $\times$ $\ET>45$\,\GeV$^{*}$, $|\eta| < 3.2$   \\  \cmidrule(lr){3-4}
&                                               & 2 $\times$\ $\ET > 55$\,\GeV$^{*}$, MV2c10, $\eff=60\%$   & 2 $\times$ $\ET>55$\,\GeV$^{*}$, $|\eta| < 3.2$   \\  \bottomrule
 
\end{tabular}
\label{tab:menu_2b2j_chains}
\end{table}

\begin{table}[htb]
\centering
\renewcommand{\arraystretch}{1.2}
\caption{Details, by year, of the lowest-threshold unprescaled
triggers giving a two-\bjet plus \HT signature.  The minimum
($b$-)jet \ET, $|\eta|$, and $b$-tagging requirements are
specified for each item.  HLT \bjets are required to be within
$|\eta| < 2.5$.  \ET thresholds denoted $^{*}$ indicate that the
GSC was applied and the value quoted is the calibrated \ET
threshold. The \HT\ is calculated at L1 by summing the \ET of the
leading five jets with $|\eta| < 2.1$, and at the HLT by summing
the \ET of all jets with $\ET > 30\,\GeV$\ and $|\eta| <
3.2$.  \vspace{2mm}}
\begin{tabular}{ l l l l } \toprule
\multirow{2}{*}{Year} & \multirow{2}{*}{L1}                                  & \multicolumn{2}{c}{HLT}                                                                        \\ \cmidrule(lr){3-4}
&                                                      & \bjets                                                     & \HT                               \\ \toprule
2016                  & 4 $\times$\ $\ET > 15$\,\GeV, $|\eta| < 3.2$         & 2 $\times$\ $\ET > 55$\,\GeV, MV2c20, $\eff=60\%$          & \multirow{3}{*}{$\HT>300\,\GeV$}  \\ \cmidrule(lr){1-3}
2017                  & \multirow{2}{*}{ $\HT > 190$\,\GeV}                  & 2 $\times$\ $\ET > 55$\,\GeV$^{*}$, MV2c10, $\eff=60\%$\    &                                   \\ \cmidrule(lr){1-1}\cmidrule(lr){3-3}
2018                  &                                                      & 2 $\times$\ $\ET > 55$\,\GeV$^{*}$, MV2c10, $\eff=50\%$\    &                                   \\ \bottomrule
 
\end{tabular}
\label{tab:menu_2b_HT_chains}
\end{table}
 
Triggers targeting specific physics processes involving \bjets were
also provided.  Triggers requiring a di-\bjet plus missing transverse
momentum (\met) signature were designed to efficiently select
pair-produced bottom squarks \cite{ATLAS-CONF-2020-031} and are
detailed in Table~\ref{tab:menu_b_MET_chains}.  Higgs bosons produced
via VBF and decaying into a pair of $b$-quarks were also able to be
efficiently selected at trigger level through the use of dedicated
triggers that require jets with a large invariant mass in the forward
region of the detector.  Additionally, some triggers required the
presence of a photon in the event (where the photon may be radiated
either from a charged weak boson or from one of the scattering
initial-state quarks that subsequently showers into a jet)
\cite{HIGG-2016-30,Aad:2020vbr}.  The photon requirements
significantly reduce the contribution from large multijet backgrounds
and allow lower \ET requirements at the trigger level to be placed on the
\bjets produced by the Higgs boson decay.  The VBF plus \bjet (plus
photon) triggers are summarised in Table \ref{tab:menu_VBF_chains}.
 
\begin{table}[htb]
\centering
\renewcommand{\arraystretch}{1.2}
\caption{Details, by year, of the lowest-threshold unprescaled
triggers giving a \bjet plus missing transverse momentum (\met)
signature.  The minimum \met\ and ($b$-)jet \ET, $|\eta|$, and
$b$-tagging requirements are specified for each item.  HLT \bjets
are required to be within $|\eta| < 2.5$. \vspace{2mm} }
\begin{tabular}{ l l l l l } \toprule
\multirow{2}{*}{Year}        & \multicolumn{2}{c}{L1}                                                    & \multicolumn{2}{c}{HLT}                                                     \\ \cmidrule(lr){2-3}\cmidrule(lr){4-5}
& Jets                                & \met                                & \bjet                                 & \met                                \\ \toprule
\multirow{2}{*}{2016}        & \multirow{4}{*}{\shortstack[l]{
2 $\times$\ $\ET > 50$\,\GeV,\\
$|\eta| < 3.2$}}                    & \multirow{4}{*}{$\met > 40\,\GeV$}  & \multirow{2}{*}{\shortstack[l]{
1 $\times$\ $\ET > 80$\,\GeV,\\
MV2c20, $\eff=60\%$}}                & \multirow{4}{*}{$\met > 60$\,\GeV}  \\
 
&                                     &                                     &                                       &                                     \\ \cmidrule(lr){1-1}\cmidrule(lr){4-4}
 
\multirow{2}{*}{2017--2018}  &                                     &                                     & \multirow{2}{*}{\shortstack[l]{
1 $\times$\ $\ET > 80$\,\GeV,\\
MV2c20, $\eff=60\%$}}                &                                     \\
&                                     &                                     &                                       &                                     \\ \bottomrule
\end{tabular}
\label{tab:menu_b_MET_chains}
\end{table}

\begin{landscape}
\begin{table}[htbp]
\centering
\renewcommand{\arraystretch}{1.2}
\caption{Details, by year, of the lowest-threshold unprescaled
triggers giving a VBF plus \bjet signature.  Additionally, some
triggers require the presence of a photon in the event, exploiting
the unique phenomenology of the VBF process to help reject
background processes and allow the use of lower jet \ET
requirements at the trigger level.  In these cases the photon is
used to seed the trigger at L1, by requiring that the summed
energy of deposits in the electromagnetic calorimeters (denoted
`EM') exceeds some minimum \ET\ and fulfils isolation
requirements.  The photon identification (ID) and isolation
working points used are described in Ref.~\cite{TRIG-2018-05}.
The minimum ($b$-)jet \ET, $|\eta|$, and $b$-tagging requirements
are also specified for each item.  HLT \bjets are required to be
within $|\eta| < 2.5$.  Some triggers additionally require that
any pair of jets in the event satisfy a minimum requirement on
their invariant mass ($m_{\mathrm{jj}}$). \ET thresholds denoted
$^{*}$ indicate that the GSC was applied and the value quoted is
the calibrated \ET threshold.  \vspace{2mm} }
\begin{tabular}{ l l l l l} \toprule
\multirow{2}{*}{Year} & \multirow{2}{*}{L1}                & \multicolumn{3}{c}{HLT}                                                                                 \\ \cmidrule(lr){3-5}
&                                    & \bjet(s)                          & Additional jet(s)                                 & Photon            \\ \toprule

\multirow{5}{*}{2016--2017} & \multirow{3}{*}{1 $\times$\ EM $>$\,22\,\GeV}  & \multirow{3}{*}{1 $\times$\ $\ET>35$\,\GeV, MV2c20, $\eff=77\%$}  & \multirow{3}{*}{\shortstack[l]{
3 $\times$\ $\ET > 35$\,\GeV,\\
$|\eta|<4.9$,\\
$m_{\mathrm{jj}} > 700$\,\GeV}}     & \multirow{3}{*}{\shortstack[l]{
1 $\times$\ $\ET>25$\,\GeV,\\
`Medium' ID}}      \\
&&&& \\
&&&& \\\cmidrule(lr){2-5}
& \multirow{2}{*}{1 $\times$\ EM $>$\,22\,\GeV}  & \multirow{2}{*}{2 $\times$\ $\ET>35$\,\GeV, MV2c20, $\eff=77\%$}  & \multirow{2}{*}{\shortstack[l]{
2 $\times$ \ $\ET > 35$\,\GeV,\\
$|\eta|<4.9$}}                  & \multirow{2}{*}{\shortstack[l]{
1 $\times$\ $\ET>25$\,\GeV,\\
`Medium' ID}}      \\
&&&& \\ \cmidrule(lr){1-5}

\multirow{9}{*}{2017--2018} & \multirow{3}{*}{
\shortstack[l]{
1 $\times$\ EM $>$\,18\,\GeV\\
$m_{\mathrm{jj}} > 300$\,\GeV}}                         & \multirow{3}{*}{1 $\times$\ $\ET>35$\,\GeV$^{*}$, MV2c10, $\eff=77\%$}  & \multirow{3}{*}{\shortstack[l]{
3 $\times$\ $\ET > 35$\,\GeV, \\
$|\eta|<4.9$,\\
$m_{\mathrm{jj}} > 500$\,\GeV}}   & \multirow{3}{*}{\shortstack[l]{
1 $\times$\ $\ET>20$\,\GeV,\\
`Tight' ID,\\
`Loose' isolation}} \\
&&&&\\
&&&&\\\cmidrule(lr){2-5}

& \multirow{4}{*}{\shortstack[l]{
1 $\times$\ $\ET > 40$\,\GeV, $|\eta| < 2.5$ \\
1 $\times$\ $\ET > 25$\,\GeV, $|\eta| < 3.2$ \\
1 $\times$\ $\ET > 20$\,\GeV, $3.1 < |\eta| < 4.9$}}   & \multirow{2}{*}{1 $\times$\ $\ET>80$\,\GeV$^{*}$, MV2c10, $\eff=70\%$}  & \multirow{4}{*}{\shortstack[l]{
1 $\times$\ $\ET > 45$\,\GeV,\\
$3.2<|\eta|<4.9$}}                  & \multirow{3}{*}{-} \\
&&&&\\ \cmidrule(lr){3-3}
&                                                        & \multirow{2}{*}{1 $\times$\ $\ET>60$\,\GeV$^{*}$, MV2c10, $\eff=85\%$}  & & \\
&&&&\\ \cmidrule(lr){2-5}

& \multirow{2}{*}{\shortstack[l]{
1 $\times$\ $\ET > 25$\,\GeV, $|\eta| < 2.3$ \\
2 $\times$\ $\ET > 15$\,\GeV, $3.1 < |\eta| < 4.9$}}   & \multirow{2}{*}{1 $\times$\ $\ET>55$\,\GeV$^{*}$, MV2c10, $\eff=70\%$}  & \multirow{2}{*}{\shortstack[l]{
2 $\times$\ $\ET > 45$\,\GeV,\\
$3.2<|\eta|<4.9$}}                 & \multirow{2}{*}{-} \\
&&&& \\ \bottomrule

\end{tabular}
\label{tab:menu_VBF_chains}
 
\end{table}
\end{landscape}

\section{Calibrations}
\label{sec:calibrations}
 
The trigger is a crucial step in the event selection of any physics
analysis, so its performance must be understood and calibrated. This
section describes the \bjet trigger efficiency measurements made using
\pp\ data collected between 2016 and 2018. In physics analyses, the
\bjet trigger is always used in tandem with offline \btagging, which
is calibrated without placing any requirements on the \bjet trigger.
A `conditional' \bjet trigger efficiency is therefore calculated
relative to the offline \btagging efficiency and defined as the
fraction of \bjets that are \btagged offline and match an HLT jet,
that also pass the \btagging requirements in the HLT.  This
conditional \bjet trigger efficiency is measured in data and evaluated
in simulated \ttbar\ events.  Simulation-to-data scale factors
(hereinafter referred to simply as scale factors) are derived to
correct for any deviation of the \bjet trigger performance in MC
simulation from that observed in data. The scale factors are applied
only to simulated events and are designed to be applied in addition to
the offline \btagging scale factors~\cite{FTAG-2018-01}. The \bjet
trigger efficiency and scale factors are measured for all combinations
of offline and online \btagging working points and only a few
representative points are included here.
 
Historically, two methods have been used to calibrate the \bjet
triggers.  A geometrical matching method similar to that described in
Ref.~\cite{PERF-2016-05} was used to provide preliminary calibrations
for Run~2 data analysis but is now superseded by the likelihood-based
method that is described here and has smaller associated uncertainties.
The same likelihood-based method is also used to calibrate the offline
reconstruction and identification of \bjets in ATLAS and is described
fully in Ref.~\cite{FTAG-2018-01}.  The results presented here closely
follow the analysis selection and method used for the offline
\btagging calibration, and only the most important features of the
likelihood-based calibration and its adaption to the online
environment, together with the results, are described.
 
Scale factors to correct for any MC-simulation mismodelling of the
rate for light-flavour jets and $c$-jets to be misidentified as \bjets
are provided for offline
\btagging~\cite{ATLAS-CONF-2018-006,ATLAS-CONF-2018-001}.  Measuring
the equivalent light-flavour and $c$-jet scale factors in the trigger
is beyond the scope of this paper, but the impact of these scale factors
is expected to be small in physics analyses that use \bjet triggers,
where background processes are typically estimated using data-driven
techniques and the signal processes, which are modelled using simulation,
have a negligible fraction of non-\bjets.
 
\subsection{Event selection}
\label{subsec:calibration_selection}
 
Top quarks are produced in abundance at the LHC and, since the
branching fraction of the top quark decay into a $W$ boson and a
$b$-quark is nearly 100\%, selecting events with pair-produced
top quarks can provide a large data sample of \bjets\ that can be used to
study the \bjet trigger efficiency.  In order to reduce the
contributions from multijet and $W/Z$+jets backgrounds, and maximise
the purity of the selection, the offline selection requires events to
have exactly one electron and one muon with opposite-sign charge and
satisfying tight identification criteria.  Furthermore, the electron and
muon provide a signature that can be used to select events at the
trigger level without using a \bjet trigger such that no bias is
introduced from online \btagging.  These `single-lepton
$b$-performance triggers' (detailed in
Table~\ref{tab:calibration_triggers}) were designed and run
specifically in order to study the performance of the \bjet triggers,
and require the presence of an electron or muon, plus two additional
jets.  The \bjet trigger software is run on the jets and all
associated \btagging information is kept, but no selection is made on
the online \btagging\ weight of the jets.  The triggers used for these
measurements were run unprescaled, but in 2016 they were only run
for part of the year and the integrated luminosity of that dataset is
13.1\,\ifb.

\begin{table}[htb]
\centering
\renewcommand{\arraystretch}{1.2}
\caption{Details of the triggers used to select a data sample to
perform the calibrations.  Electrons (muons) are required to be
isolated and pass a `Tight'~\cite{TRIG-2018-05}
(`Medium'~\cite{TRIG-2018-01}) identification working point.  Jet
\ET thresholds denoted $^{*}$\ indicate that the GSC was applied
and the value quoted is the calibrated \ET threshold.  All
triggers were run unprescaled, but in 2016 they were only run
for part of the year and the integrated luminosity of that dataset
is 13.1\,\ifb. \vspace{2mm} }
\begin{tabular}{ l l l l } \toprule
Year                    & \multicolumn{2}{c}{Lepton}          & Jets                                                                                    \\ \toprule
\multirow{2}{*}{2016}   & Electron  & $\ET > 26$\,\GeV        & \multirow{2}{*}{2 $\times$ $\ET > 35$\,\GeV}                                            \\ \cmidrule(lr){2-3}
& Muon      & $\pt > 26$\,\GeV        &                                                                                         \\ \midrule
\multirow{2}{*}{2017}   & Electron  & $\ET > 28$\,\GeV        & \multirow{2}{*}{1 $\times$ $\ET > 150$\,\GeV$^{*}$ + 1 $\times$ $\ET > 35$\,\GeV$^{*}$}  \\ \cmidrule(lr){2-3}
& Muon      & $\pt > 26$\,\GeV        &                                                                                         \\ \midrule
\multirow{2}{*}{2018}   & Electron  & $\ET > 28$\,\GeV        & \multirow{2}{*}{1 $\times$ $\ET > 150$\,\GeV$^{*}$ + 1 $\times$ $\ET > 35$\,\GeV}       \\ \cmidrule(lr){2-3}
& Muon      & $\pt > 26$\,\GeV        &                                                                                         \\ \bottomrule
\end{tabular}
\label{tab:calibration_triggers}
\end{table}

Events are required to pass the following selection:
\begin{itemize}
\item{Pass one of the single-lepton $b$-performance triggers
detailed in Table~\ref{tab:calibration_triggers}.}
\item{Contain an offline muon with $\pt \ge 28\,\GeV$, $|\eta| <
2.4$, satisfying the `Tight' identification and isolation
requirements \cite{PERF-2015-10}, and no jet with three or more
associated tracks within $\Delta R$ of 0.4.}
\item {Contain an offline electron with $\ET \ge 28\,\GeV$,\footnote{
Scale factors to correct the electron identification and
efficiencies in MC simulation to match those in data are provided only for
electrons with offline $\ET > 30$\,\GeV.  Less than 1\% of
selected electrons have offline \ET\ in the range
$28\,\GeV < \ET < 30\,\GeV$ and excellent agreement between data
and simulation is observed is this region, so any potential effect on
the calibration by not applying scale factors for these
electrons is negligible.} $|\eta| < 2.47$ excluding $1.37 \le |\eta| <
1.52$, satisfying the `Tight' identification and isolation
requirements \cite{EGAM-2018-01}.  }
\item{Leptons are required to have $|d_0|/\sigma_{d_0}$ less than 5
(3) for electrons (muons) and $|z_0\sin\theta|$ less than 0.5~mm.
These requirements ensure the selected leptons are prompt and
associated with the primary vertex, defined as the collision vertex
with the largest sum of $\pt^{2}$\ of tracks, as described in
Section~\ref{subsec:pv_finding}.}
\item{The triggered lepton must match an offline electron or muon
candidate.}
\item{Have exactly two offline \antikt $(R=0.4)$
\cite{anti_kT}\ particle flow (PFlow) jets
\cite{PERF-2015-09}\ each satisfying the following requirements:
\begin{itemize}
\item{$\pt > 35\,\GeV$ and $|\eta|< 2.5$.}
\item{Matched to an HLT jet, within $\Delta R(\mathrm{online,
offline}) < 0.2$.}
\item{Not within $\Delta R = 0.2$ of an electron.}
\item{Jets with less than three associated tracks must not be within $\Delta R = 0.4$ of a muon.}
\item{Jets with $\pt < 120\,\GeV$\ are required to pass the
`Medium' working point of the Jet Vertex Tagger (JVT)
algorithm~\cite{ATLAS-CONF-2014-018} that is used to reduce
the number of jets with large energy fractions from pile-up
collision vertices.  The JVT efficiency for jets originating
from the hard \pp scattering is 92\% in the simulation.}
\end{itemize}
}
\item{Each lepton is paired with a jet, where the pairing is
assigned by minimising ($m_{\ell_{i},\mathrm{j1}}^{2} + m_{\ell_{j},\mathrm{j2}}^{2}$) , where
$\mathrm{j1}$\ ($\mathrm{j2}$) is the leading (sub-leading) jet,
$\ell_{i,\mathrm{j}}$ are the two leptons, and
$m_{\ell,\mathrm{j1}}$\ ($m_{\ell,\mathrm{j2}}$) is the invariant
mass of the leading (sub-leading) jet and its associated lepton.
Events where either $m_{\ell,\mathrm{j1}}$ or
$m_{\ell,\mathrm{j2}}$ is more than 175\,\GeV\ are rejected.}
\end{itemize}
After applying these requirements, approximately 90\% of the selected
events in simulation contain two real \bjets.  Light-flavour and $c$-jet
backgrounds are estimated using MC simulation and included in the
likelihood fit, following the procedure described in
Ref.~\cite{FTAG-2018-01}.  Fake-lepton backgrounds are estimated from
simulation and are negligible.
 
\subsection{Calibration based on likelihood-based matching}
 
Events passing the selection described in Section
\ref{subsec:calibration_selection}\ are categorised according to the
offline jet \pT\ and the output of the online and offline \btagging
identification algorithms. Simulated events are further categorised by
the particle-level label of the jets (as described in
Section~\ref{sec:datasets}).  A maximum-likelihood fit is then
performed to extract the \btagging efficiency from data, as a
function of jet \pT.
 
As in the offline measurement~\cite{FTAG-2018-01}, a general
extended binned log-likelihood function approach is used for the
extraction of the \btagging efficiency and adapted to use only one
signal region, i.e.\ where both jets pass \btagging requirements.
This likelihood function can be written as follows:
 
\begin{equation*}
\log \mathcal{L} \left( \nu_{\textrm{tot}}, \hat \Theta \right) = -
\nu_{\textrm{tot}} + \sum_{i}^{N} n_i \log \nu_i \left(
\nu_{\textrm{tot}},\hat \Theta \right),
\end{equation*}
 
where $\nu_{\textrm{tot}}$ is the total number of expected events,
$\hat \Theta=(\Theta_1,...,\Theta_m)$ is the list of parameters to be
estimated, including the parameters of interest and the nuisance
parameters, and $\nu_i$ ($n_i$) is the expected (observed) number
of events in bin $i$ where $N$ bins are considered in total.
 
Events are divided into five categories based on offline \btagging
working points. The first category does not apply any offline \btagging
requirements, while the remaining four are based on the offline
\btagging working points, corresponding to efficiencies of 85\%,
77\%, 70\% and 60\% for true \bjets. For each category, events are
divided into bins of offline jet \pT in order to account for any
\pT-dependence of the scale factors. The \pT binning used is
[35--45, 45--60, 60--85, 85--110, 110--175, 175--600]\,\GeV.
 
A simultaneous fit of several extended binned log-likelihood functions
($\log\mathcal{L}_k)_{\ptone,\pttwo}$, each defined in a given bin
$k$\ of two jet transverse momenta (\ptone, \pttwo) is performed to
extract the online \btagging efficiency.
 
The results of this measurement are given in the form of per-jet scale
factors of two different types: online-only scale factors and
conditional scale factors. The online-only scale factors are
stand-alone results that can be applied to measure the \btagging
efficiency at the trigger level. The conditional scale factors
are intended to be combined with the existing offline \btagging scale
factors~\cite{FTAG-2018-01} in order to correct simulated events in
physics analyses where both trigger and offline \btagging are used.
 
The conditional efficiency, $\varepsilon_{b}^{\mathrm{Trig|Off}}$,
is defined as the efficiency of a jet to be tagged as a \bjet by the
online \btagging algorithm if it has also passed the offline
\btagging. Here (and elsewhere), `off' denotes the offline
\btagging, while `trig' denotes the online \btagging.
 
In order to evaluate this conditional efficiency, only
events in which both jets are already tagged by the offline \btagging
are selected and the efficiency of the online \btagging in these
events is evaluated. The ratio of the conditional efficiency
measured in data to that evaluated in MC simulation is the conditional
scale factor defined as
\[
SF_{b}^{\mathrm{Trig|Off}} =
\frac{\varepsilon_{b}^{\mathrm{Trig|Off,Data}}}{\varepsilon_{b}^{\mathrm{Trig|Off,MC}}}.
\]
 
The overall efficiency for a jet to pass both the trigger and offline
\btagging, $\varepsilon_{b}^{\mathrm{Trig} \land \mathrm{Off}}$, is
obtained for physics analysis by multiplying the conditional
efficiency, $\varepsilon_{b}^{\mathrm{Trig|Off}}$, by the corresponding
offline \btagging efficiency, $\varepsilon_{b}^{\mathrm{Off}}$\ (presented
in Ref.~\cite{FTAG-2018-01}).  As before, the scale factors are
defined as the ratio of the efficiencies measured in data and evaluated
in simulation.
 
Scale factors can also be derived in order to correct for \bjets that
have failed either the online or offline \btagging requirements (or
both).  The efficiencies of a given jet to satisfy a given combination
of passing or failing the online and offline \btagging can be computed
for all regions using the online-only
($\varepsilon_{b}^{\mathrm{Trig}}$), offline-only
($\varepsilon_{b}^{\mathrm{Off}}$), and conditional
($\varepsilon_{b}^{\mathrm{Trig|Off}}$) efficiencies, and employing
Bayes' theorem.  The efficiencies in each region can therefore be
defined in the following way:
\begin{enumerate}[label=(\roman*)]
\item{A jet that fails the trigger \btagging requirements and passes the offline \btagging requirements:
\[\varepsilon_{b}^{ \mathrm{\overline{Trig}} \land \mathrm{Off} } =
\left( 1 - \varepsilon_{b}^{\mathrm{Trig}|\mathrm{Off}} \right) \varepsilon_{b}^{\mathrm{Off}}.\]
}
\item{A jet that passes the trigger \btagging requirements and fails the offline \btagging requirements:
\[\varepsilon_{b}^{ \mathrm{Trig} \land \mathrm{\overline{Off}} } =
\varepsilon_{b}^{\mathrm{Trig}} - \varepsilon_{b}^{ \mathrm{Trig}|\mathrm{Off}} \varepsilon_{b}^{\mathrm{Off}}.\]
}
\item{A jet that fails the trigger \btagging requirements and fails the offline \btagging requirements:
\[\varepsilon_{b}^{\mathrm{\overline{Trig}} \land \mathrm{\overline{Off}}} =
1 - \varepsilon_{b}^{\mathrm{Off}} - \varepsilon_{b}^{\mathrm{Trig}} + \varepsilon_{b}^{\mathrm{Trig}|\mathrm{Off}} \varepsilon_{b}^{\mathrm{Off}}.\]
}
\end{enumerate}
In all cases the scale factors are subsequently defined as the
ratio of the  efficiencies measured in data and evaluated in simulation.

\subsection{Results}
 
The conditional \btagging efficiencies and the corresponding scale
factors as a function of offline jet \pT\ are shown in
Figures~\ref{fig:GB_eff_2016}, \ref{fig:GB_eff_2017}, and
\ref{fig:GB_eff_2018} for 2016, 2017, and 2018, respectively.
Efficiencies and scale factors are derived for all combinations of the
MV2 algorithm working points used online (40\%, 50\%, 60\%, 70\%,
77\%, 85\%) and the DL1r algorithm used offline (60\%, 70\%, 77\%,
85\%).  The \btagging conditional efficiency measurements were
carried out separately for each year and consistent results were
observed over time.  The results are shown for two representative
combinations (60\% and 85\% efficiency working points for both the
online and offline \btagging algorithms), for triggers used in 2016
(Figure~\ref{fig:GB_eff_2016}), 2017 (Figure~\ref{fig:GB_eff_2017}),
and 2018 (Figure~\ref{fig:GB_eff_2018}).  Similar behaviour is
observed for all combinations of online and offline working points.
 
\begin{figure}[htbp]
\centering
\subfloat[]{\includegraphics[width=0.5\textwidth]{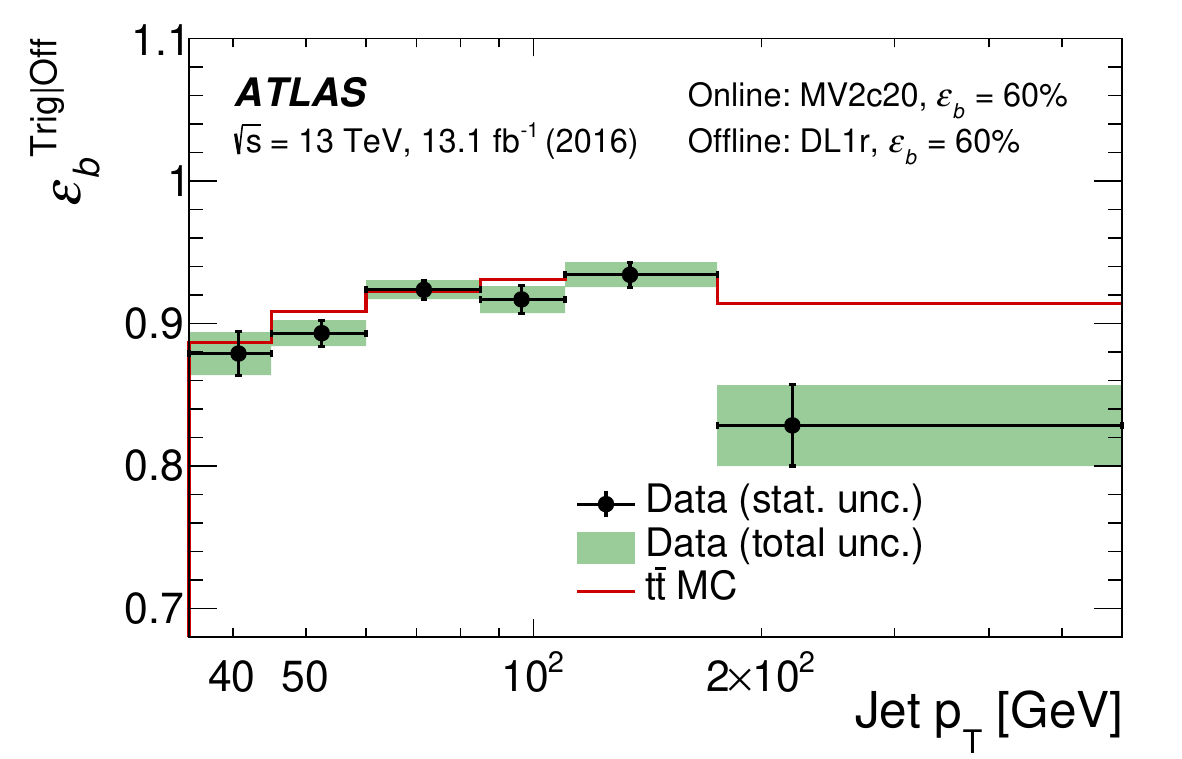}}
\subfloat[]{\includegraphics[width=0.5\textwidth]{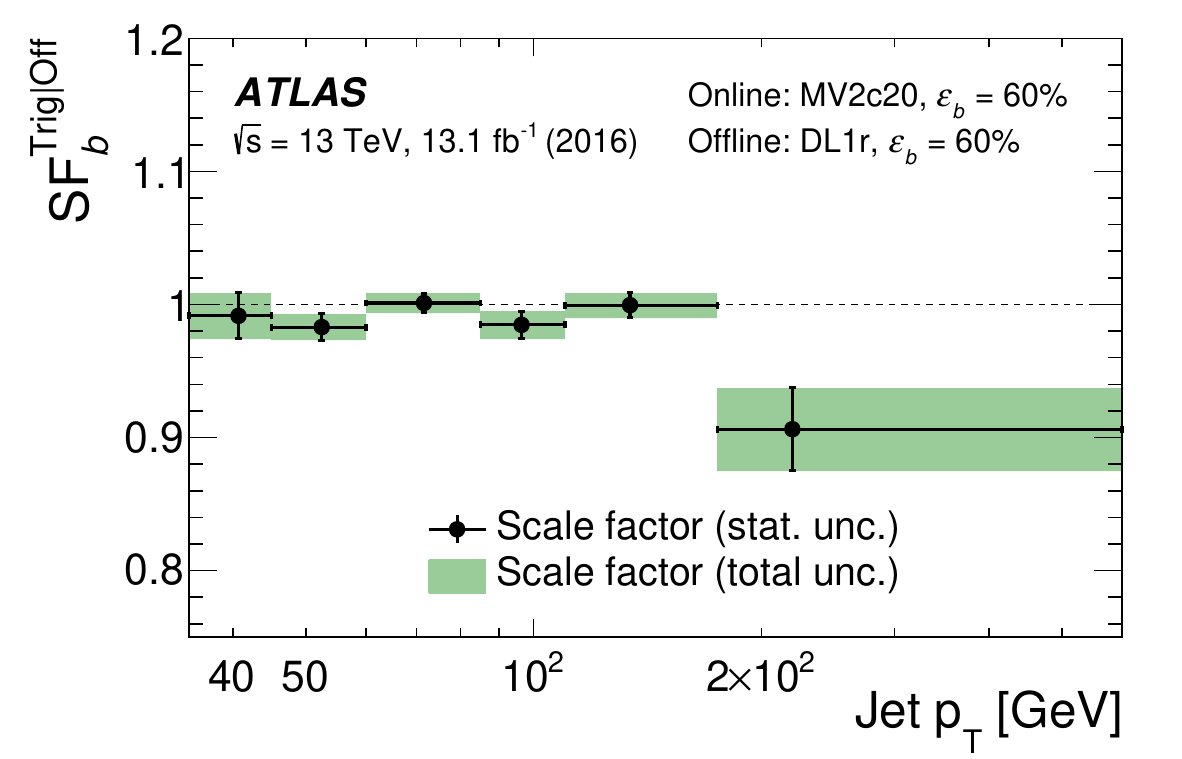}}\\
\subfloat[]{\includegraphics[width=0.5\textwidth]{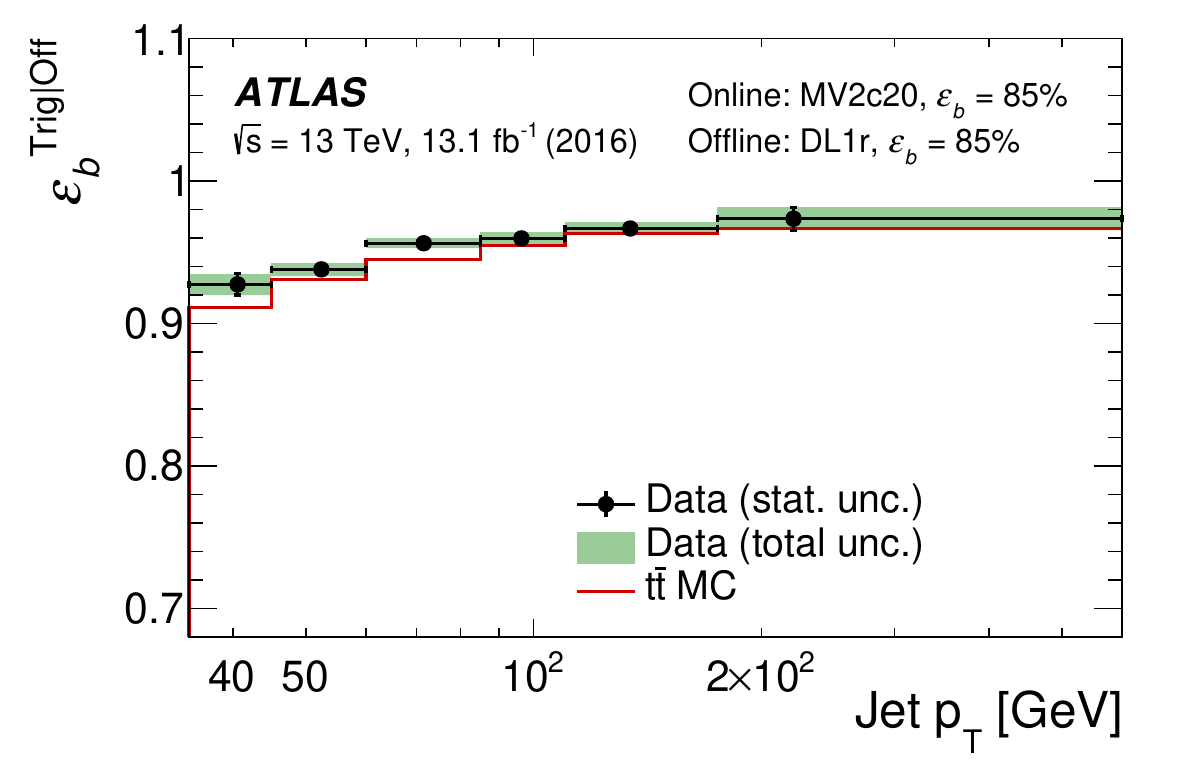}}
\subfloat[]{\includegraphics[width=0.5\textwidth]{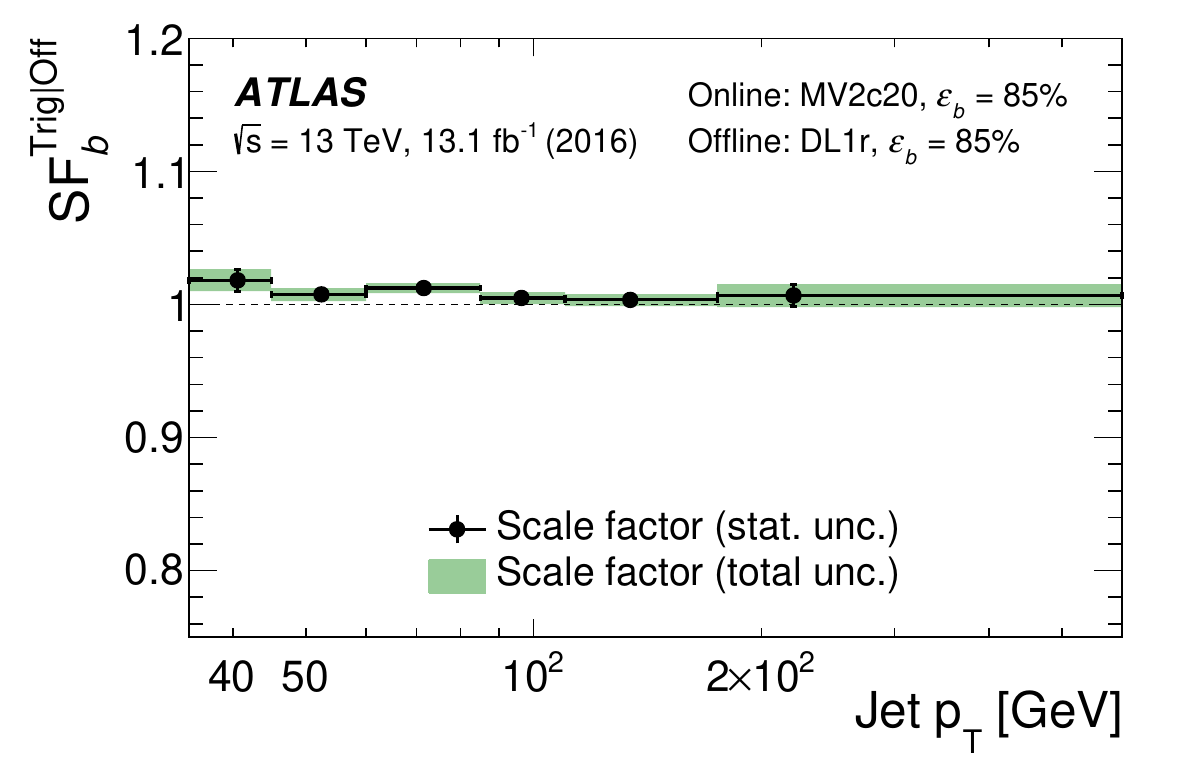}}\\
\caption {The conditional \btagging efficiency, \effcond, (a,c) and corresponding
scale factor, \sfcond, (b,d) for the online MV2c20 algorithm for the 60\%
efficiency working point (a,b) and the 85\% efficiency working point
(c,d) as measured in 2016 data.  Offline jets are required to
pass the same efficiency working points as the online jets, but
using the DL1r \btagging\ algorithm.  Vertical error bars include
data statistical uncertainties only, whereas the full band
corresponds to the sum in quadrature of all uncertainties.}
\label{fig:GB_eff_2016}
\end{figure}
 
\begin{figure}[htbp]
\centering
\subfloat[]{\includegraphics[width=0.5\textwidth]{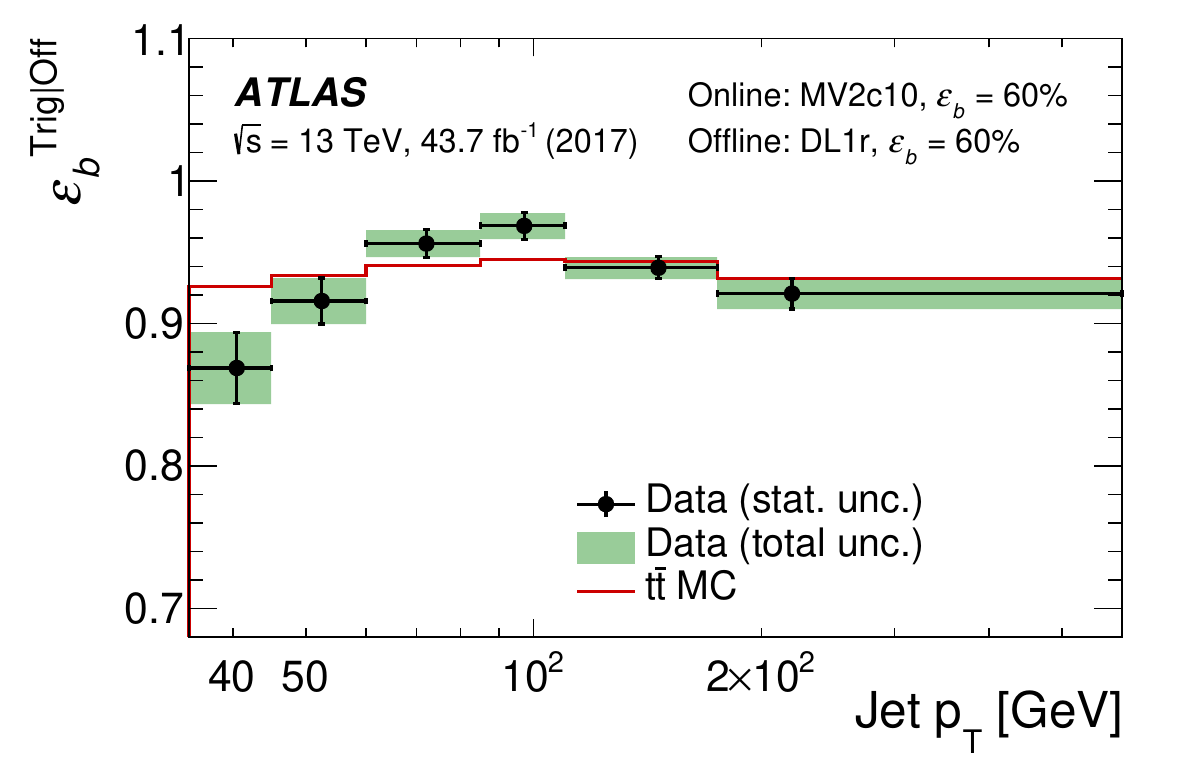}}
\subfloat[]{\includegraphics[width=0.5\textwidth]{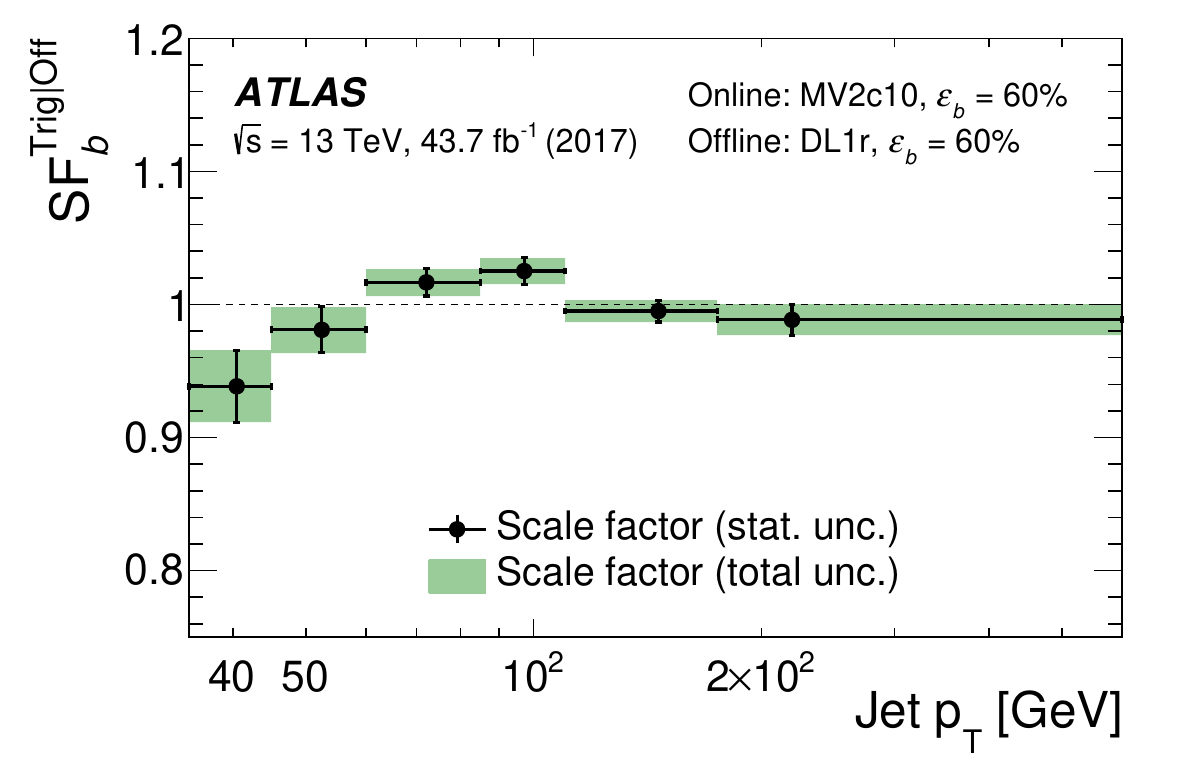}}\\
\subfloat[]{\includegraphics[width=0.5\textwidth]{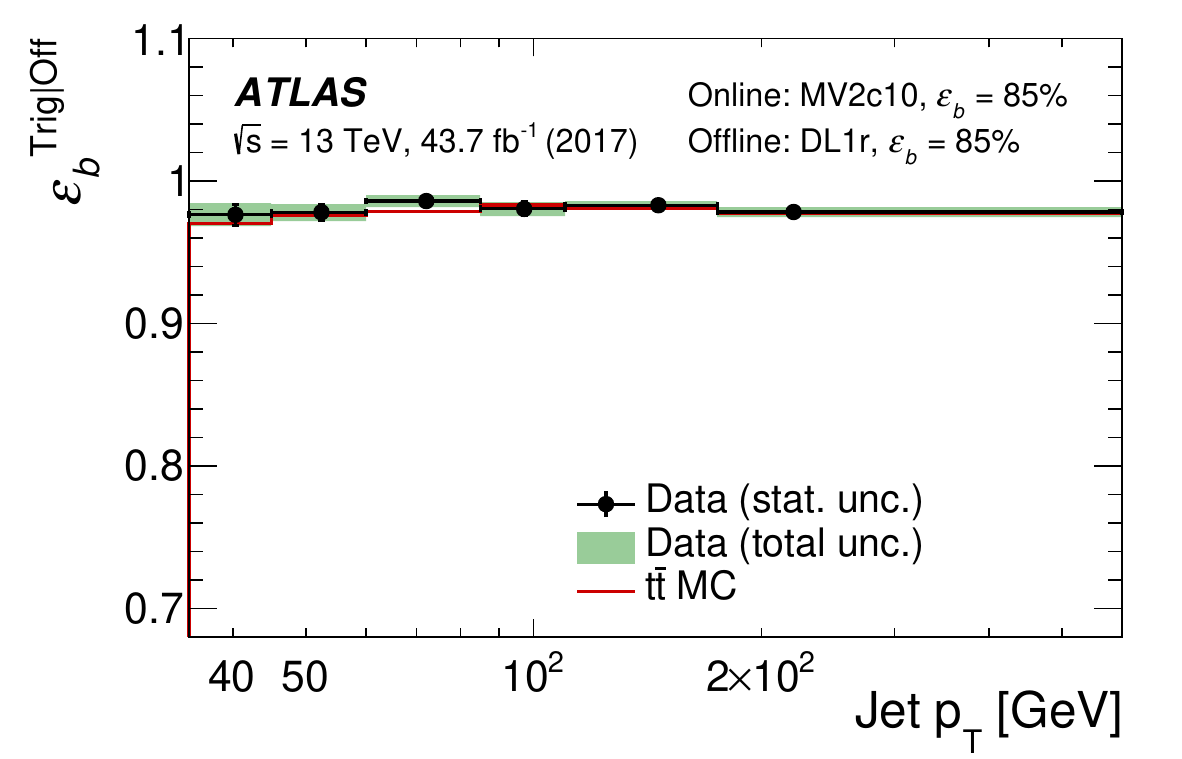}}
\subfloat[]{\includegraphics[width=0.5\textwidth]{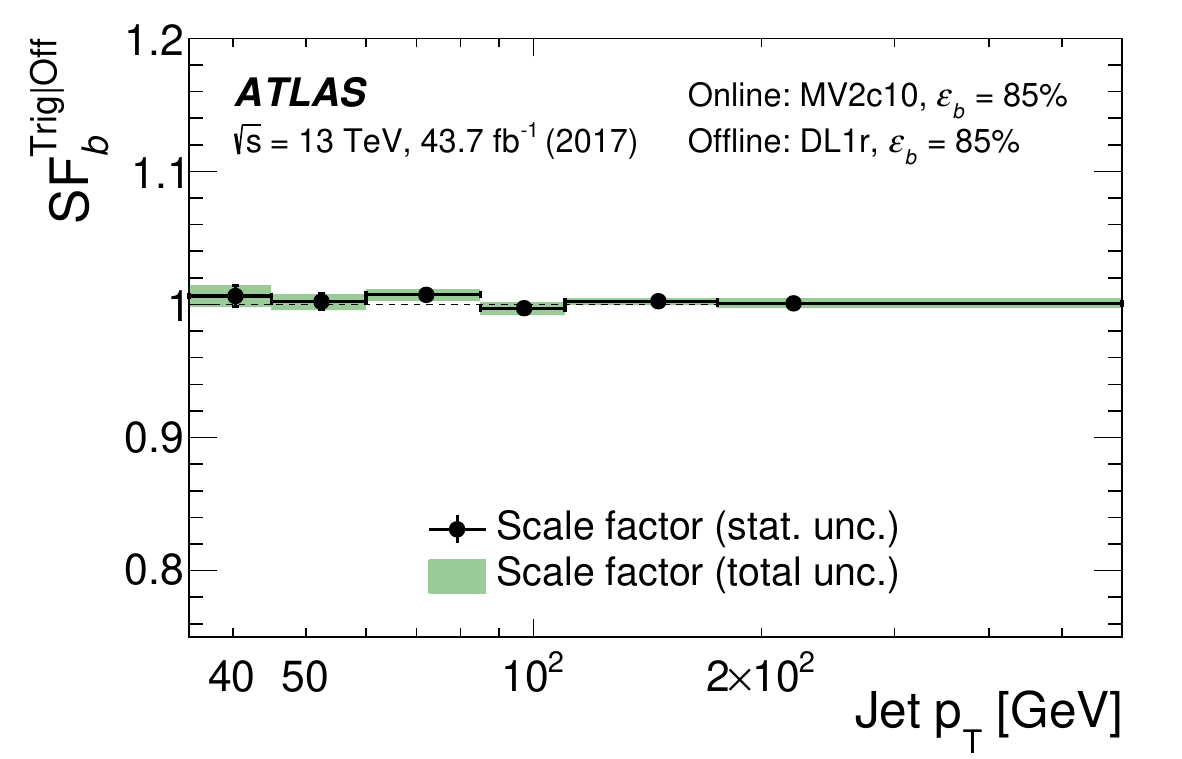}}\\
\caption {The conditional \btagging efficiency, \effcond, (a,c) and corresponding
scale factor, \sfcond, (b,d) for the online MV2c10 algorithm for the 60\%
efficiency working point (a,b) and the 85\% efficiency working point
(c,d) as measured in 2017 data.  Offline jets are required to
pass the same efficiency working points as the online jets, but
using the DL1r \btagging\ algorithm.  Vertical error bars include
data statistical uncertainties only, whereas the full band
corresponds to the sum in quadrature of all uncertainties.}
\label{fig:GB_eff_2017}
\end{figure}
 
\begin{figure}[htbp]
\centering
\subfloat[]{\includegraphics[width=0.5\textwidth]{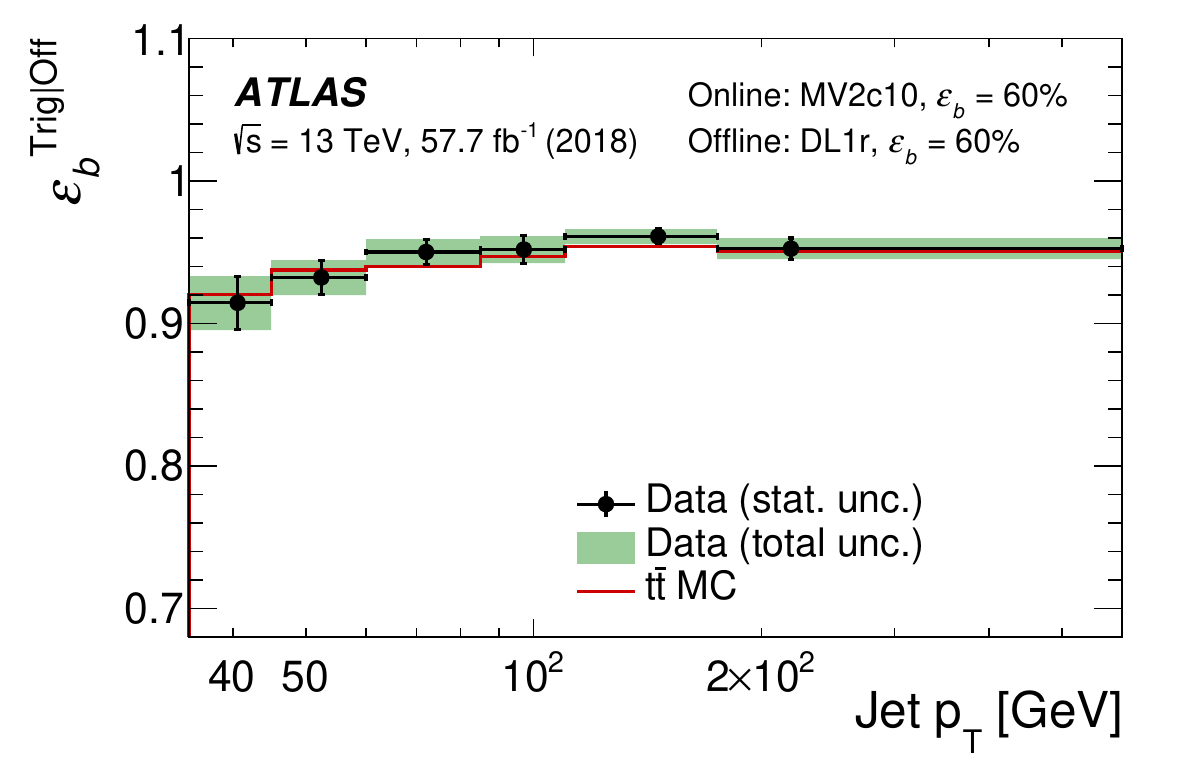}}
\subfloat[]{\includegraphics[width=0.5\textwidth]{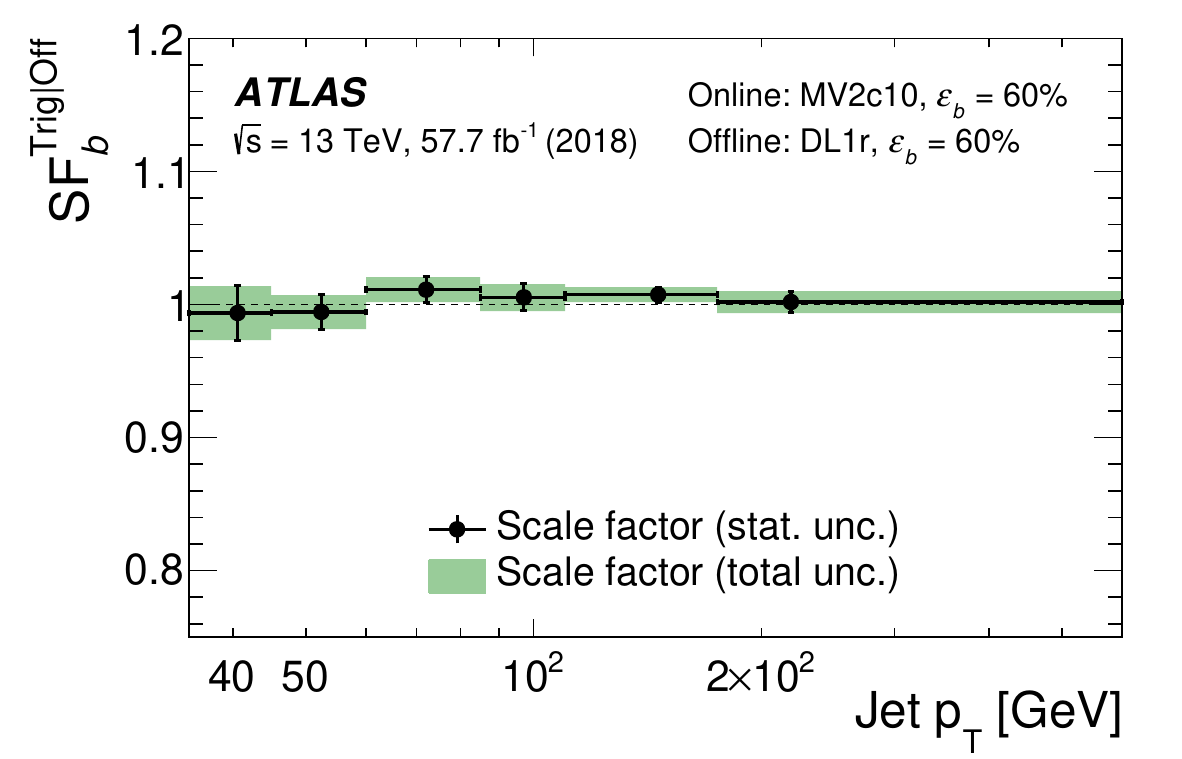}}\\
\subfloat[]{\includegraphics[width=0.5\textwidth]{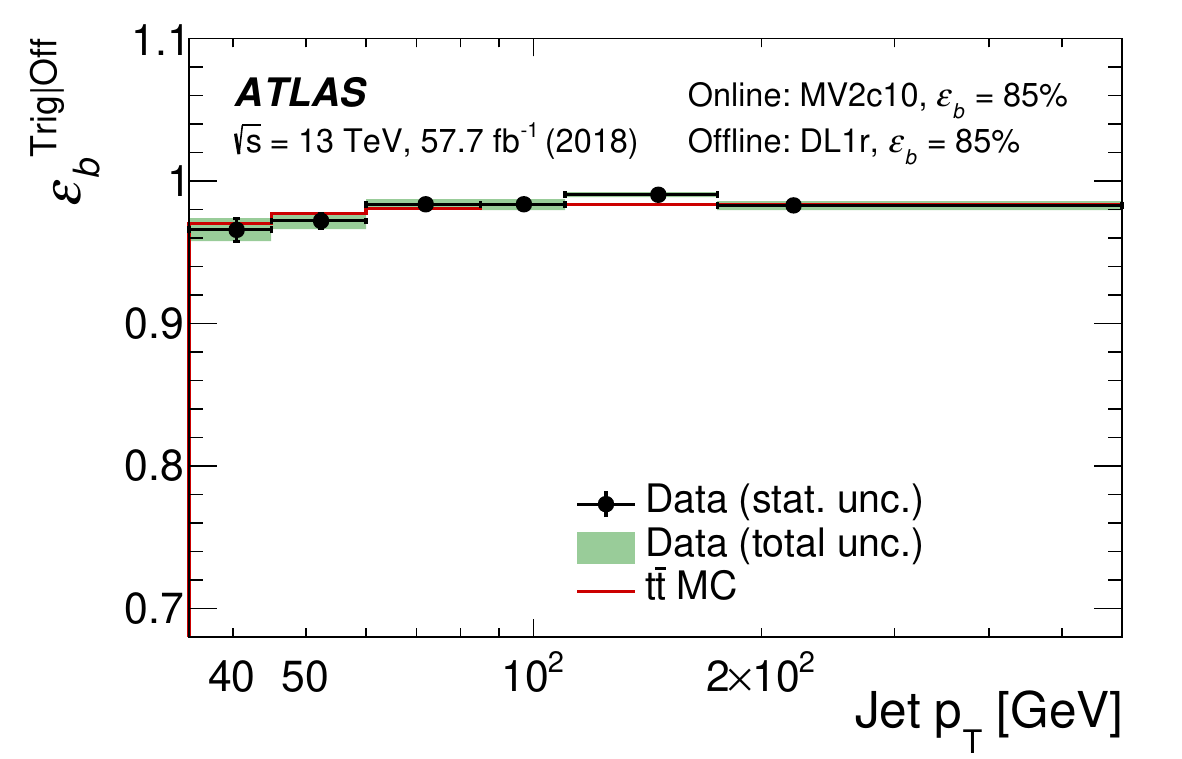}}
\subfloat[]{\includegraphics[width=0.5\textwidth]{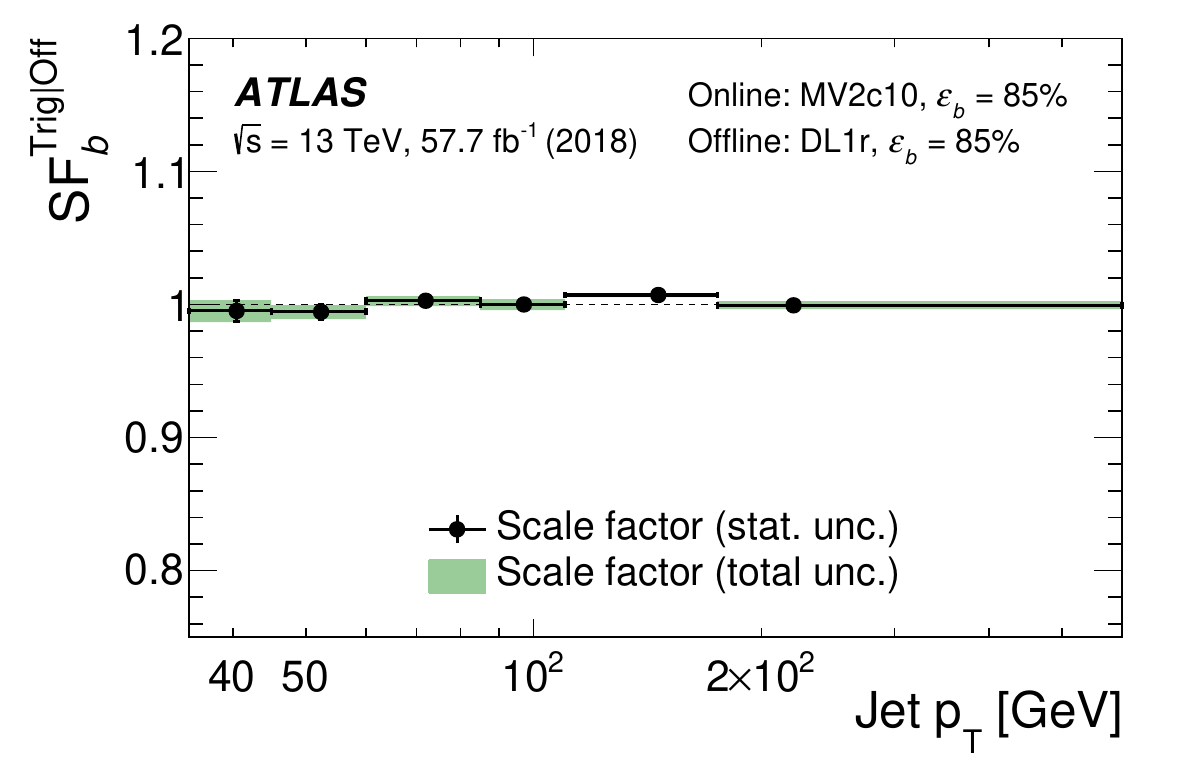}}\\
\caption {The conditional \btagging efficiency, \effcond, (a,c) and corresponding
scale factor, \sfcond, (b,d) for the online MV2c10 algorithm for the 60\%
efficiency working point (a,b) and the 85\% efficiency working point
(c,d) as measured in 2018 data.  Offline jets are required to
pass the same efficiency working points as the online jets, but
using the DL1r \btagging\ algorithm.  Vertical error bars include
data statistical uncertainties only, whereas the full band
corresponds to the sum in quadrature of all uncertainties.}
\label{fig:GB_eff_2018}
\end{figure}
 
The conditional efficiency obtained using the equivalent online and offline
working points ranges from approximately 85\% in the lowest \pt\ bins
(33--45\,\GeV) to approximately 98\% for higher-\pt\ jets.  The
conditional efficiency measured in data falls to $\sim$80\% for jets
with \pt~$>$~200\,\GeV\ that were recorded in 2016 data and are
required to pass the 60\% efficiency working point both online and
offline, as shown in Figure~\ref{fig:GB_eff_2016}(a).  It is noted that
the efficiency measured in this region in data is lower than the MC
prediction. Similar effects are observed for other combinations of
working points in 2016 data, with the efficiencies being lowest for
the tightest combinations of working points and recovered for the
loosest combinations of working points, for example when the 85\%
efficiency working point is used both online and offline in
Figure~\ref{fig:GB_eff_2016}(c).  The scale factors have values
consistent with unity in most other regions of jet \pt and in data
taken in other years, illustrating the generally good modelling of the
online \btagging performance, although differences in the scale factors
of up to $\sim$10\% are observed in some bins.
 
Uncertainties in the measurements are calculated following the same
procedures as described in Ref.~\cite{FTAG-2018-01} and any additional
sources of uncertainty specific to the trigger were found to be
negligible.  The total uncertainty in the measurement ranges from $<
1\%$ to about 5\% across the full jet \pT\ range.  Modelling
uncertainties are present in both the numerator and the denominator of
the conditional efficiency and so tend to cancel out, leaving the
statistical uncertainty to dominate the measurement.  Due to this
cancellation effect, uncertainties in the offline efficiency scale
factors tend to dominate over the uncertainties in the online
conditional efficiency scale factors in physics analysis.
Furthermore, uncertainties on the offline efficiency scale factors
tend to be dominated by uncertainties on the light-flavour jet
backgrounds, which have a smaller impact on the conditional efficiency
scale factors where the tighter requirements on the denominator mean
that the contamination from light-flavour jets is correspondingly
smaller.  Since the online conditional efficiency measurements use a
subset of the dataset used for the offline efficiency measurements,
and the same systematic variations are considered for both,
uncertainties in the two sets of scale factors are correlated.  Few
data events satisfy all of the selection criteria described in
Section~\ref{subsec:calibration_selection} at very high jet \pT, and
the statistical uncertainties associated with the results are largest
in this region.
 
For the online-only efficiencies with the tightest working points, the
scale of the systematic uncertainty approaches that of the statistical
uncertainty.  In these cases the total uncertainty on the efficiency
measurements reaches 10\% in some bins, while the total uncertainty on
the scale factors can be up to 20\% in some bins.  The total uncertainties
on the efficiencies and scale factors for looser working points are
typically less than 5\%.  The largest source of systematic uncertainty
comes from the modelling of top-quark events, in particular the impact
of using a different parton shower and hadronisation model for
simulated \ttbar\ events.  This uncertainty was evaluated by comparing
the nominal \ttbar sample with another event sample configured with
the same setup to produce the matrix elements, but interfaced with
\HERWIG~7.04~\cite{Bahr:2008pv,Bellm:2015jjp}, using the H7UE set of
tuned parameters~\cite{Bellm:2015jjp} and the MMHT2014\textsc{lo}\ PDF
set~\cite{Harland-Lang:2014zoa}.  All other systematic uncertainties
have a very small impact.
 
A method for reducing the total number of uncertainties while
preserving the bin-by-bin correlations is provided for use in physics
analyses by performing an eigenvector decomposition.  Versions of the
scale factors that have been smoothed in jet \pt\ are also provided in
order to prevent distortions in the variables of interest induced by
the application of the scale factors. Both the eigenvector
decomposition and the smoothing procedure are applied using the method
described in Ref.~\cite{PERF-2016-05}.
 
Conditional efficiencies and scale factors are also provided for jets
\btagged offline with the MV2c10 algorithm, using the same method, but
not presented in this work as the MV2c10 algorithm is now superseded
by DL1r.  As expected, the conditional efficiencies are up to a few
percent higher and the uncertainties are slightly reduced for the
tightest combinations of working points when MV2c10 rather than DL1r
is used offline, due to increased correlation between the online and
offline \btagging algorithms.  Any decrease in the degree of
correlation between the taggers when moving from MV2c10 to DL1r for
offline \btagging is more than compensated for in analyses by the
improved performance that DL1r offers.
 
The \bjet trigger conditional efficiency as a function of pile-up is
shown for data and simulated \ttbar\ events in
Figure~\ref{fig:eff_pileup}.  Offline jets are required to pass the
70\% efficiency working point of the DL1r algorithm.  Only the
statistical uncertainties (which dominate the measurement) are
included.  The MV2 and DL1r \btagging algorithms are tuned to provide
constant efficiency under conditions of increasing pile-up.
Consequently, the conditional efficiencies typically fall by less than
5\% over the full range of pile-up conditions for all combinations of
online and offline working points, although drops of up to 10\% are
observed in events with the least pile-up in 2016 data.
 
\begin{figure}[htbp]
\centering
\subfloat[]{\includegraphics[width=0.5\textwidth]{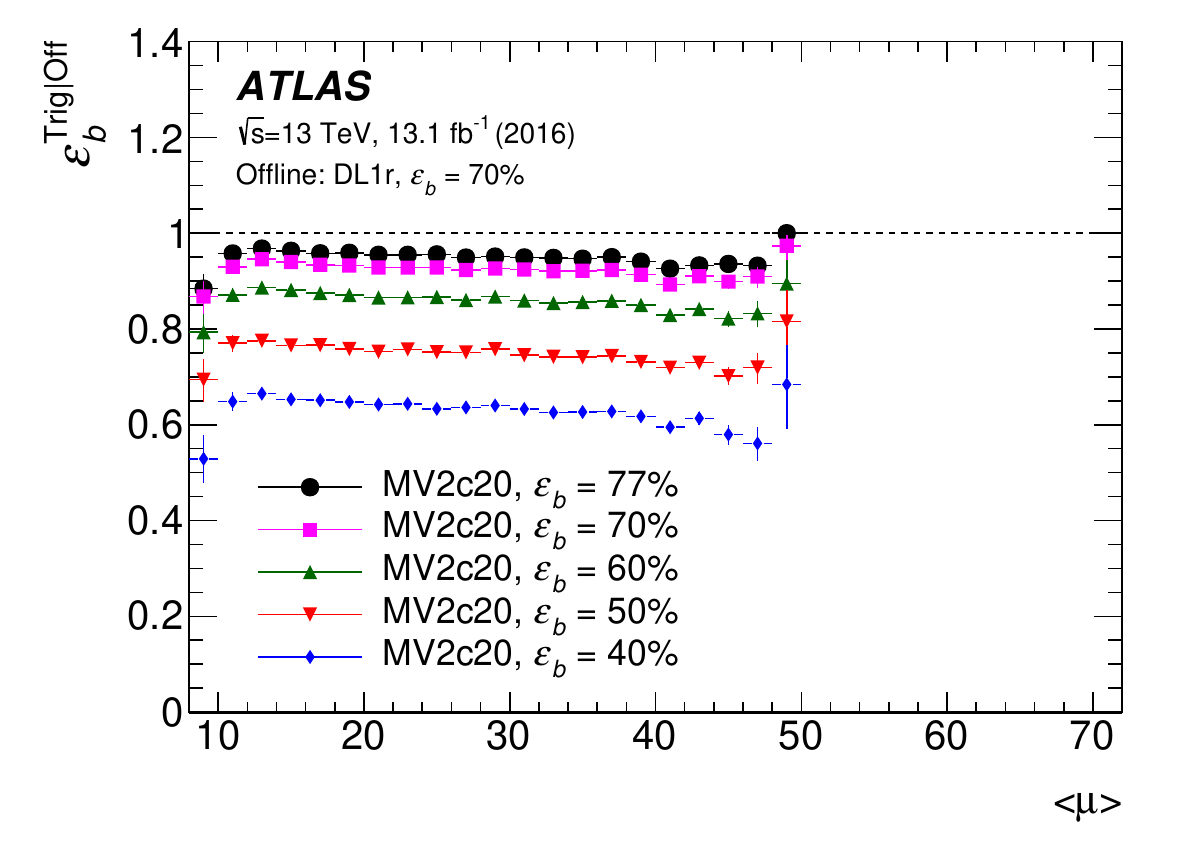}}
\subfloat[]{\includegraphics[width=0.5\textwidth]{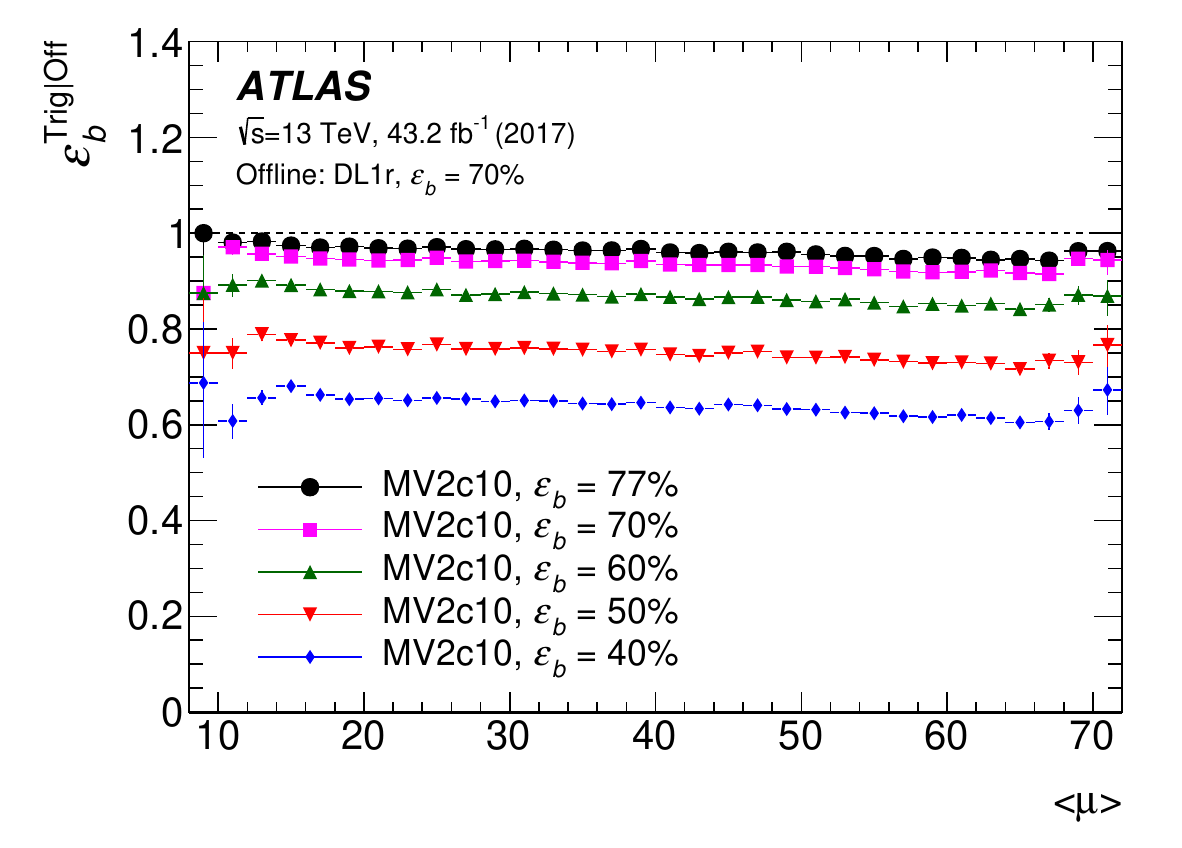}}\\
\subfloat[]{\includegraphics[width=0.5\textwidth]{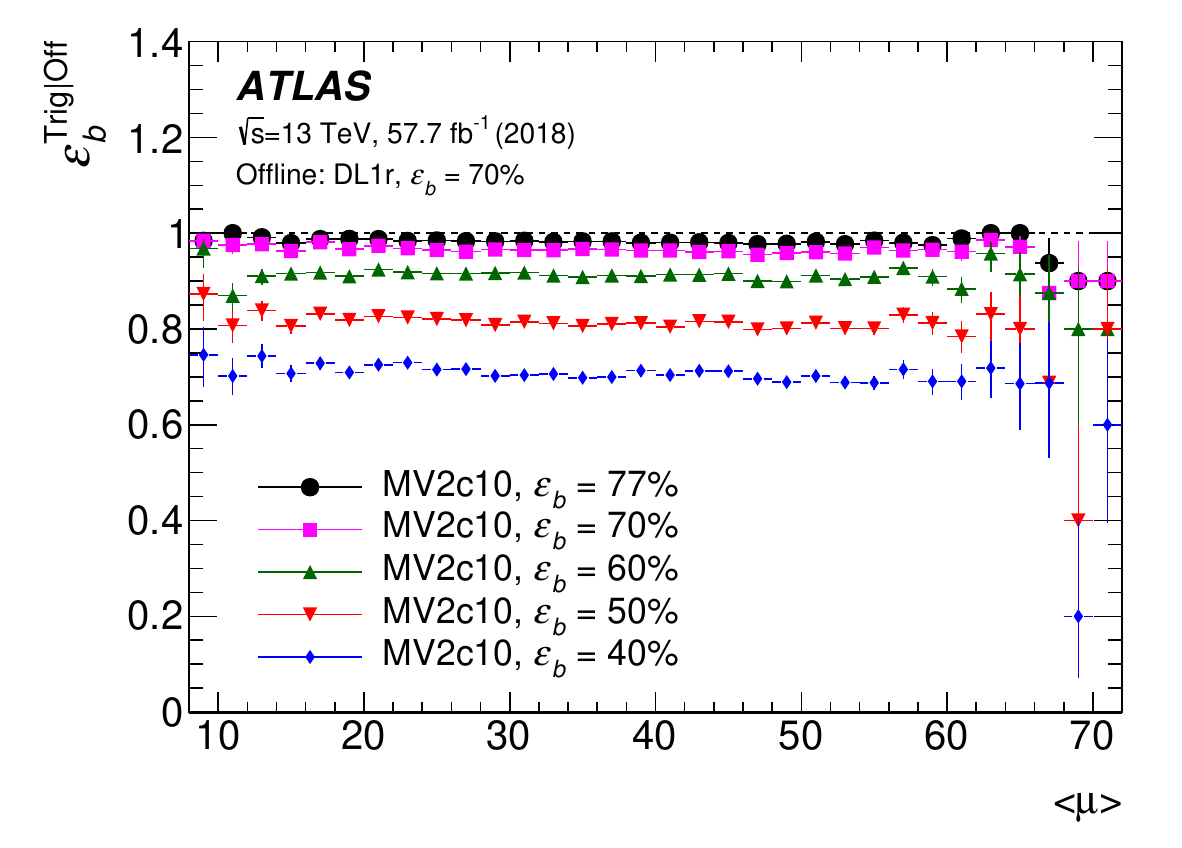}}
\subfloat[]{\includegraphics[width=0.5\textwidth]{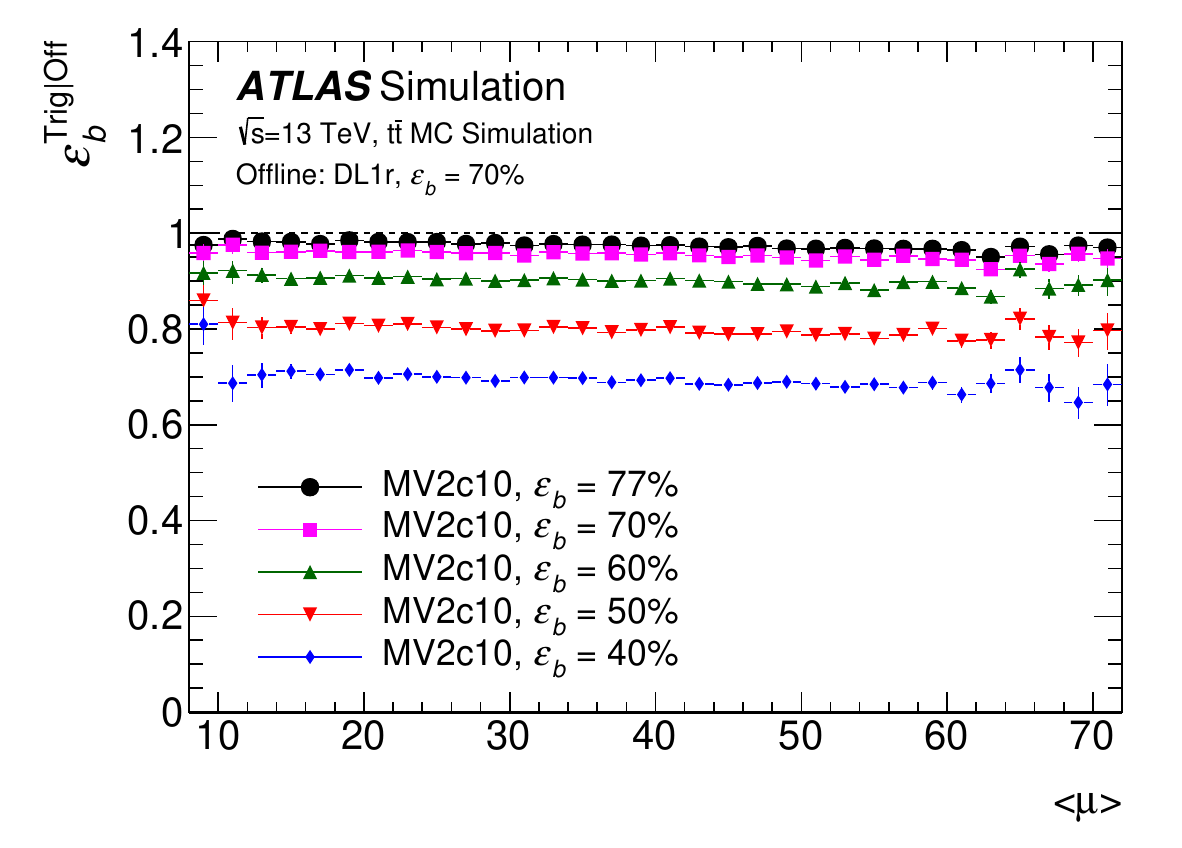}}
\caption {The conditional \btagging efficiency, \effcond, for data
taken in (a) 2016, (b) 2017, (c) 2018, and (d) simulated
\ttbar\ events, as a function of pile-up, \pileupmu.  All
working points of the MV2 \btagging algorithm used online are
shown in combination with offline jets that are required to pass
the 70\% efficiency working point of the DL1r algorithm.  Only
statistical uncertainties are shown.}
\label{fig:eff_pileup}
\end{figure}

\section{Muon-jet triggers}
\label{sec:muon_jets}
 
Approximately 20\% of \bjets contain a muon from the decay chain of
the $b$-hadron.  These muons are typically soft and produced at small
angles relative to the axis of the jet (typically within $\Delta R =
0.5$).  The low \pt of these leptons plus the additional hadronic
activity around them mean that they cannot be triggered on using the
standard ATLAS lepton triggers~\cite{TRIG-2016-01} which include
isolation requirements for all but the highest-\ET\ items, in order to
reject fake-lepton backgrounds.  Dedicated triggers are therefore
designed to select low-\pt muons that are geometrically matched to a
jet -- a `muon-jet'.  Requiring the presence of a muon-jet in the
event increases the rejection power against light-flavour jet
backgrounds and allows these semileptonic \bjet triggers to reach
lower in jet \ET\ than the standard \bjet triggers.
 
Muon-jet triggers are used to provide a sample of \bjet-enriched data
used to calibrate the \btagging algorithms used offline, and also have
potential to enhance the acceptance efficiency for processes containing
a large number of \bjets and/or ones with low \pT (described in
Section~\ref{subsec:muon_jets_pp}).  They also provide the only way to
select events containing \bjets during lead-ion collision runs,
where events typically have a large number of jets and high track
multiplicity, and running the standard \bjet triggers becomes
unfeasible due to the high rates and high CPU cost of running tracking
on all jets.

\subsection{Muon-jet triggers for heavy-ion collisions}
 
One of the open questions regarding the quark--gluon plasma (QGP)
created in heavy-ion (HI) collisions at the LHC is the energy loss
mechanisms that partons experience while traversing the hot and dense
QCD medium~\cite{Blaizot:2015lma}.
Heavy quarks are produced at the early stages of the ion collisions in
scattering processes that involve large momentum transfers, $Q$, so
their formation time, of the order $1/Q < 0.1~\mathrm{fm}/c$, is much smaller
than the lifetime of the QGP, estimated to be 10--11~$\mathrm{fm}/c$\ at the
LHC~\cite{Aamodt:2011mr}.  The energy loss of heavy quarks in the QGP
is predicted to be smaller than that of light-flavour quarks, due to the gluon
radiation suppression at small angles -- the so called `dead cone'
effect~\cite{Dokshitzer:2001zm}.
 
In 2018, ATLAS collected 1.42~nb$^{-1}$ of data from collisions of lead
ions with a nucleon--nucleon centre-of-mass energy $\sqrt{s_{_\text{NN}}}~=~5.02\,\TeV$.
Dedicated triggers were necessary not only to fulfil the specific
physics requirements, but also to accommodate the different detector
environment during \PbPb\ data-taking, resulting from the intrinsic
geometry of the nuclear overlap leading to large variations of both
track multiplicity and energy density, compared with \pp\ runs.
During \PbPb\ data-taking, it would be prohibitive to run
the \bjet triggers developed for \pp collisions, owing to the high rates
and large CPU cost of triggering in the relevant jet \ET\ range.
Muon-jet triggers that require a muon and jet that are geometrically
matched within $\Delta R < 0.5$\ are used instead to provide a sample
of data events that are enriched in semileptonic $b$-hadron decays.
 
Several different muon-jet triggers imposing various combinations of
muon \pt\ and jet \ET\ thresholds were provided.  In most cases these
were seeded at L1 by a single muon with $\pt > 4$ or 6\,\GeV, although
in one instance a L1 jet was additionally required.  In the HLT, a
muon with $\pt > 4$\ or 6\,\GeV\ within $\Delta R = 0.5$\ of a jet with
$\ET > 30$, 40, 50, or 60\,\GeV\ was required.  Jets were
reconstructed using the \antikt\ algorithm with radius parameter $R =
0.2$, 0.3 or 0.4, and corrected for the underlying event produced in
heavy-ion collisions, as detailed in Ref.~\cite{HION-2011-02}.  The
list of triggers was designed to be optimal within the allocated
trigger acceptance rate of approximately 80~Hz and is summarised in
Table~\ref{tab:menu_heavyions}.  In order to accommodate the
increasing instantaneous luminosity during the data-taking period and
ensure that that output rate remained within the rate allocation, the
set of triggers that required a muon with \pT$~>~$4\,\GeV\ and applied
no additional jet requirements at L1 were prescaled for some runs.
The prescale factors were applied coherently to all of the triggers
and the values ranged from 1.0 (i.e.\ unprescaled) to 1.307.  The
average prescale factor across the entire \PbPb\ data-taking period in
2018 was 1.065.
 
\begin{table}[ht]
\begin{center}
\renewcommand{\arraystretch}{1.2}
\caption{Details of the semileptonic $b$-decay (muon-jet) triggers
used during the \PbPb\ collision run in 2018.  In all cases, the
muon and jet selected by the HLT are required to be within $\Delta
R = 0.5$ of each other.  Prescale factors were applied coherently
to all prescaled triggers and the values ranged from 1.0 (i.e.\ unprescaled) to 1.307.  The
average prescale factor across the entire \PbPb\ data-taking period
in 2018 was 1.065.\vspace{2mm} }
\begin{tabular}{ c c c c c c c } \toprule
\multicolumn{3}{c}{L1}                    & \multicolumn{3}{c}{HLT}              & \multirow{2}{*}{Prescaled?} \\ \cmidrule(lr){1-3} \cmidrule(lr){4-6}
Muon \pT     & Jet \ET   & Jet $|\eta|$   & Muon \pT  & Jet \ET    & Jet Radius  &                             \\ \toprule
4\,\GeV      & \multicolumn{2}{c}{-}      & 4\,\GeV   & 40\,\GeV   & 0.4         & Yes                         \\
4\,\GeV      & \multicolumn{2}{c}{-}      & 4\,\GeV   & 50\,\GeV   & 0.4         & Yes                         \\
4\,\GeV      & 15\,\GeV  & $<$3.2         & 4\,\GeV   & 60\,\GeV   & 0.4         & No                          \\
4\,\GeV      & \multicolumn{2}{c}{-}      & 4\,\GeV   & 30\,\GeV   & 0.2         & Yes                         \\
4\,\GeV      & \multicolumn{2}{c}{-}      & 4\,\GeV   & 30\,\GeV   & 0.3         & Yes                         \\
4\,\GeV      & \multicolumn{2}{c}{-}      & 4\,\GeV   & 40\,\GeV   & 0.2         & Yes                         \\
4\,\GeV      & \multicolumn{2}{c}{-}      & 4\,\GeV   & 40\,\GeV   & 0.3         & Yes                         \\
6\,\GeV      & \multicolumn{2}{c}{-}      & 6\,\GeV   & 40\,\GeV   & 0.4         & No                          \\
6\,\GeV      & \multicolumn{2}{c}{-}      & 6\,\GeV   & 50\,\GeV   & 0.4         & No                          \\ \bottomrule
\end{tabular}
\label{tab:menu_heavyions}
\end{center}
\end{table}
 
The HLT conditional muon-jet trigger efficiency is defined as the
number of offline muon-jet objects satisfying the muon-jet trigger
requirements, divided by the total number of offline muon-jets that
fired a single-muon trigger with the same \pt\ threshold as the
muon-jet trigger used in the numerator.  The offline muon-jets are
constructed from muons passing a `Tight' identification working point
(corresponding to an efficiency of approximately 90\% for true muons
in MC events) \cite{PERF-2015-10}, matched to a jet within a distance
of $\Delta R < 0.5$.  Both the muon and the jet are required to have
$|\eta| < 2.4$.  The conditional muon-jet trigger efficiency for a
given HLT muon-jet trigger is therefore defined as
\begin{equation}
\varepsilon^{\mu\mathrm{j}|\mu} =
\frac{N_{\mu\mathrm{j}}^{\mathrm{Trig}+\mathrm{Off}}}
{N_{\mu}^{\mathrm{Trig}+\mathrm{Off}}},
\label{eq:mujet_trig_eff_cond} 
\end{equation}
where $N_{\mu\mathrm{j}}^{\mathrm{Trig}+\mathrm{Off}}$\ is the number
of muon-jet objects passing the HLT and offline muon-jet selections, and
$N_{\mu}^{\mathrm{Trig}+\mathrm{Off}}$\ is the number of muon-jets passing
the HLT and offline muon requirements.
 
The events passing the muon-jet trigger are an exact subset of events
that pass the single-muon trigger with the same \pT\ threshold and so
the absolute muon-jet trigger efficiency can be defined as the product
of the conditional trigger efficiency given in
Eq.~(\ref{eq:mujet_trig_eff_cond}), and the single-muon trigger
efficiency ($\varepsilon^{\mu}$) which was measured using the
method described in Ref.~\cite{HION-2016-07}:
\begin{equation}
\varepsilon^{\mu\mathrm{j}} =
\varepsilon^{\mu\mathrm{j}|\mu}\,\varepsilon^{\mu}.
\label{eq:mujet_trig_eff_abs}
\end{equation}
 
The performance of the muon-jet trigger is constrained by the limited
acceptance of the L1 trigger, based on the information received from
the calorimeters and muon trigger chambers.  The geometric coverage of
the latter is $\sim$99\% in the endcap ($1.05 < |\eta| < 2.40$)
regions and $\sim$80\% in the barrel region ($|\eta| < 1.05$)
\cite{TRIG-2018-01}.  The measurements are therefore made separately
in the two pseudorapidity ranges.
The efficiency is also measured for different categories of collision
centrality, in order to account for a possible decrease in
performance due to the characteristics of \PbPb\ collisions.  The
centrality of a collision is assessed on an event-by-event basis using
the \ET\ deposited in the forward calorimeters, \ETFCal\ in
$3.2\leq|\eta|<4.9$.  The Glauber MC model~\cite{Glauber} is
used to obtain a correspondence between the \ETFCal\ and the sampling
fraction of the total inelastic \PbPb\ cross-section, allowing
centrality percentiles to be set~\cite{HION-2011-05}.  In this
analysis, central collisions are defined as those in the 0--40\%
centrality interval where the contribution from underlying-event
effects is the largest.  Peripheral collisions are those within the
40--80\% centrality interval.
 
The performance of muon-jet triggers where the muon \pt~threshold is
4\,\GeV\ and the muon must be within $\Delta R=0.5$ of a jet passing an
\ET~threshold of 40, 50 or 60\,\GeV\ is presented relative to the
single-muon trigger that requires a muon with $\pT > 4$\,\GeV\ at L1
and in the HLT.\footnote{The efficiency is calculated in the same way for the
muon-jet trigger that requires a $>4\,\GeV$\ muon and $>60\,\GeV$\
jet and additionally includes a jet requirement in its L1
seed.}  The efficiency of this single-muon trigger was measured in
Ref.~\cite{HION-2016-07} to be approximately 80\% and 85\% in the
barrel region, for central and peripheral collisions,
respectively. This low efficiency is a consequence of the lower
acceptance of the L1 trigger. In the
endcap region the efficiency is noticeably higher, reaching 97\%, and
less sensitive to the centrality of the collision.
 
Figure~\ref{fig:1dim_mu4_j456} compares the efficiency of the three
muon-jet triggers as a function of the offline jet \pt for events
passing the single-muon trigger and containing an offline muon with
\pt $>$ 12\,\GeV.  In peripheral collisions and in the barrel region
the efficiency is above 99\% for offline jets with \pt\ larger than
46, 59, and 66\,\GeV\ (for triggers with 40, 50, and 60\,\GeV\ jet
\ET\ thresholds, respectively). The efficiency saturates at slightly
higher jet \pt\ values in the endcap region.  In central collisions
the turn-on is slower than in peripheral collisions and the range
with full efficiency starts at higher \pt\ values. This sensitivity to the
centrality of the collisions is also observed in inclusive jet trigger
efficiency measurements.
 
\begin{figure}[htbp]
\centering
\subfloat[]{\includegraphics[width=0.5\textwidth]{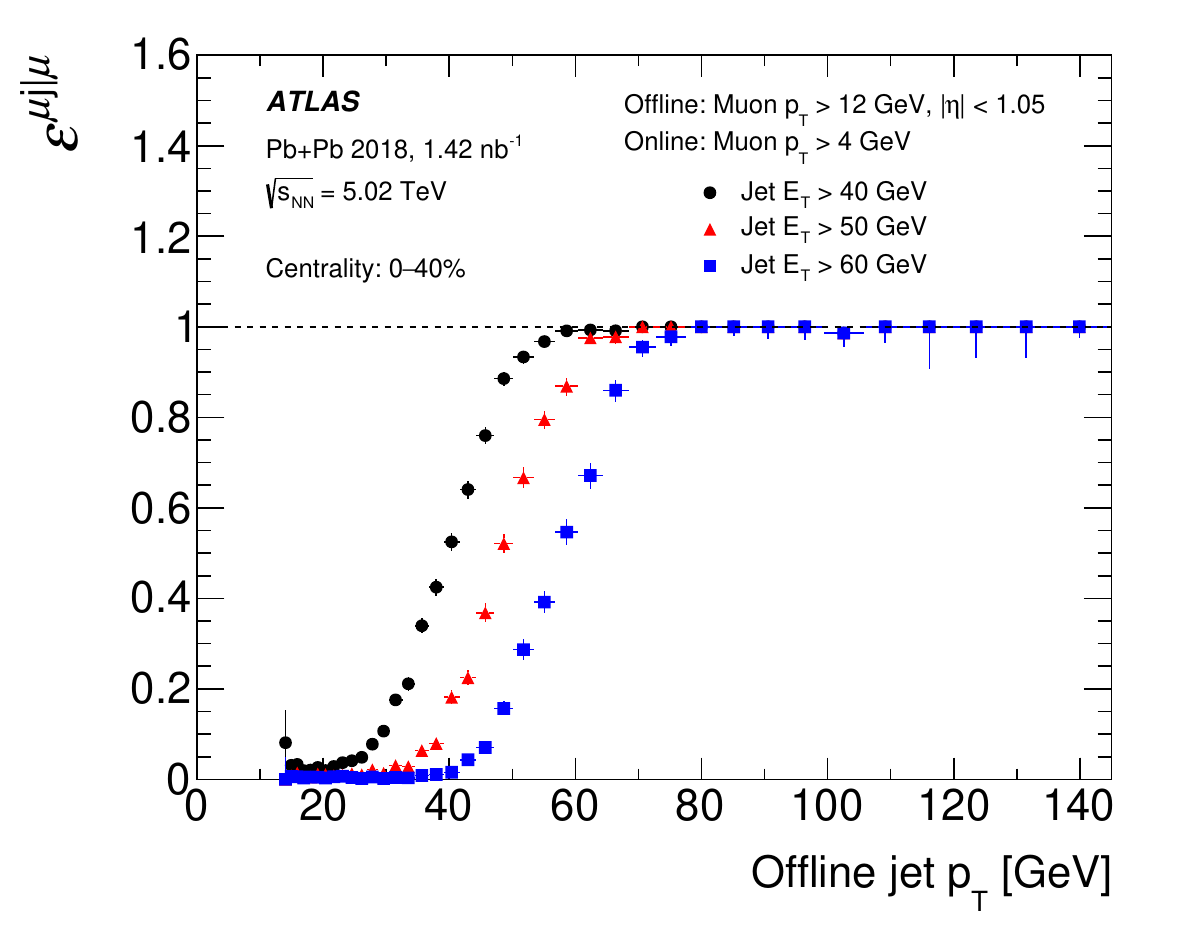}}
\subfloat[]{\includegraphics[width=0.5\textwidth]{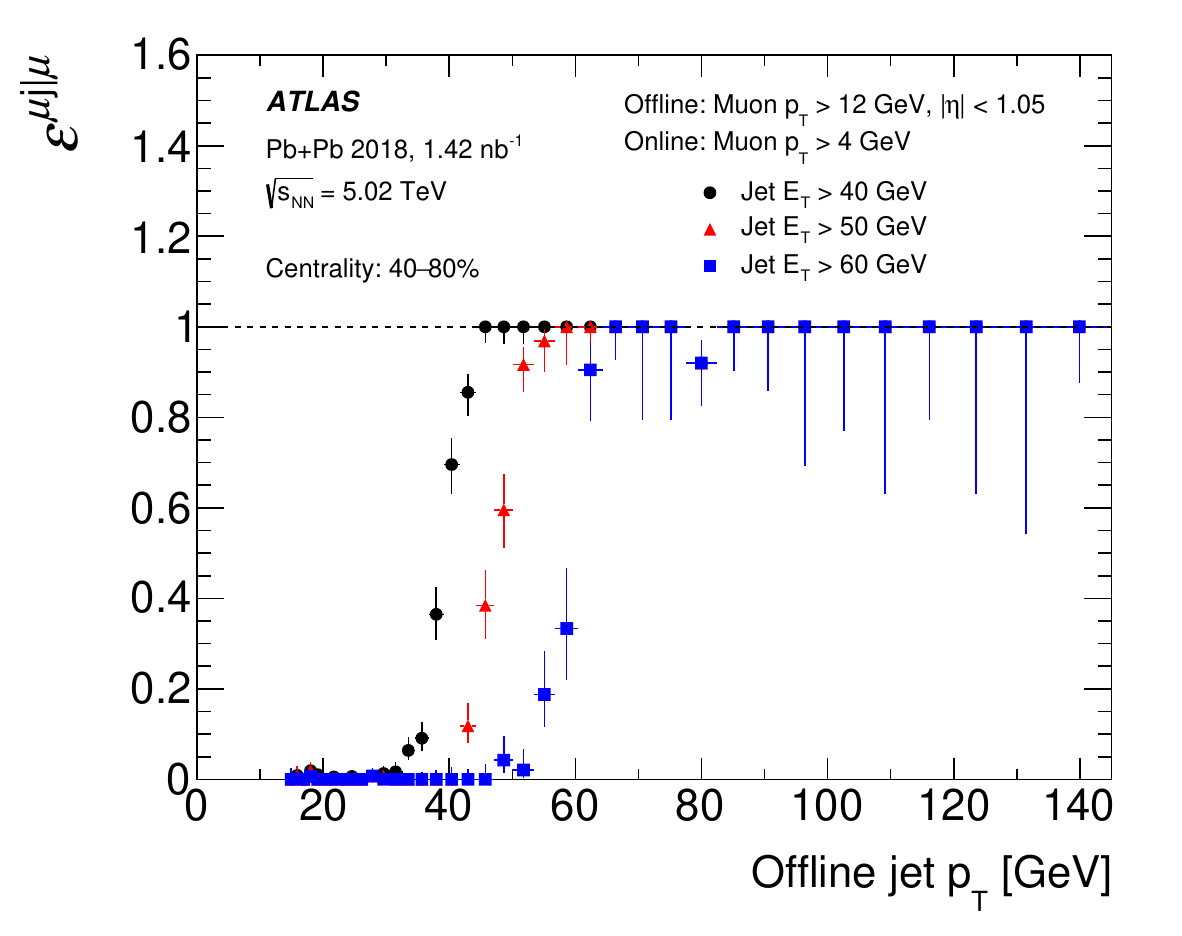}}\\
\subfloat[]{\includegraphics[width=0.5\textwidth]{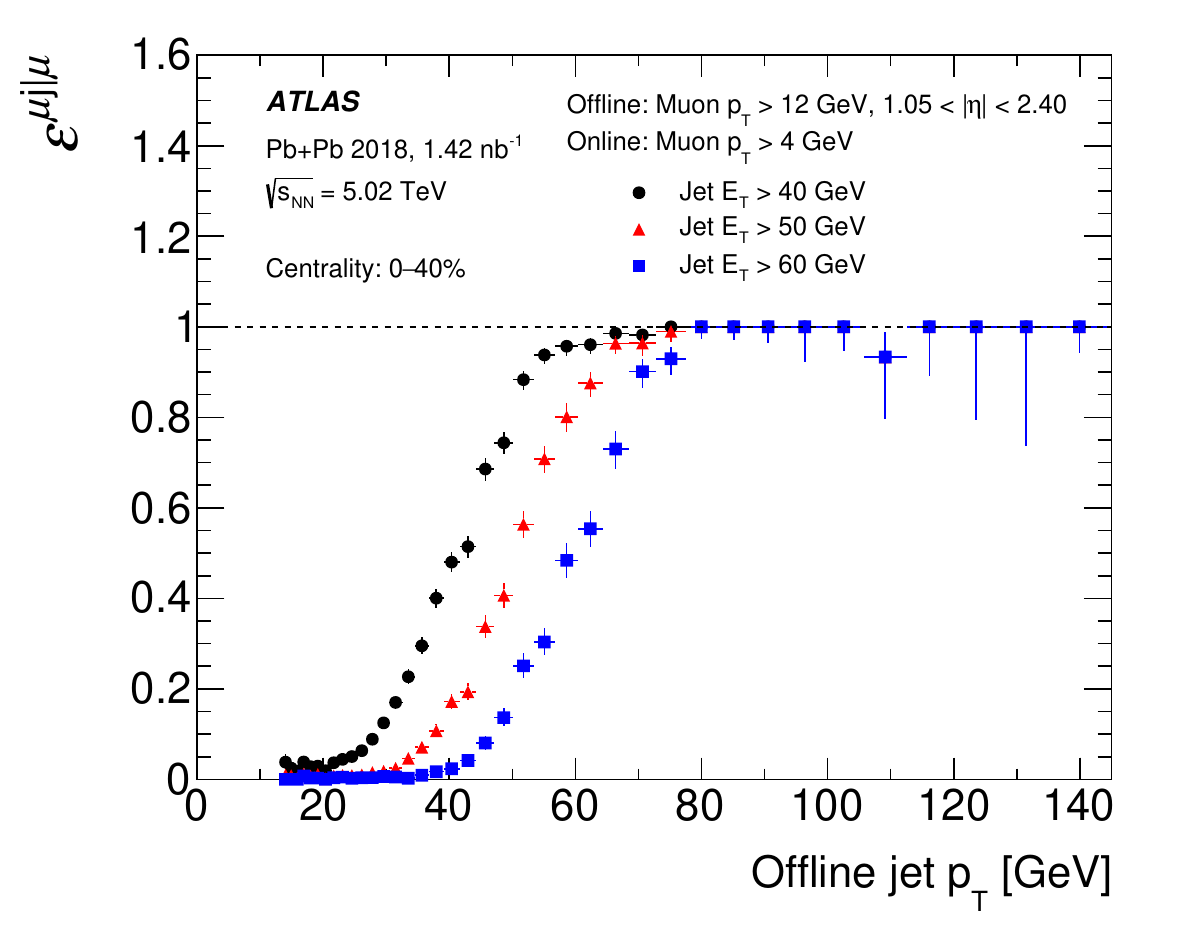}}
\subfloat[]{\includegraphics[width=0.5\textwidth]{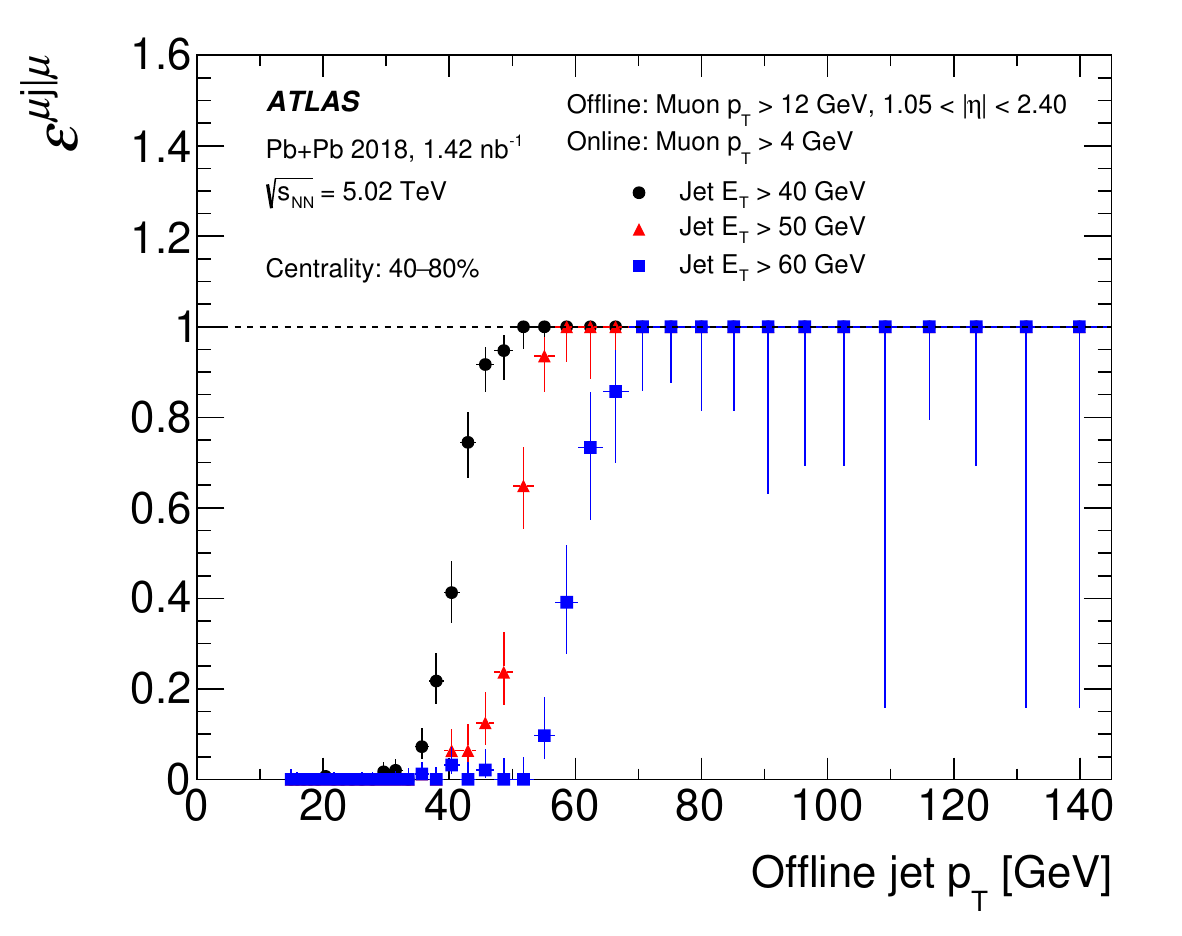}}
\caption{The conditional muon-jet trigger efficiency, $\varepsilon^{\mu\mathrm{j}|\mu}$, as defined in
Eq.~(\ref{eq:mujet_trig_eff_cond}), as a function of offline jet
\pt.  The muon-jet triggers shown require a muon with \pt $>$
4\,\GeV\ and jet \ET\ thresholds of 40, 50, and 60\,\GeV. The
muon-jet trigger efficiency is estimated relative to the
single-muon trigger requiring a muon with $\pt > 4$\,\GeV.  Events are
also required to have an offline muon with \pt\ more than
12\,\GeV.  The last bin contains overflow.  The performance is
shown in the (a,b) barrel and (c,d) endcap regions of the
detector, for (a,c) central collisions (0--40\%) and (b,d)
peripheral collisions (40--80\%).}
\label{fig:1dim_mu4_j456}
\end{figure}
 
Figure~\ref{fig:2dim_abs} shows the two-dimensional absolute trigger
efficiency, as defined in Eq.~(\ref{eq:mujet_trig_eff_abs}), for a
muon-jet trigger requiring a muon with $\pt >4$\,\GeV\ and jet with
$\ET > 40$\,\GeV, and is shown as a function of the offline muon \pt
and jet \pt.  The efficiency of this trigger reaches a maximum for
offline jet $\pt \gtrsim 60$\,\GeV\, but does not reach 100\% in
most regions.  This lower efficiency, particularly in the barrel
region, compared with the conditional efficiency shown in Figure
\ref{fig:1dim_mu4_j456}, reflects the inefficiency of the muon
trigger.

\begin{figure}[htb]
\centering
\subfloat[]{\includegraphics[width=0.5\textwidth]{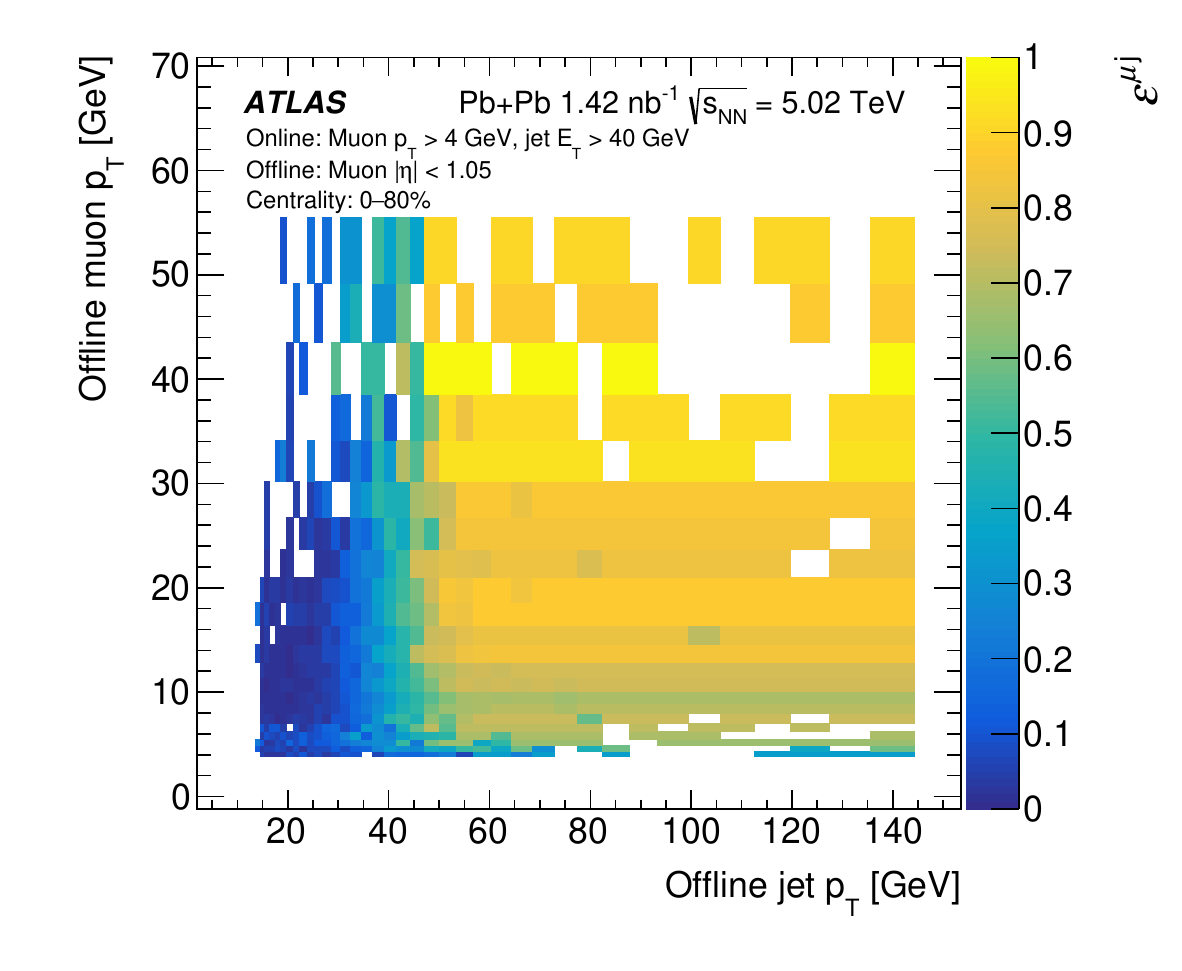}}
\subfloat[]{\includegraphics[width=0.5\textwidth]{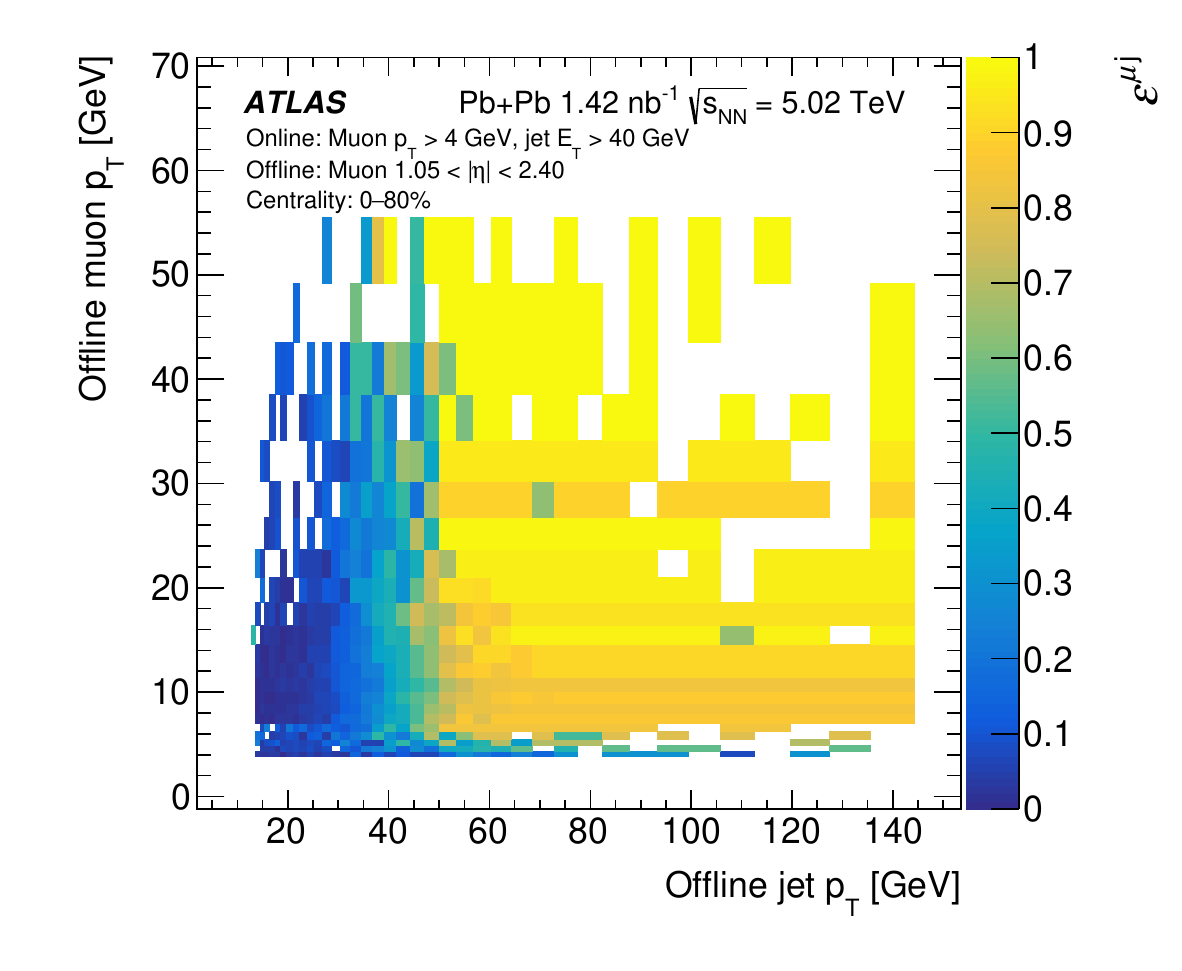}}
\caption{The two-dimensional absolute trigger efficiency, $\varepsilon^{\mu\mathrm{j}}$, as defined
in Eq.~(\ref{eq:mujet_trig_eff_abs}), of the muon-jet trigger that
requires a muon with $\pt > 4$\,\GeV\ and a jet with
$\ET > 40$\,\GeV, as a function of the offline muon \pt and jet \pt.
The measurements are performed inclusively across the full
collision centrality range (0--80\%) for the (a) barrel and (b)
endcap regions.  The last bins of both the $x$- and $y$-axes contain
overflow.}
\label{fig:2dim_abs}
\end{figure}
 
The fraction of selected jets that originate from a $b$-hadron decay
is determined using a template fit to the
$p_{\mathrm{T}}^{\mathrm{rel}}$\ distribution, defined as the
transverse momentum of the muon with respect to the combined muon-jet
axis.  Muons from a $b$-hadron decay typically have a harder
$p_{\mathrm{T}}^{\mathrm{rel}}$\ spectrum than those originating from
light-flavour or $c$-jets.  A template for the
$p_{\mathrm{T}}^{\mathrm{rel}}$\ distribution of light-flavour jets is
derived from a light-flavour-enriched data sample, obtained by
selecting pairs of tracks and jets from data collected with a
single-jet trigger with an $\ET$\ threshold of 50\,\GeV\ in the HLT
and using the method described in Ref.~\cite{STDM-2010-15}.  Templates
for the $b$-jet and $c$-jet distributions are obtained from simulated
di-jet events, generated with \PYTHIA~8.230~\cite{Sjostrand:2014zea}
using the \nnpdftwo PDF set and the A14~\cite{ATL-PHYS-PUB-2014-021}
set of tuned parameters.  The samples are filtered to require one muon
with $\pt > 3$\,\GeV\ at generator level.
The templates for $b$-, $c$-, and light-flavour jets are then fitted
to a data sample selected
using a muon-jet trigger that
requires a muon with $\pt > 4$\,\GeV\ and a jet with $R = 0.4$\ and
$\ET > 40$\,\GeV\ that is within $\Delta R=0.5$ of the muon in the HLT.
Offline muons are required to have $\pt > 4$\,\GeV, $|\eta| < 2.4$, to
pass the `Tight' identification working point~\cite{PERF-2015-10}, and
to be geometrically matched to the trigger muon within $\Delta R < 0.01$.
Offline jets are reconstructed with radius parameter $R = 0.2$, and are
required to have $|\eta| < 2.1$, be within $\Delta R < 0.2$\ of the
muon, and have $\pt > 58\,\GeV$ (where this minimum \pt\ requirement
is determined by the point where the trigger is 99\% efficient).  The
templates for $b$-, $c$-, and light-flavour jets are then fitted to
the data sample.
In this way, the relative fraction of each jet
flavour is determined and a $b$-jet fraction of 30\%--35\% is
measured, with no centrality-dependence observed.  Comparable
$b$-fractions were measured when repeating the analysis on
48.8\,\ipb\ of $pp$ collision data with $\sqrt{s} =
5.02\,\TeV$\ collected during a special run in 2017, and using offline
jets with radii of 0.2 and 0.4.  The muon and jet are estimated to
come from different hard scatter interactions in 1\% of events in
central \PbPb\ collisions that produce the most demanding
environments.  This fake-rate is smaller in peripheral collisions.

\subsection{Muon-jet triggers for proton--proton collisions}
\label{subsec:muon_jets_pp}
 
Triggers with similar design and thresholds to those detailed in Table
\ref{tab:menu_heavyions} were run prescaled during \pp collision
data-taking in order to collect a sample of data enriched with
$b\bar{b}$ decays that are used to calibrate the offline flavour-tagging
algorithms.  In these cases, muon-jet triggers are seeded from
either a single-muon or a muon-plus-jet requirement at L1.  In
the HLT, muons are required to satisfy $\Delta R(\mu,\mathrm{jet}) <$
0.5 and $\Delta z(\mu, \mathrm{jet}) <$ 2 mm (where the $z$-position
of the jet is taken to be the primary-vertex $z$-position), in order
to be considered as `matched' to a jet.
 
Muon-jet triggers to select interesting physics processes were also
provided during 2016 data-taking, but were discontinued due to their
prohibitively large CPU cost.  For these triggers it was desirable to
exploit other characteristic features of the process of interest, for
example by placing additional requirements on the multiplicity, \ET,
and \btagging weight of other jets in the event.  In these cases, only
jets that failed the matching requirements with the muon were
considered for further processing (e.g.\ \btagging) by the
\bjet trigger software.  The muon-jet can therefore form one component
of a more complex trigger, for example by requiring that an event
contains some combination of muon-jet(s), \btagged jet(s), untagged
(light-flavour) jet(s), or any other object that ATLAS is able to
trigger on.  These muon-jet triggers have the potential to be beneficial
for analyses using \pp collision data that have large \bjet
multiplicity (e.g.\ $HH\rightarrow b\bbar b\bbar$), and/or for those
that only have low-\pt \bjets, e.g.\ $b\phi(\phi\rightarrow b\bbar)$.
 
\FloatBarrier
\section{Summary}
 
ATLAS has successfully operated \bjet triggers throughout Runs~1 and 2
of the LHC.  The \bjet trigger software was completely redesigned
during the long shutdown period that followed Run~1, was validated
during 2015 data-taking, and became fully operational in 2016.  The
software uses a two-stage approach to improve primary-vertex finding
and ensure stability under increasingly harsh pile-up conditions, and
deploys state-of-the-art offline \btagging algorithms in the HLT.
These changes, together with improved tracking performance in the
trigger and the installation of the insertable B-layer for Run~2, lead
to significantly improved performance compared to Run~1.
Light-flavour jet rejection was improved by an order of magnitude for
the same \bjet selection efficiency in 2016 compared with the \bjet
triggers used in Run~1.  An additional factor of $\sim$1.5 in
light-flavour jet rejection was achieved in 2017 and 2018 by further
optimising the use of the MV2 algorithm in the HLT, while
simultaneously reoptimising the software to reduce the total CPU
processing time by $\sim$30\%.  These improvements allowed ATLAS to
maintain the \ET\ thresholds and \btagging working points of \bjet
triggers throughout Run~2, in spite of the increasingly harsh pile-up
conditions.
 
The same likelihood-based method that is used to calibrate the offline
\btagging algorithms in ATLAS is adapted for use with the \bjet
triggers for the first time.  Conditional efficiencies are measured in
data and evaluated in simulation for different combinations of online
and offline working points for each year of data-taking (2016--2018).
The conditional efficiencies are typically in the range 85\%--97\%,
depending on the combination of working points considered.  Good
agreement of MC simulation with data is generally observed, and scale factors are
provided to correct the simulation to data.  The use of the likelihood
method provides a substantial reduction in uncertainties compared to
the geometrical matching approaches used previously, enabling the
conditional efficiencies to be measured with a typical accuracy of
a few percent.
 
Specially designed \bjet triggers were also deployed for the first
time during \PbPb\ data-taking in 2018, by adapting the \bjet trigger
software to identify semileptonic $b$-hadron decays by selecting muons
geometrically matched to a jet.  These triggers reach an efficiency of
$>99$\% with respect to both the single muon trigger and offline
requirements above the jet \ET turn-on region, and provide a mechanism
to study the flavour-dependence of radiative quark energy loss in the
quark--gluon plasma, where the busy detector environment made it
unfeasible to run the standard \bjet triggers.

 
\section*{Acknowledgements}
 

We thank CERN for the very successful operation of the LHC, as well as the
support staff from our institutions without whom ATLAS could not be
operated efficiently.
 
We acknowledge the support of
ANPCyT, Argentina;
YerPhI, Armenia;
ARC, Australia;
BMWFW and FWF, Austria;
ANAS, Azerbaijan;
SSTC, Belarus;
CNPq and FAPESP, Brazil;
NSERC, NRC and CFI, Canada;
CERN;
ANID, Chile;
CAS, MOST and NSFC, China;
Minciencias, Colombia;
MSMT CR, MPO CR and VSC CR, Czech Republic;
DNRF and DNSRC, Denmark;
IN2P3-CNRS and CEA-DRF/IRFU, France;
SRNSFG, Georgia;
BMBF, HGF and MPG, Germany;
GSRI, Greece;
RGC and Hong Kong SAR, China;
ISF and Benoziyo Center, Israel;
INFN, Italy;
MEXT and JSPS, Japan;
CNRST, Morocco;
NWO, Netherlands;
RCN, Norway;
MNiSW and NCN, Poland;
FCT, Portugal;
MNE/IFA, Romania;
JINR;
MES of Russia and NRC KI, Russian Federation;
MESTD, Serbia;
MSSR, Slovakia;
ARRS and MIZ\v{S}, Slovenia;
DSI/NRF, South Africa;
MICINN, Spain;
SRC and Wallenberg Foundation, Sweden;
SERI, SNSF and Cantons of Bern and Geneva, Switzerland;
MOST, Taiwan;
TAEK, Turkey;
STFC, United Kingdom;
DOE and NSF, United States of America.
In addition, individual groups and members have received support from
BCKDF, CANARIE, Compute Canada and CRC, Canada;
COST, ERC, ERDF, Horizon 2020 and Marie Sk{\l}odowska-Curie Actions, European Union;
Investissements d'Avenir Labex, Investissements d'Avenir Idex and ANR, France;
DFG and AvH Foundation, Germany;
Herakleitos, Thales and Aristeia programmes co-financed by EU-ESF and the Greek NSRF, Greece;
BSF-NSF and GIF, Israel;
Norwegian Financial Mechanism 2014-2021, Norway;
La Caixa Banking Foundation, CERCA Programme Generalitat de Catalunya and PROMETEO and GenT Programmes Generalitat Valenciana, Spain;
G\"{o}ran Gustafssons Stiftelse, Sweden;
The Royal Society and Leverhulme Trust, United Kingdom.
 
The crucial computing support from all WLCG partners is acknowledged gratefully, in particular from CERN, the ATLAS Tier-1 facilities at TRIUMF (Canada), NDGF (Denmark, Norway, Sweden), CC-IN2P3 (France), KIT/GridKA (Germany), INFN-CNAF (Italy), NL-T1 (Netherlands), PIC (Spain), ASGC (Taiwan), RAL (UK) and BNL (USA), the Tier-2 facilities worldwide and large non-WLCG resource providers. Major contributors of computing resources are listed in Ref.~\cite{ATL-SOFT-PUB-2021-003}.
 
 
\clearpage
 
\printbibliography

@article{Bertelsen_2016,
	doi = {10.1088/1748-0221/11/02/c02020},
	url = {https://doi.org/10.1088/1748-0221/11/02/c02020},
	year = 2016,
	month = {feb},
	publisher = {{IOP} Publishing},
	volume = {11},
	number = {02},
	pages = {C02020--C02020},
	author = {H. Bertelsen and G. Carrillo Montoya and P.-O. Deviveiros and T. Eifert and G. Galster and J. Glatzer and S. Haas and A. Marzin and M.V. Silva Oliveira and T. Pauly and K. Schmieden and R. Spiwoks and J. Stelzer},
	title = {Operation of the upgraded {ATLAS} Central Trigger Processor during the {LHC} Run 2},
	journal = {Journal of Instrumentation}
}

@article{LHC,
    editor = "Evans, Lyndon and Bryant, Philip",
    title = "{LHC Machine}",
    doi = "10.1088/1748-0221/3/08/S08001",
    journal = "JINST",
    volume = "3",
    pages = "S08001",
    year = "2008"
}

@article{Aad:2020vbr,
	author = "{ATLAS Collaboration}",
	collaboration = "ATLAS",
	title = "{Search for Higgs boson production in association with a high-energy photon via vector-boson fusion with decay into bottom quark pairs at $\sqrt{s}$=13 TeV with the ATLAS detector}",
	eprint = "2010.13651",
	archivePrefix = "arXiv",
	primaryClass = "hep-ex",
	reportNumber = "CERN-EP-2020-179",
	month = "10",
	year = "2020"
}

@article{newt,
    author         = "Cornelissen, T. and Elsing, M. and Gavrilenko, I. and Liebig, W. and Moyse, E. and Salzburger, A.",
    title          = "{The new ATLAS track reconstruction (NEWT)}",
    booktitle      = "{Proceedings, 16th International Conference on Computing in High Energy and Nuclear Physics (CHEP 2007): Victoria, Canada, September 2-7, 2007}",
    journal        = "J. Phys. Conf. Ser.",
    volume         = "119",
    year           = "2008",
    pages          = "032014",
    doi            = "10.1088/1742-6596/119/3/032014",
    SLACcitation   = "%%CITATION = 00462,119,032014;%%"
}

@article{ATLAS_L1Calo,
	doi = {10.1088/1748-0221/3/03/p03001},
	year = 2008,
	month = {3},
	publisher = {{IOP} Publishing},
	volume = {3},
	number = {03},
	pages = {P03001--P03001},
	author = {R Achenbach and P Adragna and V Andrei and P Apostologlou and B {\AA}sman and C Ay and B M Barnett and B Bauss and M Bendel and C Bohm and J R A Booth and I P Brawn and P Bright Thomas and D G Charlton and N J Collins and C J Curtis and A Dahlhoff and A O Davis and S Eckweiler and J P Edwards and E Eisenhandler and P J W Faulkner and J Fleckner and F Föhlisch and J Garvey and C N P Gee and A R Gillman and P Hanke and R P Hatley and S Hellman and A Hidv{\'{e}}gi and S J Hillier and K Jakobs and M Johansen and E -E Kluge and M Landon and V Lendermann and J N Lilley and K Mahboubi and G Mahout and A Mass and K Meier and T Moa and E Moyse and F Müller and A Neusiedl and C Nöding and B Oltmann and J M Pentney and V J O Perera and U Pfeiffer and D P F Prieur and W Qian and D L Rees and S Rieke and F Rühr and D P C Sankey and U Schäfer and K Schmitt and H -C Schultz-Coulon and C Schumacher and S Silverstein and R J Staley and R Stamen and M C Stockton and S Tapprogge and J P Thomas and T Trefzger and P M Watkins and A Watson and P Weber and E -E Woehrling},
	title = {The {ATLAS} Level-1 Calorimeter Trigger},
	journal = {JINST}
	
}

@phdthesis{LaurieThesis,
      author         = "McClymont, L. I.",
      title          = "Searches for new physics using pairs of jets containing b-quarks at the ATLAS detector",
      school         = "University College London",
      year           = "2018",
      note           = "\url{http://discovery.ucl.ac.uk/id/eprint/10052993}"
}

@article{Frixione:2007vw,
      author         = "Frixione, Stefano and Nason, Paolo and Oleari, Carlo",
      title          = "{Matching NLO QCD computations with parton shower\\
                        simulations: the POWHEG method}",
      journal        = "JHEP",
      volume         = "11",
      year           = "2007",
      pages          = "070",
      doi            = "10.1088/1126-6708/2007/11/070",
      eprint         = "0709.2092",
      archivePrefix  = "arXiv",
      primaryClass   = "hep-ph",
      reportNumber   = "BICOCCA-FT-07-9, GEF-TH-21-2007",
      SLACcitation   = "%%CITATION = ARXIV:0709.2092;%%"
}

@Article{Lange:2001uf,
     author    = "Lange, D. J.",
     title     = "The EvtGen particle decay simulation package",
     journal   = "Nucl. Instrum. Meth. A\\",
     volume    = "462",
     year      = "2001",
     pages     = "152-155",
     doi       = "10.1016/S0168-9002(01)00089-4",
     SLACcitation  = "%%CITATION = NUIMA,A462,152;%%"
}

@Article{anti_kT,
     author = "{M. Cacciari, G. P. Salam and G. Soyez}",
     title  = "{The anti-$k_t$ jet clustering algorithm}",
      journal        = "JHEP",
      hidevolume         = "0804",
      volume         = "04",
      pages          = "063",
      doi            = "10.1088/1126-6708/2008/04/063",
      year           = "2008",
      eprint         = "0802.1189",
      archivePrefix  = "arXiv",
      primaryClass   = "hep-ph",
      reportNumber   = "LPTHE-07-03",
      SLACcitation   = "%%CITATION = ARXIV:0802.1189;%%",
}

@article{Czakon:2011xx,
      author         = "Czakon, Michal and Mitov, Alexander",
      title          = "{Top++: a program for the calculation of the top-pair cross-section at hadron colliders}",
      journal        = "Comput. Phys. Commun.",
      volume         = "185",
      year           = "2014",
      pages          = "2930",
      doi            = "10.1016/j.cpc.2014.06.021",
      eprint         = "1112.5675",
      archivePrefix  = "arXiv",
      primaryClass   = "hep-ph",
      reportNumber   = "CERN-PH-TH-2011-303, TTK-11-58",
      SLACcitation   = "%%CITATION = ARXIV:1112.5675;%%"
}

@article{Kalman,
      author         = "Fr{\"u}hwirth, R.",
      title          = "{Application of Kalman filtering to track and vertex
                        fitting}",
      journal        = "Nucl. Instrum. Meth. A",
      volume         = "262",
      year           = "1987",
      pages          = "444-450",
      doi            = "10.1016/0168-9002(87)90887-4",
      reportNumber   = "HEPHY-PUB-87-503",
      SLACcitation   = "%%CITATION = NUIMA,A262,444;%%"
}

@article{l2kalman,
    author = {Kalman, R. E.},
    title = "{A New Approach to Linear Filtering and Prediction Problems}",
    journal = {J. Fluids Eng.},
    volume = {82},
    number = {1},
    pages = {35-45},
    year = {1960},
    month = {03},
    issn = {0098-2202},
    doi = {10.1115/1.3662552},
}

@article{Billoir,
      author         = "Billoir, Pierre",
      title          = "{Track Fitting With Multiple Scattering: A New Method}",
      journal        = "Nucl. Instrum. Meth. A",
      volume         = "225",
      year           = "1984",
      pages          = "352-366",
      doi            = "10.1016/0167-5087(84)90274-6",
      reportNumber   = "LPC-83-12",
      SLACcitation   = "%%CITATION = NUIMA,A225,352;%%"
}

@article{LUCID2,
      author={G. Avoni and others},
      title={The new LUCID-2 detector for luminosity measurement and monitoring in ATLAS},
      journal={JINST},
      volume={13},
      number={07},
      pages={P07017},
      doi="10.1088/1748-0221/13/07/P07017",
      year={2018},
}

@article{Blaizot:2015lma,
      author         = "Blaizot, Jean-Paul and Mehtar-Tani, Yacine",
      title          = "{Jet Structure in Heavy Ion Collisions}",
      journal        = "Int. J. Mod. Phys. E",
      volume         = "24",
      year           = "2015",
      number         = "11",
      pages          = "1530012",
      doi            = "10.1142/S021830131530012X",
      eprint         = "1503.05958",
      archivePrefix  = "arXiv",
      primaryClass   = "hep-ph",
      reportNumber   = "INT-PUB-15-006",
      SLACcitation   = "%%CITATION = ARXIV:1503.05958;%%"
}

@article{Aamodt:2011mr,
      author         = "{ALICE Collaboration}",
      title          = "{Two-pion Bose-Einstein correlations in central Pb-Pb
                        collisions at $\sqrt{{s}_{NN}} =$ 2.76 TeV}",
      journal        = "Phys. Lett. B",
      volume         = "696",
      year           = "2011",
      pages          = "328-337",
      doi            = "10.1016/j.physletb.2010.12.053",
      eprint         = "1012.4035",
      archivePrefix  = "arXiv",
      primaryClass   = "nucl-ex",
      reportNumber   = "CERN-PH-EP-ALICE-2010-006, CERN-PH-EP-2010-078",
      SLACcitation   = "%%CITATION = ARXIV:1012.4035;%%"
}

@article{Dokshitzer:2001zm,
      author         = "Dokshitzer, Yuri L. and Kharzeev, D. E.",
      title          = "{Heavy quark colorimetry of QCD matter}",
      journal        = "Phys. Lett. B",
      volume         = "519",
      year           = "2001",
      pages          = "199-206",
      doi            = "10.1016/S0370-2693(01)01130-3",
      eprint         = "hep-ph/0106202",
      archivePrefix  = "arXiv",
      primaryClass   = "hep-ph",
      reportNumber   = "LPT-ORSAY-01-58, BNL-NT-01-9",
      SLACcitation   = "%%CITATION = HEP-PH/0106202;%%"
}

@Article{Glauber,    
     author = "Miller, Michael L. and Reygers, Klaus and Sanders, Stephen J. and Steinberg, Peter",
     title = "{Glauber modeling in high energy nuclear collisions}",
     eprint = "nucl-ex/0701025",
     archivePrefix = "arXiv",
     doi = "10.1146/annurev.nucl.57.090506.123020",
     journal = "Ann. Rev. Nucl. Part. Sci.",
     volume = "57",
     pages = "205--243",
     year = "2007"
     
}

@article{Aad:2020wji,
	author = "{ATLAS Collaboration}",
	collaboration = "ATLAS",
	title = "{Operation of the ATLAS trigger system in Run 2}",
	eprint = "2007.12539",
	archivePrefix = "arXiv",
	primaryClass = "physics.ins-det",
	reportNumber = "CERN-EP-2020-109",
	doi = "10.1088/1748-0221/15/10/P10004",
	journal = "JINST",
	volume = "15",
	number = "10",
	pages = "P10004",
	year = "2020"

}

@article{TrigID2020,
      title={The ATLAS Inner Detector Trigger performance in pp collisions at 13 TeV during LHC Run 2}, 
      author="{ATLAS Collaboration}",
      year={2021},
      eprint={2107.02485},
      archivePrefix={arXiv},
      primaryClass={hep-ex}
}

@Article{Fastjet,
      author         = "Cacciari, Matteo and Salam, Gavin P. and Soyez, Gregory",
      title          = "{FastJet user manual}",
      journal        = "Eur. Phys. J. C",
      volume         = "72",
      year           = "2012",
      pages          = "1896",
      doi            = "10.1140/epjc/s10052-012-1896-2",
      eprint         = "1111.6097",
      archivePrefix  = "arXiv",
      primaryClass   = "hep-ph",
      reportNumber   = "CERN-PH-TH-2011-297",
      SLACcitation   = "%%CITATION = ARXIV:1111.6097;%%"
}

@article{Harland-Lang:2014zoa,
      author         = "Harland-Lang, L.A. and Martin, A.D. and Motylinski, P.
                        and Thorne, R.S.",
      title          = "{Parton distributions in the LHC era: MMHT 2014 PDFs}",
      journal        = "Eur. Phys. J. C",
      number         = "5",
      volume         = "75",
      pages          = "204",
      doi            = "10.1140/epjc/s10052-015-3397-6",
      year           = "2015",
      eprint         = "1412.3989",
      archivePrefix  = "arXiv",
      primaryClass   = "hep-ph",
      reportNumber   = "LCTS-2014-47, IPPP-14-97, DCPT-14-194",
      SLACcitation   = "%%CITATION = ARXIV:1412.3989;%%",
}

@article{Ball:2012cx,
      author         = "Ball, Richard D. and others",
      title          = "{Parton distributions with LHC data}",
      journal        = "Nucl. Phys. B",
      volume         = "867",
      year           = "2013",
      pages          = "244",
      doi            = "10.1016/j.nuclphysb.2012.10.003",
      eprint         = "1207.1303",
      archivePrefix  = "arXiv",
      primaryClass   = "hep-ph",
      reportNumber   = "EDINBURGH-2012-08, IFUM-FT-997, FR-PHENO-2012-014,
                        RWTH-TTK-12-25, CERN-PH-TH-2012-037, SFB-CPP-12-47\,
                        --CERN-PH-TH-2012-037",
      SLACcitation   = "%%CITATION = ARXIV:1207.1303;%%"
}

@article{Ball:2014uwa,
      author         = "Ball, Richard D. and others",
      title          = "{Parton distributions for the LHC run II}",
      collaboration  = "NNPDF",
      journal        = "JHEP",
      volume         = "04",
      year           = "2015",
      pages          = "040",
      doi            = "10.1007/JHEP04(2015)040",
      eprint         = "1410.8849",
      archivePrefix  = "arXiv",
      primaryClass   = "hep-ph",
      reportNumber   = "EDINBURGH-2014-15, IFUM-1034-FT, CERN-PH-TH-2013-253,
                        OUTP-14-11P, CAVENDISH-HEP-14-11",
      SLACcitation   = "%%CITATION = ARXIV:1410.8849;%%"
}

@article{Czakon:2012pz,
      author         = "Czakon, Michal and Mitov, Alexander",
      title          = "{NNLO corrections to top pair production at hadron
                        colliders: the quark-gluon reaction}",
      journal        = "JHEP",
      volume         = "01",
      pages          = "080",
      doi            = "10.1007/JHEP01(2013)080",
      year           = "2013",
      eprint         = "1210.6832",
      archivePrefix  = "arXiv",
      primaryClass   = "hep-ph",
      SLACcitation   = "%%CITATION = ARXIV:1210.6832;%%",
}

@article{Sjostrand:2014zea,
      author         = "Sj{\"o}strand, Torbj{\"o}rn and Ask, Stefan and Christiansen,
                        Jesper R. and Corke, Richard and Desai, Nishita and Ilten,
                        Philip and Mrenna, Stephen and Prestel, Stefan and
                        Rasmussen, Christine O. and Skands, Peter Z.",
      title          = "{An introduction to PYTHIA 8.2}",
      journal        = "Comput. Phys. Commun.",
      volume         = "191",
      year           = "2015",
      pages          = "159",
      doi            = "10.1016/j.cpc.2015.01.024",
      eprint         = "1410.3012",
      archivePrefix  = "arXiv",
      primaryClass   = "hep-ph",
      reportNumber   = "LU-TP-14-36, MCNET-14-22, CERN-PH-TH-2014-190,
                        FERMILAB-PUB-14-316-CD, DESY-14-178, SLAC-PUB-16122,
                        --FERMILAB-PUB-14-316-CD",
      SLACcitation   = "%%CITATION = ARXIV:1410.3012;%%"
}

@Article{Bahr:2008pv,
      author         = "B{\"a}hr, M. and others",
      title          = "{Herwig++ physics and manual}",
      journal        = "Eur. Phys. J. C",
      volume         = "58",
      year           = "2008",
      pages          = "639",
      doi            = "10.1140/epjc/s10052-008-0798-9",
      eprint         = "0803.0883",
      archivePrefix  = "arXiv",
      primaryClass   = "hep-ph",
      reportNumber   = "CERN-PH-TH-2008-038, CAVENDISH-HEP-08-03, KA-TP-05-2008,
                        DCPT-08-22, IPPP-08-11, CP3-08-05",
      SLACcitation   = "%%CITATION = ARXIV:0803.0883;%%"
}

@Article{Bellm:2015jjp,
      author         = "Bellm, Johannes and others",
      title          = "{Herwig 7.0/Herwig++ 3.0 release note}",
      journal        = "Eur. Phys. J. C",
      volume         = "76",
      year           = "2016",
      number         = "4",
      pages          = "196",
      doi            = "10.1140/epjc/s10052-016-4018-8",
      eprint         = "1512.01178",
      archivePrefix  = "arXiv",
      primaryClass   = "hep-ph",
      reportNumber   = "CERN-PH-TH-2015-289, MAN-HEP-2015-15, IFJPAN-IV-2015-13,
                        HERWIG-2015-01, KA-TP-18-2015, DCPT-15-142, MCNET-15-28,
                        IPPP-15-71, --HERWIG-2015-01",
      SLACcitation   = "%%CITATION = ARXIV:1512.01178;%%"
}

@Article{Nason:2004rx,
      author         = "Nason, Paolo",
      title          = "{A new method for combining NLO QCD with shower Monte Carlo algorithms}",
      journal        = "JHEP",
      volume         = "11",
      pages          = "040",
      doi            = "10.1088/1126-6708/2004/11/040",
      year           = "2004",
      eprint         = "hep-ph/0409146",
      archivePrefix  = "arXiv",
}

@Article{Alioli:2010xd,
      author         = "Alioli, Simone and Nason, Paolo and Oleari, Carlo and Re,
                        Emanuele",
      title          = "{A general framework for implementing NLO calculations in
                        shower Monte Carlo programs: the POWHEG BOX}",
      journal        = "JHEP",
      volume         = "06",
      pages          = "043",
      doi            = "10.1007/JHEP06(2010)043",
      year           = "2010",
      eprint         = "1002.2581",
      archivePrefix  = "arXiv",
      primaryClass   = "hep-ph",
}

@article{Beneke:2011mq,
      author         = "Beneke, M. and Falgari, P. and Klein, S. and Schwinn, C.",
      title          = "{Hadronic top-quark pair production with NNLL threshold
                        resummation}",
      journal        = "Nucl. Phys. B",
      volume         = "855",
      year           = "2012",
      pages          = "695-741",
      doi            = "10.1016/j.nuclphysb.2011.10.021",
      eprint         = "1109.1536",
      archivePrefix  = "arXiv",
      primaryClass   = "hep-ph",
      reportNumber   = "TTK-11-38, ITP-UU-11-26, SPIN-11-19, FR-PHENO-2011-015,
                        SFB-CPP-11-49",
      SLACcitation   = "%%CITATION = ARXIV:1109.1536;%%"
}

@article{Cacciari:2011hy,
      author         = "Cacciari, Matteo and Czakon, Michal and Mangano,
                        Michelangelo and Mitov, Alexander and Nason, Paolo",
      title          = "{Top-pair production at hadron colliders with
                        next-to-next-to-leading logarithmic soft-gluon
                        resummation}",
      journal        = "Phys. Lett. B",
      volume         = "710",
      pages          = "612-622",
      doi            = "10.1016/j.physletb.2012.03.013",
      year           = "2012",
      eprint         = "1111.5869",
      primaryClass   = "hep-ph",
      archivePrefix  = "arXiv",
      reportNumber   = "CERN-PH-TH-2011-277, TTK-11-54",
      SLACcitation   = "%%CITATION = ARXIV:1111.5869;%%",
}

@article{Czakon:2012zr,
      author         = "Czakon, Michal and Mitov, Alexander",
      title          = "{NNLO corrections to top-pair production at hadron
                        colliders: the all-fermionic scattering channels}",
      journal        = "JHEP",
      volume         = "12",
      pages          = "054",
      doi            = "10.1007/JHEP12(2012)054",
      year           = "2012",
      eprint         = "1207.0236",
      archivePrefix  = "arXiv",
      primaryClass   = "hep-ph",
      SLACcitation   = "%%CITATION = ARXIV:1207.0236;%%",
}

@article{Czakon:2013goa,
      author         = "Czakon, Michal and Fiedler, Paul and Mitov, Alexander",
      title          = "{Total Top-Quark Pair-Production Cross Section at
                        Hadron Colliders Through \(O(\alpha_S^4)\)}",
      journal        = "Phys. Rev. Lett.",
      volume         = "110",
      pages          = "252004",
      doi            = "10.1103/PhysRevLett.110.252004",
      year           = "2013",
      eprint         = "1303.6254",
      archivePrefix  = "arXiv",
      primaryClass   = "hep-ph",
      reportNumber   = "CERN-PH-TH-2013-056, TTK-13-08",
      SLACcitation   = "%%CITATION = ARXIV:1303.6254;%%",
}

@article{Baernreuther:2012ws,
      author         = "B{\"a}rnreuther, Peter and Czakon, Michal and Mitov,
                        Alexander",
      title          = "{Percent-Level-Precision Physics at the
                  Tevatron: Next-to-Next-to-Leading Order QCD
                  Corrections to \(q \bar{q} \to t \bar{t} + X\)}",
      journal        = "Phys. Rev. Lett.",
      volume         = "109",
      pages          = "132001",
      doi            = "10.1103/PhysRevLett.109.132001",
      year           = "2012",
      eprint         = "1204.5201",
      archivePrefix  = "arXiv",
      primaryClass   = "hep-ph",
      SLACcitation   = "%%CITATION = ARXIV:1204.5201;%%",
}

@article{Frixione:2007nw,
      author         = "Frixione, Stefano and Nason, Paolo and Ridolfi, Giovanni",
      title          = "{A positive-weight next-to-leading-order Monte Carlo for
                        heavy flavour hadroproduction}",
      journal        = "JHEP",
      volume         = "09",
      pages          = "126",
      doi            = "10.1088/1126-6708/2007/09/126",
      year           = "2007",
      eprint         = "0707.3088",
      archivePrefix  = "arXiv",
      primaryClass   = "hep-ph",
      SLACcitation   = "%%CITATION = ARXIV:0707.3088;%%",
}

@article{Agostinelli:2002hh,
      author         = "{GEANT4 Collaboration} and Agostinelli, S. and others",
      title          = "{\textsc{Geant4} -- a simulation toolkit}",
      journal        = "Nucl. Instrum. Meth. A",
      volume         = "506",
      year           = "2003",
      pages          = "250",
      doi            = "10.1016/S0168-9002(03)01368-8",
      reportNumber   = "SLAC-PUB-9350, FERMILAB-PUB-03-339",
      SLACcitation   = "%%CITATION = NUIMA,A506,250;%%"
}

@Article{PERF-2007-01,
    author         = "{ATLAS Collaboration}",
    title          = "{The ATLAS Experiment at the CERN Large Hadron Collider}",
    journal        = "JINST",
    volume         = "3",
    year           = "2008",
    pages          = "S08003",
    doi            = "10.1088/1748-0221/3/08/S08003",
    primaryClass   = "hep-ex",
}

@Article{SOFT-2010-01,
    author         = "{ATLAS Collaboration}",
    title          = "{The ATLAS Simulation Infrastructure}",
    journal        = "Eur. Phys. J. C",
    volume         = "70",
    year           = "2010",
    pages          = "823",
    doi            = "10.1140/epjc/s10052-010-1429-9",
    eprint         = "1005.4568",
    archivePrefix  = "arXiv",
    primaryClass   = "physics.ins-det",
}

@Article{STDM-2010-15,
    author         = "{ATLAS Collaboration}",
    title          = "{Measurement of the inclusive and dijet cross-sections of \(b\)-jets in \(pp\) collisions at \(\sqrt{s} = 7\,\text{TeV}\) with the ATLAS detector}",
    journal        = "Eur. Phys. J. C",
    volume         = "71",
    year           = "2011",
    pages          = "1846",
    doi            = "10.1140/epjc/s10052-011-1846-4",
    reportNumber   = "CERN-PH-EP-2011-146",
    eprint         = "1109.6833",
    archivePrefix  = "arXiv",
    primaryClass   = "hep-ex",
}

@Article{HION-2011-02,
    author         = "{ATLAS Collaboration}",
    title          = "{Measurement of the jet radius and transverse momentum dependence of inclusive jet suppression in lead--lead collisions at \(\sqrt{s_{\text{NN}}} = 2.76\,\text{TeV}\) with the ATLAS detector}",
    journal        = "Phys. Lett. B",
    volume         = "719",
    year           = "2013",
    pages          = "220",
    doi            = "10.1016/j.physletb.2013.01.024",
    reportNumber   = "CERN-PH-EP-2012-134",
    eprint         = "1208.1967",
    archivePrefix  = "arXiv",
    primaryClass   = "hep-ex",
}

@Article{HION-2011-05,
    author         = "{ATLAS Collaboration}",
    title          = "{Measurement of the pseudorapidity and transverse momentum dependence of the elliptic flow of charged particles in lead--lead collisions at \(\sqrt{s_{\text{NN}}}=2.76\,\text{TeV}\) with the ATLAS detector}",
    journal        = "Phys. Lett. B",
    volume         = "707",
    year           = "2012",
    pages          = "330",
    doi            = "10.1016/j.physletb.2011.12.056",
    reportNumber   = "CERN-PH-EP-2011-124",
    eprint         = "1108.6018",
    archivePrefix  = "arXiv",
    primaryClass   = "hep-ex",
}

@Article{TRIG-2011-02,
    author         = "{ATLAS Collaboration}",
    title          = "{Performance of the ATLAS Trigger System in 2010}",
    journal        = "Eur. Phys. J. C",
    volume         = "72",
    year           = "2012",
    pages          = "1849",
    doi            = "10.1140/epjc/s10052-011-1849-1",
    reportNumber   = "CERN-PH-EP-2011-078",
    eprint         = "1110.1530",
    archivePrefix  = "arXiv",
    primaryClass   = "hep-ex",
}

@Article{TRIG-2012-01,
    author         = "{ATLAS Collaboration}",
    title          = "{The performance of the jet trigger for the ATLAS detector during 2011 data taking}",
    journal        = "Eur. Phys. J. C",
    volume         = "76",
    year           = "2016",
    pages          = "526",
    doi            = "10.1140/epjc/s10052-016-4325-0",
    reportNumber   = "CERN-EP-2016-031",
    eprint         = "1606.07759",
    archivePrefix  = "arXiv",
    primaryClass   = "hep-ex",
}

@Article{PERF-2014-02,
    author         = "{ATLAS Collaboration}",
    title          = "{Determination of jet calibration and energy resolution in proton--proton collisions at \(\sqrt{s} = 8\,\text{TeV}\) using the ATLAS detector}",
    journal        = "Eur. Phys. J. C",
    volume         = "80",
    year           = "2020",
    pages          = "1104",
    doi            = "10.1140/epjc/s10052-020-08477-8",
    reportNumber   = "CERN-EP-2019-057",
    eprint         = "1910.04482",
    archivePrefix  = "arXiv",
    primaryClass   = "hep-ex",
}

@Article{PERF-2014-07,
    author         = "{ATLAS Collaboration}",
    title          = "{Topological cell clustering in the ATLAS calorimeters and its performance in LHC Run~1}",
    journal        = "Eur. Phys. J. C",
    volume         = "77",
    year           = "2017",
    pages          = "490",
    doi            = "10.1140/epjc/s10052-017-5004-5",
    reportNumber   = "CERN-PH-EP-2015-304",
    eprint         = "1603.02934",
    archivePrefix  = "arXiv",
    primaryClass   = "hep-ex",
}

@Article{HIGG-2015-05,
    author         = "{ATLAS Collaboration}",
    title          = "{Search for the Standard Model Higgs boson decaying into \(b\bar{b}\) produced in association with top quarks decaying hadronically in \(pp\) collisions at \(\sqrt{s} = 8\,\text{TeV}\) with the ATLAS detector}",
    journal        = "JHEP",
    volume         = "05",
    year           = "2016",
    pages          = "160",
    doi            = "10.1007/JHEP05(2016)160",
    reportNumber   = "CERN-EP-2016-058",
    eprint         = "1604.03812",
    archivePrefix  = "arXiv",
    primaryClass   = "hep-ex",
}

@Article{PERF-2015-01,
    author         = "{ATLAS Collaboration}",
    title          = "{Reconstruction of primary vertices at the ATLAS experiment in Run~1 proton--proton collisions at the LHC}",
    journal        = "Eur. Phys. J. C",
    volume         = "77",
    year           = "2017",
    pages          = "332",
    doi            = "10.1140/epjc/s10052-017-4887-5",
    reportNumber   = "CERN-EP-2016-150",
    eprint         = "1611.10235",
    archivePrefix  = "arXiv",
    primaryClass   = "hep-ex",
}

@Article{PERF-2015-09,
    author         = "{ATLAS Collaboration}",
    title          = "{Jet reconstruction and performance using particle flow with the ATLAS Detector}",
    journal        = "Eur. Phys. J. C",
    volume         = "77",
    year           = "2017",
    pages          = "466",
    doi            = "10.1140/epjc/s10052-017-5031-2",
    reportNumber   = "CERN-EP-2017-024",
    eprint         = "1703.10485",
    archivePrefix  = "arXiv",
    primaryClass   = "hep-ex",
}

@Article{PERF-2015-10,
    author         = "{ATLAS Collaboration}",
    title          = "{Muon reconstruction performance of the ATLAS detector in proton--proton collision data at \(\sqrt{s} = 13\,\text{TeV}\)}",
    journal        = "Eur. Phys. J. C",
    volume         = "76",
    year           = "2016",
    pages          = "292",
    doi            = "10.1140/epjc/s10052-016-4120-y",
    reportNumber   = "CERN-EP-2016-033",
    eprint         = "1603.05598",
    archivePrefix  = "arXiv",
    primaryClass   = "hep-ex",
}

@Article{EXOT-2016-31,
    author         = "{ATLAS Collaboration}",
    title          = "{Search for pair production of Higgs bosons in the \(b\bar{b}b\bar{b}\) final state using proton--proton collisions at \(\sqrt{s} = 13\,\text{TeV}\) with the ATLAS detector}",
    journal        = "JHEP",
    volume         = "01",
    year           = "2019",
    pages          = "030",
    doi            = "10.1007/JHEP01(2019)030",
    reportNumber   = "CERN-EP-2018-029",
    eprint         = "1804.06174",
    archivePrefix  = "arXiv",
    primaryClass   = "hep-ex",
}

@Article{HIGG-2016-30,
    author         = "{ATLAS Collaboration}",
    title          = "{Search for Higgs bosons produced via vector-boson fusion and decaying into bottom quark pairs in \(\sqrt{s} = 13\,\text{TeV}\) \(pp\) collisions with the ATLAS detector}",
    journal        = "Phys. Rev. D",
    volume         = "98",
    year           = "2018",
    pages          = "052003",
    doi            = "10.1103/PhysRevD.98.052003",
    reportNumber   = "CERN-EP-2018-140",
    eprint         = "1807.08639",
    archivePrefix  = "arXiv",
    primaryClass   = "hep-ex",
}

@Article{HIGG-2016-32,
    author         = "{ATLAS Collaboration}",
    title          = "{Search for heavy neutral Higgs bosons produced in association with \(b\)-quarks and decaying into \(b\)-quarks at \(\sqrt{s} = 13\,\text{TeV}\) with the ATLAS detector}",
    journal        = "Phys. Rev. D",
    volume         = "102",
    year           = "2020",
    pages          = "032004",
    doi            = "10.1103/PhysRevD.102.032004",
    reportNumber   = "CERN-EP-2019-092",
    eprint         = "1907.02749",
    archivePrefix  = "arXiv",
    primaryClass   = "hep-ex",
}

@Article{HION-2016-07,
    author         = "{ATLAS Collaboration}",
    title          = "{Prompt and non-prompt \(J/\psi\) and \(\psi(2\text{S})\) suppression at high transverse momentum in \(5.02\,\text{TeV}\) Pb+Pb collisions with the ATLAS experiment}",
    journal        = "Eur. Phys. J. C",
    volume         = "78",
    year           = "2018",
    pages          = "762",
    doi            = "10.1140/epjc/s10052-018-6219-9",
    reportNumber   = "CERN-EP-2018-049",
    eprint         = "1805.04077",
    archivePrefix  = "arXiv",
    primaryClass   = "hep-ex",
}

@Article{PERF-2016-05,
    author         = "{ATLAS Collaboration}",
    title          = "{Measurements of \(b\)-jet tagging efficiency with the ATLAS detector using \(t\bar{t}\) events at \(\sqrt{s} = 13\,\text{TeV}\)}",
    journal        = "JHEP",
    volume         = "08",
    year           = "2018",
    pages          = "089",
    doi            = "10.1007/JHEP08(2018)089",
    reportNumber   = "CERN-EP-2018-047",
    eprint         = "1805.01845",
    archivePrefix  = "arXiv",
    primaryClass   = "hep-ex",
}

@Article{TRIG-2016-01,
    author         = "{ATLAS Collaboration}",
    title          = "{Performance of the ATLAS trigger system in 2015}",
    journal        = "Eur. Phys. J. C",
    volume         = "77",
    year           = "2017",
    pages          = "317",
    doi            = "10.1140/epjc/s10052-017-4852-3",
    reportNumber   = "CERN-EP-2016-241",
    eprint         = "1611.09661",
    archivePrefix  = "arXiv",
    primaryClass   = "hep-ex",
}

@Article{EGAM-2018-01,
    author         = "{ATLAS Collaboration}",
    title          = "{Electron and photon performance measurements with the ATLAS detector using the 2015--2017 LHC proton--proton collision data}",
    journal        = "JINST",
    volume         = "14",
    year           = "2019",
    pages          = "P12006",
    doi            = "10.1088/1748-0221/14/12/P12006",
    reportNumber   = "CERN-EP-2019-145",
    eprint         = "1908.00005",
    archivePrefix  = "arXiv",
    primaryClass   = "hep-ex",
}

@Article{FTAG-2018-01,
    author         = "{ATLAS Collaboration}",
    title          = "{ATLAS \(b\)-jet identification performance and efficiency measurement with \(t\bar{t}\) events in \(pp\) collisions at \(\sqrt{s} = 13\,\text{TeV}\)}",
    journal        = "Eur. Phys. J. C",
    volume         = "79",
    year           = "2019",
    pages          = "970",
    doi            = "10.1140/epjc/s10052-019-7450-8",
    reportNumber   = "CERN-EP-2019-132",
    eprint         = "1907.05120",
    archivePrefix  = "arXiv",
    primaryClass   = "hep-ex",
}

@Article{HDBS-2018-18,
    author         = "{ATLAS Collaboration}",
    title          = "{Search for the \(HH \to b\bar{b}b\bar{b}\) process via vector-boson fusion production using proton--proton collisions at \(\sqrt{s} = 13\,\text{TeV}\) with the ATLAS detector}",
    journal        = "JHEP",
    volume         = "07",
    year           = "2020",
    pages          = "108",
    doi            = "10.1007/JHEP07(2020)108",
    reportNumber   = "CERN-EP-2019-267",
    eprint         = "2001.05178",
    archivePrefix  = "arXiv",
    primaryClass   = "hep-ex",
    related        = "HDBS-2018-18-err",
    relatedstring  = "Erratum:",
}

@Article{JETM-2018-05,
    author         = "{ATLAS Collaboration}",
    title          = "{Jet energy scale and resolution measured in proton--proton collisions at \(\sqrt{s} = 13\,\text{TeV}\) with the ATLAS detector}",
    year           = "2020",
    reportNumber   = "CERN-EP-2020-083",
    eprint         = "2007.02645",
    archivePrefix  = "arXiv",
    primaryClass   = "hep-ex",
}

@Article{PIX-2018-001,
    author         = "Abbott, B. and others",
    title          = "{Production and integration of the ATLAS Insertable B-Layer}",
    journal        = "JINST",
    volume         = "13",
    year           = "2018",
    pages          = "T05008",
    doi            = "10.1088/1748-0221/13/05/T05008",
    eprint         = "1803.00844",
    archivePrefix  = "arXiv",
    primaryClass   = "physics.ins-det",
}

@Article{TRIG-2018-01,
    author         = "{ATLAS Collaboration}",
    title          = "{Performance of the ATLAS muon triggers in Run~2}",
    journal        = "JINST",
    volume         = "15",
    year           = "2020",
    pages          = "P09015",
    doi            = "10.1088/1748-0221/15/09/p09015",
    reportNumber   = "CERN-EP-2020-031",
    eprint         = "2004.13447",
    archivePrefix  = "arXiv",
    primaryClass   = "hep-ex",
}

@Article{TRIG-2018-05,
    author         = "{ATLAS Collaboration}",
    title          = "{Performance of electron and photon triggers in ATLAS during LHC Run~2}",
    journal        = "Eur. Phys. J. C",
    volume         = "80",
    year           = "2020",
    pages          = "47",
    doi            = "10.1140/epjc/s10052-019-7500-2",
    reportNumber   = "CERN-EP-2019-169",
    eprint         = "1909.00761",
    archivePrefix  = "arXiv",
    primaryClass   = "hep-ex",
}

@Article{TDAQ-2019-01,
    author         = "{ATLAS Collaboration}",
    title          = "{Performance of the upgraded PreProcessor of the ATLAS Level-1 Calorimeter Trigger}",
    journal        = "JINST",
    volume         = "15",
    year           = "2020",
    pages          = "P11016",
    doi            = "10.1088/1748-0221/15/11/P11016",
    reportNumber   = "CERN-EP-2020-042",
    eprint         = "2005.04179",
    archivePrefix  = "arXiv",
    primaryClass   = "hep-ex",
}

@Article{TRIG-2019-04,
    author         = "{ATLAS Collaboration}",
    title          = "{Operation of the ATLAS trigger system in Run~2}",
    journal        = "JINST",
    volume         = "15",
    year           = "2020",
    pages          = "P10004",
    doi            = "10.1088/1748-0221/15/10/P10004",
    reportNumber   = "CERN-EP-2020-109",
    eprint         = "2007.12539",
    archivePrefix  = "arXiv",
    primaryClass   = "hep-ex",
}

@Booklet{ATL-SOFT-PUB-2021-001,
    author         = "{ATLAS Collaboration}",
    title          = "{The ATLAS Collaboration Software and Firmware}",
    howpublished   = "{ATL-SOFT-PUB-2021-001}",
    url            = "https://cds.cern.ch/record/2767187",
    year           = "2021",
}

@Booklet{ATL-SOFT-PUB-2021-003,
    author         = "{ATLAS Collaboration}",
    title          = "{ATLAS Computing Acknowledgements}",
    howpublished   = "{ATL-SOFT-PUB-2021-003}",
    url            = "https://cds.cern.ch/record/2776662",
}

@Report{ATLAS-TDR-2010-19,
    author         = "{ATLAS Collaboration}",
    title          = "{ATLAS Insertable B-Layer Technical Design Report}",
    type           = "ATLAS-TDR-19; CERN-LHCC-2010-013",
    year           = "2010",
    url            = "https://cds.cern.ch/record/1291633",
    related        = "ATLAS-TDR-2010-19-addm",
    relatedstring  = "Addendum:",
}

@Booklet{ATLAS-CONF-2014-018,
    author         = "{ATLAS Collaboration}",
    title          = "{Tagging and suppression of pileup jets with the ATLAS detector}",
    howpublished   = "{ATLAS-CONF-2014-018}",
    url            = "https://cds.cern.ch/record/1700870",
    year           = "2014",
}

@Booklet{ATLAS-CONF-2018-001,
    author         = "{ATLAS Collaboration}",
    title          = "{Measurement of \(b\)-tagging efficiency of \(c\)-jets in \(t\bar{t}\) events using a likelihood approach with the ATLAS detector}",
    howpublished   = "{ATLAS-CONF-2018-001}",
    url            = "https://cds.cern.ch/record/2306649",
    year           = "2018",
}

@Booklet{ATLAS-CONF-2018-006,
    author         = "{ATLAS Collaboration}",
    title          = "{Calibration of light-flavour \(b\)-jet mistagging rates using ATLAS proton--proton collision data at \(\sqrt{s} = 13~\text{TeV}\)}",
    howpublished   = "{ATLAS-CONF-2018-006}",
    url            = "https://cds.cern.ch/record/2314418",
    year           = "2018",
}

@Booklet{ATLAS-CONF-2019-021,
    author         = "{ATLAS Collaboration}",
    title          = "{Luminosity determination in \(pp\) collisions at \(\sqrt{s} = 13\,\text{TeV}\) using the ATLAS detector at the LHC}",
    howpublished   = "{ATLAS-CONF-2019-021}",
    url            = "https://cds.cern.ch/record/2677054",
    year           = "2019",
}

@Booklet{ATLAS-CONF-2020-031,
    author         = "{ATLAS Collaboration}",
    title          = "{Search for bottom-squark pair production in \(pp\) collision events at \(\sqrt{s} = 13\,\text{TeV}\) with hadronically decaying tau leptons, \(b\)-jets and missing transverse momentum using the ATLAS detector}",
    howpublished   = "{ATLAS-CONF-2020-031}",
    url            = "https://cds.cern.ch/record/2725737",
    year           = "2020",
}

@Booklet{ATL-PHYS-PUB-2014-021,
    author         = "{ATLAS Collaboration}",
    title          = "{ATLAS Pythia~8 tunes to \(7~\text{TeV}\) data}",
    howpublished   = "{ATL-PHYS-PUB-2014-021}",
    url            = "https://cds.cern.ch/record/1966419",
    year           = "2014",
}

@Booklet{ATL-PHYS-PUB-2016-017,
    author         = "{ATLAS Collaboration}",
    title          = "{The Pythia~8 A3 tune description of ATLAS minimum bias and inelastic measurements incorporating the Donnachie--Landshoff diffractive model}",
    howpublished   = "{ATL-PHYS-PUB-2016-017}",
    url            = "https://cds.cern.ch/record/2206965",
    year           = "2016",
}

@Booklet{ATL-PHYS-PUB-2016-020,
    author         = "{ATLAS Collaboration}",
    title          = "{Studies on top-quark Monte Carlo modelling for Top2016}",
    howpublished   = "{ATL-PHYS-PUB-2016-020}",
    url            = "https://cds.cern.ch/record/2216168",
    year           = "2016",
}

@Booklet{ATL-PHYS-PUB-2017-011,
    author         = "{ATLAS Collaboration}",
    title          = "{Secondary vertex finding for jet flavour identification with the ATLAS detector}",
    howpublished   = "{ATL-PHYS-PUB-2017-011}",
    url            = "https://cds.cern.ch/record/2270366",
    year           = "2017",
}

@Booklet{ATL-PHYS-PUB-2017-013,
    author         = "{ATLAS Collaboration}",
    title          = "{Optimisation and performance studies of the ATLAS \(b\)-tagging algorithms for the 2017-18 LHC run}",
    howpublished   = "{ATL-PHYS-PUB-2017-013}",
    url            = "https://cds.cern.ch/record/2273281",
    year           = "2017",
}

@Booklet{ATL-PHYS-PUB-2018-025,
    author         = "{ATLAS Collaboration}",
    title          = "{Topological \(b\)-hadron decay reconstruction and identification of \(b\)-jets with the JetFitter package in the ATLAS experiment at the LHC}",
    howpublished   = "{ATL-PHYS-PUB-2018-025}",
    url            = "https://cds.cern.ch/record/2645405",
    year           = "2018",
}
 
\clearpage
 
\begin{flushleft}
\hypersetup{urlcolor=black}
{\Large The ATLAS Collaboration}

\bigskip

\AtlasOrcid[0000-0002-6665-4934]{G.~Aad}$^\textrm{\scriptsize 101}$,    
\AtlasOrcid[0000-0002-5888-2734]{B.~Abbott}$^\textrm{\scriptsize 128}$,    
\AtlasOrcid[0000-0002-7248-3203]{D.C.~Abbott}$^\textrm{\scriptsize 102}$,    
\AtlasOrcid[0000-0002-2788-3822]{A.~Abed~Abud}$^\textrm{\scriptsize 36}$,    
\AtlasOrcid[0000-0002-1002-1652]{K.~Abeling}$^\textrm{\scriptsize 53}$,    
\AtlasOrcid[0000-0002-2987-4006]{D.K.~Abhayasinghe}$^\textrm{\scriptsize 93}$,    
\AtlasOrcid[0000-0002-8496-9294]{S.H.~Abidi}$^\textrm{\scriptsize 29}$,    
\AtlasOrcid[0000-0002-8279-9324]{O.S.~AbouZeid}$^\textrm{\scriptsize 40}$,    
\AtlasOrcid{N.L.~Abraham}$^\textrm{\scriptsize 156}$,    
\AtlasOrcid[0000-0001-5329-6640]{H.~Abramowicz}$^\textrm{\scriptsize 161}$
% ...
\bigskip
\\

% ...
$^{29}$Physics Department, Brookhaven National Laboratory, Upton NY; United States of America.\\
% ...
$^{36}$CERN, Geneva; Switzerland.\\
% ...
$^{40}$Niels Bohr Institute, University of Copenhagen, Copenhagen; Denmark.\\
% ...
$^{53}$II. Physikalisches Institut, Georg-August-Universit\"{a}t G\"ottingen, G\"ottingen; Germany.\\
% ...
$^{93}$Department of Physics, Royal Holloway University of London, Egham; United Kingdom.\\
% ...
$^{101}$CPPM, Aix-Marseille Universit\'e, CNRS/IN2P3, Marseille; France.\\
$^{102}$Department of Physics, University of Massachusetts, Amherst MA; United States of America.\\
% ...
$^{128}$Homer L. Dodge Department of Physics and Astronomy, University of Oklahoma, Norman OK; United States of America.\\
% ...
$^{156}$Department of Physics and Astronomy, University of Sussex, Brighton; United Kingdom.\\
% ...
$^{161}$Raymond and Beverly Sackler School of Physics and Astronomy, Tel Aviv University, Tel Aviv; Israel.\\
% ...

\end{flushleft}

 
\clearpage

\end{document}